\documentclass[twocolumn]{aastex631}

\newcommand{\LAGN}{\citetalias{Lebowitz2025}}

\usepackage{xcolor}
\usepackage{graphicx}
\usepackage{hyperref}
\usepackage{amsmath}
\usepackage{txfonts}

\begin{document}
\defcitealias{Lebowitz2025}{L25} 

\title{Spatially Resolved AGN Ionization and Star Formation at Cosmic Noon with JWST/JEMS}

\author[0009-0002-0710-194X]{Sophie Lebowitz}
\affiliation{Steward Observatory, University of Arizona, 933 N. Cherry Ave., Tucson, AZ 85721, USA}

\author[0000-0003-4565-8239]{Kevin N.\ Hainline}
\affiliation{Steward Observatory, University of Arizona, 933 N. Cherry Ave., Tucson, AZ 85721, USA}

\author[0000-0002-0000-2394]{St\'{e}phanie Juneau}
\affiliation{NSF NOIRLab, 950 N. Cherry Ave., Tucson, AZ, 85719, USA}

\author[0000-0003-2919-7495]{Christina C.\ Williams}
\affiliation{NSF NOIRLab, 950 N. Cherry Ave., Tucson, AZ, 85719, USA}
\affiliation{Steward Observatory, University of Arizona, 933 N. Cherry Ave., Tucson, AZ 85721, USA}

\author[0000-0002-5854-7426]{Swayamtrupta Panda}
\altaffiliation{Gemini Science Fellow}
\affiliation{International Gemini Observatory/NSF NOIRLab, Casilla 603, La Serena, Chile\\}

\author[0000-0002-6221-1829]{Jianwei Lyu}
\affiliation{Steward Observatory, University of Arizona, 933 N. Cherry Ave., Tucson, AZ 85721, USA}

\author[0000-0003-0695-4414]{Michael V.\ Maseda}
\affiliation{Department of Astronomy, University of Wisconsin-Madison, 475 N. Charter St. Madison, WI 53706, USA}

\author[0000-0002-8224-4505]{Sandro Tacchella}
\affiliation{Kavli Institute for Cosmology, Cambridge, Madingley Road, Cambridge CB3 0HA, United Kingdom}

\author[0000-0003-3307-7525]{Yongda Zhu}
\affiliation{Steward Observatory, University of Arizona, 933 N. Cherry Ave., Tucson, AZ 85721, USA}

\author{Jessica L.\ Aguayo}
\affiliation{Steward Observatory, University of Arizona, 933 N. Cherry Ave., Tucson, AZ 85721, USA}

\begin{abstract}
At Cosmic Noon ($z\approx 2-3$), both star formation and Active Galactic Nuclei (AGN) activity peaked, each playing a significant role in ionizing interstellar gas on galaxy-wide scales. The spatial distribution of this ionized gas provides a direct probe of how AGN and stellar ionization shape the gaseous reservoirs of their host galaxies. Using JWST/NIRCam imaging from the JWST Extragalactic Medium-band Survey (JEMS) we spatially map two complementary tracers of ionized gas, [\ion{O}{3}]$+\mathrm{H}\beta$ and Pa$\beta$, in $\sim200$ galaxies at $2.5 < z < 2.9$. We apply multiwavelength AGN diagnostics to divide the sample into AGN hosts (33 galaxies), Pa$\beta$-detected systems (32 galaxies), and control objects (175 galaxies). We measure the [\ion{O}{3}]$+\mathrm{H}\beta$ and Pa$\beta$ spatial extents in each population and relate them to AGN and host properties derived from Spectral Energy Distribution (SED) modeling. Both tracers exhibit systematically larger maximum radial extents in AGN hosts than in control galaxies (by $\sim0.3$ dex), with [\ion{O}{3}]$+\mathrm{H}\beta$ emission modestly more extended than Pa$\beta$ by $\sim0.1$ dex. With this statistically robust AGN sample, we measure the [\ion{O}{3}]$+\mathrm{H}\beta$ radial extent–AGN luminosity relation at $z\sim3$ and derive a slope of $\sim0.2$, consistent with the shallow end of values reported at low redshift. The larger ionized gas extents among AGN hosts relative to the control sample, combined with the strong correlation between [\ion{O}{3}]$+\mathrm{H}\beta$ extent and AGN luminosity suggest that AGN activity may dominate gas ionization in galaxies with mixed AGN and star-forming activity at Cosmic Noon, although stellar processes can still contribute significantly on kiloparsec scales.
\end{abstract}

\section{Introduction} \label{sec:intro}
The period $2-3$ billion years after the Big Bang, commonly called Cosmic Noon ($z \sim 2$–3), was a pivotal era in galaxy evolution, during which the most massive galaxies seen in the local Universe assembled the bulk of their stellar mass. During this epoch, both star formation and black hole accretion peaked, with galaxies forming roughly half of their present-day stellar content \citep{Madau2014}. Following Cosmic Noon, the cosmic star formation rate density has been observed to decline sharply, approximately as $\propto (1+z)^{2.7}$, as galaxies rapidly transitioned toward quiescence by $z = 0$ \citep{Schreiber2020}. Albeit with a different normalization, the shape of the evolution of black hole accretion closely tracks the cosmic star formation history \citep{Madau2014, Heckman&Best2014, Fiore2017}. This shared trend implies a strong connection between star formation and actively accreting supermassive black holes, or Active Galactic Nuclei (AGN), with both processes coupled to the gaseous reservoirs of their host galaxies.

At these redshifts, abundant cold gas reservoirs, elevated accretion rates, and substantial dust obscuration give rise to systems where AGN activity and vigorous star formation frequently co-occur \citep{Hickox&Alexander2018, Feltre2016}. The intense radiation produced by both massive stars and AGN can heat, ionize, and/or disrupt the surrounding gas, altering the physical conditions of the interstellar and circumgalactic gas across a wide range of temperature ($10~\mathrm{K}-10^{8}~\mathrm{K}$) and spatial scales  \citep[$1 ~\mathrm{pc}-10^{6}~\mathrm{pc}$;][]{Harrison2017}. Through this radiative coupling between energy sources and the ambient gas, both stellar- and AGN-driven processes have been invoked in simulations as mechanisms for regulating galaxy growth by suppressing star formation \citep{Springel2005, DiMatteo2005, Somerville2015}. When both stellar and AGN-driven processes act on the same gas reservoirs, their effects can overlap spatially and spectrally, often producing similar observational signatures and complicating efforts to disentangle their relative contributions. 

The rest-frame optical [\ion{O}{3}]$\lambda5007$ line is one of the strongest and most commonly detected nebular emission lines, and it has long served as a practical tracer of ionized gas in both AGN and star-forming galaxies \citep{Baldwin1981}. It is a collisionally excited, forbidden transition of doubly ionized oxygen (O$^{++}$) that arises in hot, low-density environments. Because producing O$^{++}$ requires photons with energies above 35.1 eV, [\ion{O}{3}] emission traces regions exposed to relatively hard ionizing radiation fields \citep{Osterbrock2006}. In stellar regions, [\ion{O}{3}] emission traces ionization from hot, massive O stars, operating as the main cooling line in the surrounding star-forming nebulae \citep{Peterson1997}. In AGN-dominated systems, [\ion{O}{3}] emission is primarily produced by ultraviolet photons from the accretion disk that ionize gas on kiloparsec scales \citep{Antonucci1993}. This emission traces the narrow-line region (NLR), named for the narrow emission-line widths ($\sim10^{2}$–$10^{3}$ km s$^{-1}$) observed in AGN spectra \citep{Bennert2002}. Collimation of the ionizing radiation by dust surrounding the disk (often invoked as a torus-like structure) leads to an anisotropic escape of photons, producing a conical or biconical [\ion{O}{3}] morphology that extends in the AGN polar direction, commonly referred to as an AGN ``ionization cone'' or "bicone" \citep{Antonucci1993, Netzer2015}, which can extend to several kpc with sharp edges \citep[e.g.,][]{Durre&Mould2018,Lopez-coba,Juneau2022}.  

In AGN NLRs, since the production of [\ion{O}{3}] photons is closely tied to the strength of the underlying ionizing radiation field from the accretion disk, the radial extent of the NLR provides a natural link between an AGN’s radiative output and its impact on the surrounding host galaxy. At low redshift ($z<0.5$), numerous studies have demonstrated a correlation between the radial extent of [\ion{O}{3}] emission and AGN luminosity, with characteristic NLR sizes of $\sim$2 kpc for low-to-moderate luminosity AGN, increasing to $\sim$20 kpc for the most luminous quasars ($L_{\mathrm{bol}}>10^{46}$ erg s$^{-1}$) \citep{Liu2013a, Liu2014, Hainline2013, Bennert2006, Haineline2014, Husemann2014, Sun2017}. The slope of this relation encodes information about how ionizing radiation interacts with the ambient gas in the NLR: a steeper slope indicates that the maximum NLR extent is set primarily by the ionizing photon density (the so-called ``ionization-bounded’’ regime), whereas a shallower slope suggests that gas availability limits the NLR size (the ``matter-bounded’’ regime) \citep{Bennert2006}. Thus, measuring both ionized gas extents and the slope of the [\ion{O}{3}] extent-luminosity trend at high redshift provides a direct probe of how efficiently AGN radiation interacts with the gas-rich environments of galaxies at Cosmic Noon.

At high redshift, however, existing measurements remain limited by observational biases and small sample sizes, with most studies targeting spatially resolved [\ion{O}{3}] focusing on luminous quasars ($L_{\mathrm{bol}}>10^{46}$ erg s$^{-1}$). These works typically report modest [\ion{O}{3}] spatial extents of $\sim$0.7–3 kpc, yet the quoted sizes often correspond to the radii of AGN-driven outflows rather than the full extent of the [\ion{O}{3}]-emitting gas \citep{Harrison2012, Carniani2015, Lau2024}. Because ionized outflows are generally confined to smaller spatial scales than the quiescent NLR, such measurements may not accurately capture the true NLR size \citep{Fischer2018, Lau2024}. Moreover, most high-redshift studies rely on ground-based IFU observations, which may lack the spatial resolution and surface-brightness sensitivity required to detect extended, low-surface-brightness [\ion{O}{3}] emission. 

\citet{Lebowitz2025} (hereafter \LAGN{}) attempted the first space-based measurement of this relation at $z\sim3$ from NIRCam imaging of nine moderate-luminosity AGN ($L_{\mathrm{bol}}\sim10^{43-45}$ erg s$^{-1}$). These authors found tentative hints of a linear relationship in agreement with what was seen with local AGN, but with a small sample size the scatter was significant. Observational degeneracies further complicate efforts to link ionized gas extents to AGN activity, as stellar processes can also generate extended photoionized regions---particularly at this epoch, when higher star formation rate densities are expected to contribute more to the total gas ionization \citep{Madau2014}. As a result, it remains unclear what dominates the ionization of oxygen at Cosmic Noon, how elevated gas availability and mixed AGN activity and star formation alter the distribution of ionized gas in the host galaxy, and how these extents scale with luminosity relative to low redshift. Addressing these questions requires spatially resolved measurements of ionized gas that are both sensitive to low-surface-brightness emission and applicable to statistically representative samples at Cosmic Noon.

One promising approach is the use of deep medium-band imaging to isolate strong emission lines and spatially map their extent across statistically meaningful samples, as demonstrated in \LAGN{}. The James Webb Space Telescope (JWST) provides a powerful new opportunity to to investigate ionization sources at earlier cosmic epochs in a statistically robust sample \citep{Gardner2023}. In particular, the JWST Extragalactic Medium-band Survey (JEMS) \citep[JEMS;][]{Williams2023}, conducted in the Great Observatories Origins Deep Survey South (GOODS-S) field \citep{Giavalisco2004}, offers the medium-band filter coverage and observation depth required to spatially map [\ion{O}{3}]$+\mathrm{H}\beta$ emission at Cosmic Noon ($z \sim 3$). Importantly, this same dataset allows concurrent coverage of an additional ionized gas tracer: in the redshift range $z = 2.5–2.9$, the hydrogen recombination line Paschen $\beta$ (Pa$\beta$) can be simultaneously observed, providing an ancillary probe of ionized gas at this epoch.

As a near-IR hydrogen recombination line, Pa$\beta$ traces ionizing ($\mathrm{E}>13.6~\mathrm{eV}$) radiation produced by massive OB stars and/or AGN, and the line is relatively insensitive to dust attenuation, making it a valuable tracer of ionized gas and star formation in dust-enshrouded environments \citep{Cleri2022, Lamperti2017, Reddy2023}. Compared to the collisionally excited [\ion{O}{3}]$\lambda5007$ line, Pa$\beta$ is intrinsically weaker because hydrogen recombination emission is distributed among many transitions. However, Pa$\beta$ may reveal compact, high-surface-brightness regions associated with dense ionized gas reservoirs. Joint mapping of the [\ion{O}{3}]$+\mathrm{H}\beta$ and Pa$\beta$ emission therefore provides complementary constraints on AGN- and stellar-ionized gas that trace different physical conditions, enabling a more complete assessment of how these processes shape the spatial distribution of ionized gas across host galaxy environments.

In this paper, we use JWST/NIRCam observations from JEMS to spatially map the [\ion{O}{3}]$+\mathrm{H}\beta$ and Pa$\beta$ emission in a large sample of $\sim200$ Cosmic Noon galaxies at $z = 2.5–2.9$. While imaging alone cannot uniquely identify the dominant ionization mechanism, it enables robust measurements of the morphology and spatial extent of ionized gas. To help isolate the roles of AGN- and stellar-driven ionization, we apply multi-wavelength AGN diagnostics to divide the sample into AGN hosts, Pa$\beta$-detected systems, and control galaxies. We then measure the [\ion{O}{3}]$+\mathrm{H}\beta$ and Pa$\beta$ spatial extents across these populations and relate them to AGN and host-galaxy luminosities derived from spectral energy distribution (SED) fitting. By combining spatial diagnostics with photometric and host-galaxy properties, we link the observed emission to its underlying physical origin.

We describe the data and sample selection in Section \ref{section:data}. In Section \ref{sec:methods}, we outline our methodology, including the construction of emission-line maps from NIRCam observations, measurements of ionized gas spatial extents, spectral energy distribution (SED) modeling, and AGN identification. The results of our analysis, including the [\ion{O}{3}]$+\mathrm{H}\beta$ and Pa$\beta$ spatial extents and the associated observed and derived galaxy properties, are presented in Section \ref{sec:results}. In Section \ref{sec:discussion}, we discuss the implications of these measurements and place our findings in the context of both low- and high-redshift studies. We summarize our main conclusions in Section \ref{sec:conclusion}. Throughout this paper, we adopt a standard $\Lambda$CDM cosmology with $H_{0} = 70$ km s$^{-1}$ Mpc$^{-1}$, $\Omega_{\mathrm{m}} = 0.3$, and $\Omega_{\Lambda} = 0.7$.

\section{Data} \label{section:data}
To implement the medium-band imaging strategy described above and spatially map ionized gas at Cosmic Noon, we used JWST/NIRCam observations of $\sim$200 galaxies at $z\sim3$ in the Great Observatories Origins Deep Survey South (GOODS-S) field \citep{Giavalisco2004}, taken as part of the JWST Extragalactic Medium-band Survey \citep[JEMS;][]{Williams2023}. JEMS observations use NIRCam medium-band filters (F182M, F210M, F430M, F460M, and F480M) covering a 15.6 arcmin$^{2}$ field-of-view in the Hubble Ultra Deep Field, reaching a $5\sigma$ point-source limit (AB mag) of $\sim29.3-29.4$ in the $2\mu m$ filters and $\sim28.2-28.7$ in the $4\mu m$ filters \citep{Williams2023}. This analysis also benefits from additional deep medium- and wide-filter NIRCam observations from the First Reionization Epoch Spectroscopically Complete Observations \citep[FRESCO, ][]{Oesch2023}, the JWST Advanced Deep Extragalactic Survey (JADES) \citep{Hainline2023, Rieke2023b, Eisenstein2023a, Eisenstein2023b}, and MIRI observations from the Systematic Mid-infrared Instrument (MIRI) Legacy Extragalactic Survey \citep[SMILES, ][]{Alberts2024}. In the JADES data products, the F182M and F210M mosaics incorporate co-added imaging from both JADES and FRESCO, although the latter is comparatively shallower. Additional data from JWST/MIRI \citep{Alberts2026}, HST/ACS \citep{Whitaker2019}, Chandra 7Ms (SNR$>5$ in the soft ($0.5-2.0$ keV) or hard ($2.0-7.0$ keV) bands) \citep{Luo2017}, and the VLA at 3 and 6 GHz \citep{Alberts2020, Lyu2022} enables robust modeling of the Spectral Energy Distributions (SED) of our sample and the derivation of AGN and host galaxy properties (see Section \ref{subsec:SED_modeling}).

\begin{figure}
        \centering
        \includegraphics[width=0.48\textwidth]{"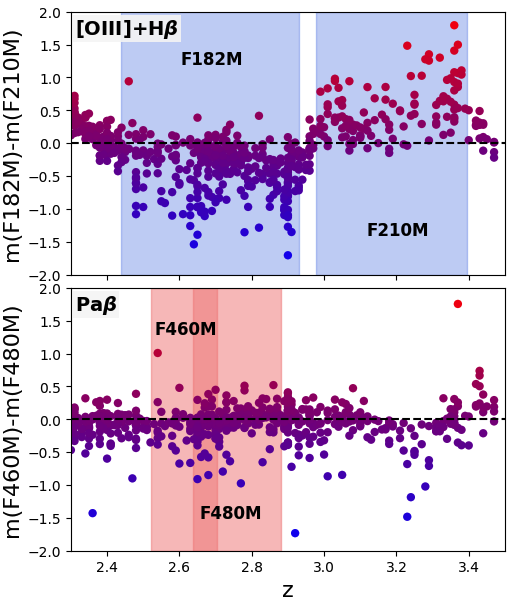"}
        \caption{Color-redshift plots showing $m(\mathrm{F182M}) - m(\mathrm{F210M})$ (top) and $m(\mathrm{F460M}) - m(\mathrm{F480M})$ (bottom) colors for objects in JEMS as a function of redshift. Points are colored by their y-axis values, with red corresponding to more positive values and blue to more negative values. This demonstrates that [\ion{O}{3}]$+\mathrm{H}\beta$ and Pa$\beta$ can be targeted simultaneously in the range $2.52 < z < 2.88$.}
        \label{fig:color-z_plot}
\end{figure}

To ensure that our medium-band observations sample both [\ion{O}{3}]$+\mathrm{H}\beta$ and Pa$\beta$, we compile redshift information for each source using photometric redshifts derived from \texttt{EAZY} template fits to Kron aperture photometry \citep{Brammer2008, Robertson2026, Hainline2026}. To account for uncertainties in the photometric redshift estimates, we adopt values from both the DR5 \citep{Robertson2026} and DR3 \citep{DEugenio2025} catalogs. Spectroscopic redshifts from JADES NIRSpec observations are used when available ($n=41$, where $n$ is the number of objects) \citep{Curtis-Lake2025, Scholtz2025, Robertson2026, DEugenio2025}. Using these redshift measurements, we select sources within the range $2.52 < z < 2.88$, where [\ion{O}{3}]$+\mathrm{H}\beta$ and Pa$\beta$ fall within the NIRCam medium-band filters F182M and F460M/F480M, respectively, as used in JEMS. The effectiveness of this filter configuration across the adopted redshift window is illustrated in Figure \ref{fig:color-z_plot}, where color excess appears as bluer colors in F182M and F460M and redder colors in F210M and F480M. To construct emission-line maps, we require a signal-to-noise ratio of $\mathrm{SNR} > 5$ on the integrated photometry in each of the relevant filters (F182M for [\ion{O}{3}]$+\mathrm{H}\beta$, and F460M and F480M for Pa$\beta$), resulting in an initial sample of 307 objects.

We then construct [\ion{O}{3}]$+\mathrm{H}\beta$ and Pa$\beta$ maps for each object and confirm their detections by performing a visual inspection according to the procedure detailed in Sections \ref{subsec:emission_maps} and \ref{subsec:size_measurements}. Given that Pa$\beta$ is intrinsically weaker than [\ion{O}{3}], we require only a visual detection of [\ion{O}{3}]$+\mathrm{H}\beta$ for an object to be included in the sample, while objects with visually identified Pa$\beta$ emission form a smaller subset, hereafter referred to as the Pa$\beta$-detected sample. After applying the initial SNR and redshift cuts, we removed objects lacking significant [\ion{O}{3}]$+\mathrm{H}\beta$ emission based on visual inspection ($n=73$), as well as objects with contamination from a nearby point source ($n=2$), duplicate objects ($n=15$), and unresolved point sources ($n=2$). Finally, we excluded six additional objects for which spectroscopic redshifts from NIRSpec or HST grism observations place [\ion{O}{3}]$+\mathrm{H}\beta$ outside the F182M bandpass. This yields a final sample of 208 galaxies, of which 32 comprise the Pa$\beta$-detected sample.

\section{Methods} \label{sec:methods}
To quantify the spatial distribution of ionized gas and assess its connection to AGN activity and host galaxy properties, we adopt the following methodology. In Section \ref{subsec:emission_maps}, we detail our procedure for constructing [\ion{O}{3}]$+\mathrm{H}\beta$ and Pa$\beta$ maps from NIRCam medium-band imaging. We describe our methods for measuring the spatial extents of these emission lines in Section \ref{subsec:size_measurements}. In Section \ref{subsec:SED_modeling}, we describe our derivation of AGN and host galaxy parameters via SED modeling. In Section \ref{subsec:AGN_selection}, we discuss our AGN selection methodology. 

\subsection{Emission-line Maps} \label{subsec:emission_maps}
To investigate the morphologies and spatial extents of ionized emission, we construct continuum-subtracted emission-line maps using pairs of medium-band filters. For each target emission line, one filter is chosen to overlap the expected wavelength of the line, while a neighboring filter samples the adjacent continuum. Subtracting the continuum image isolates the flux from the targeted emission line. 

\begin{figure}
        \centering
        \includegraphics[width=0.48\textwidth]{"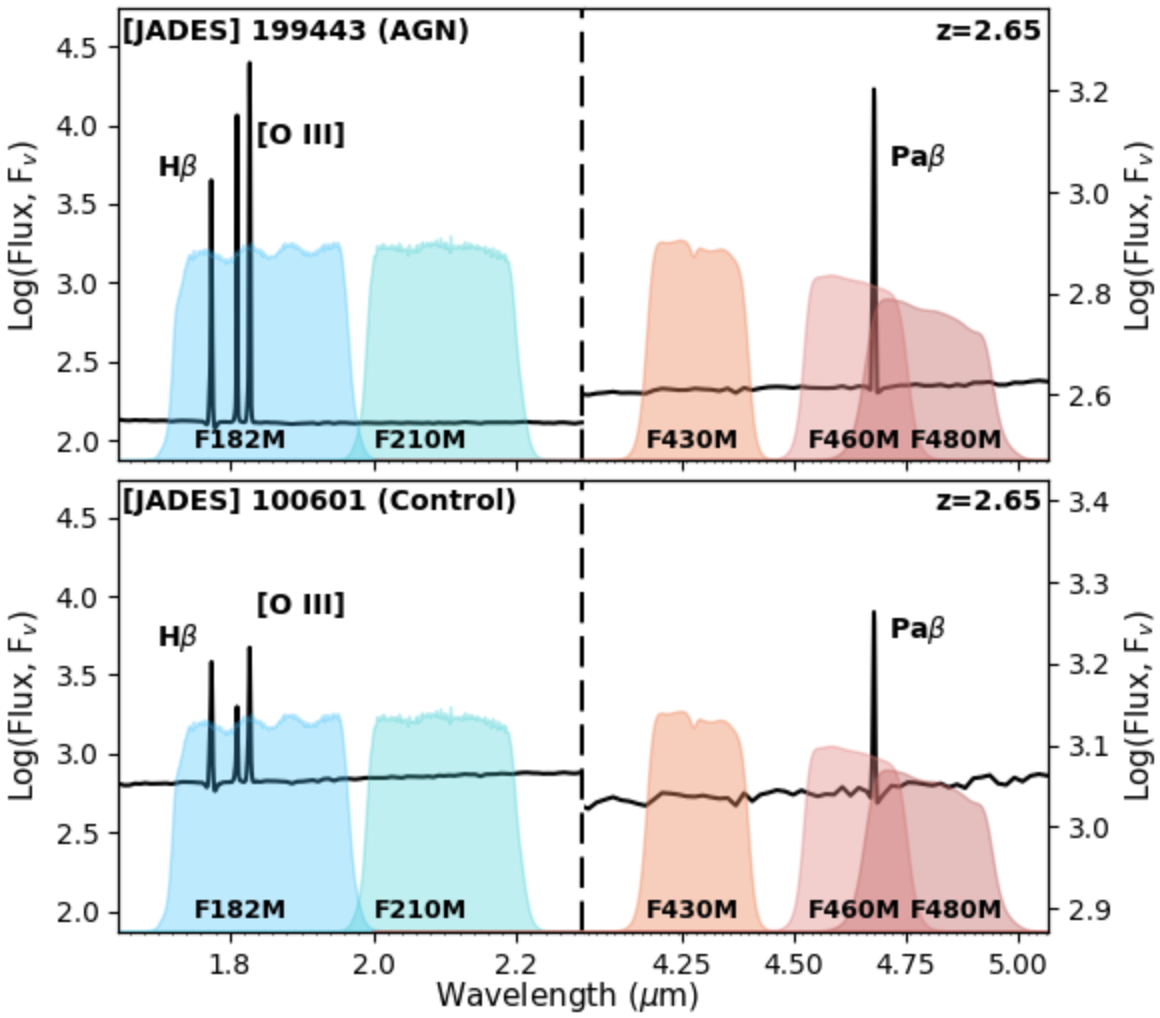"}
        \caption{Example \texttt{EAZY} (left panels) and CIGALE (right panels) SEDs for an AGN and control sample galaxy at $z=2.7$ with NIRCam F182M, F210M, F430M, F460M, and F480M filter curves overlaid. At this redshift, [\ion{O}{3}]$+\mathrm{H}\beta$ falls into F182M, while Pa$\beta$ is targeted by F460M. The adjacent F210M and F430M filters capture the continuum emission without contamination from other lines.}
        \label{fig:spec_w_filter}
\end{figure}

\begin{figure*}[t]
        \centering
        \includegraphics[width=0.9\textwidth]{"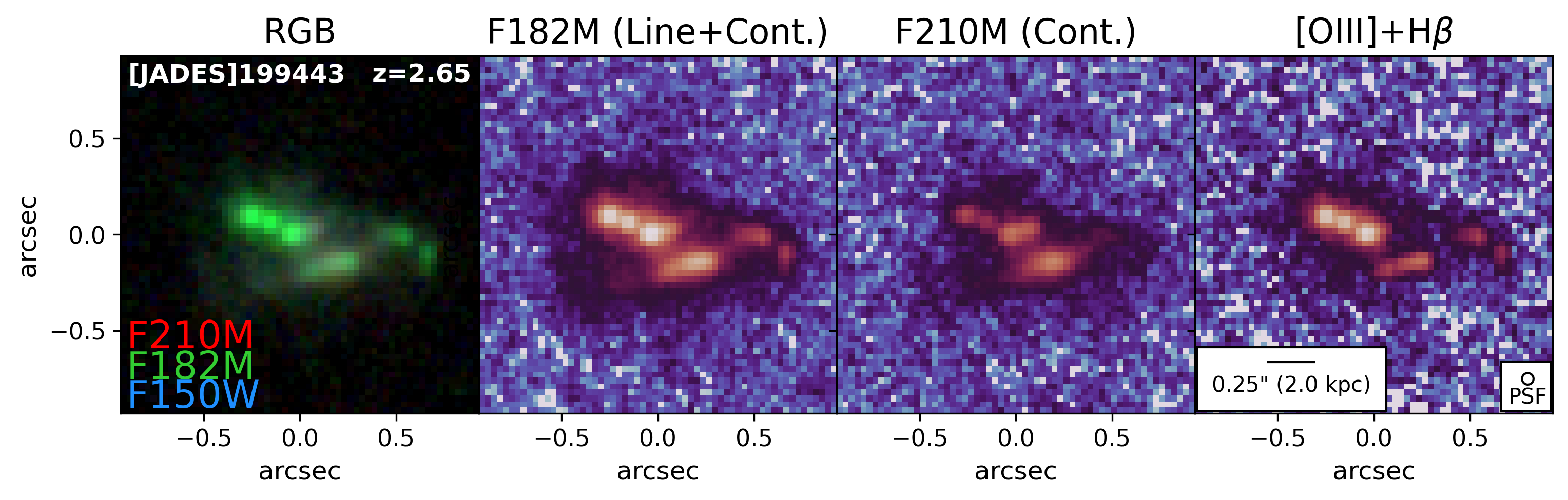"}
        \includegraphics[width=0.9\textwidth]{"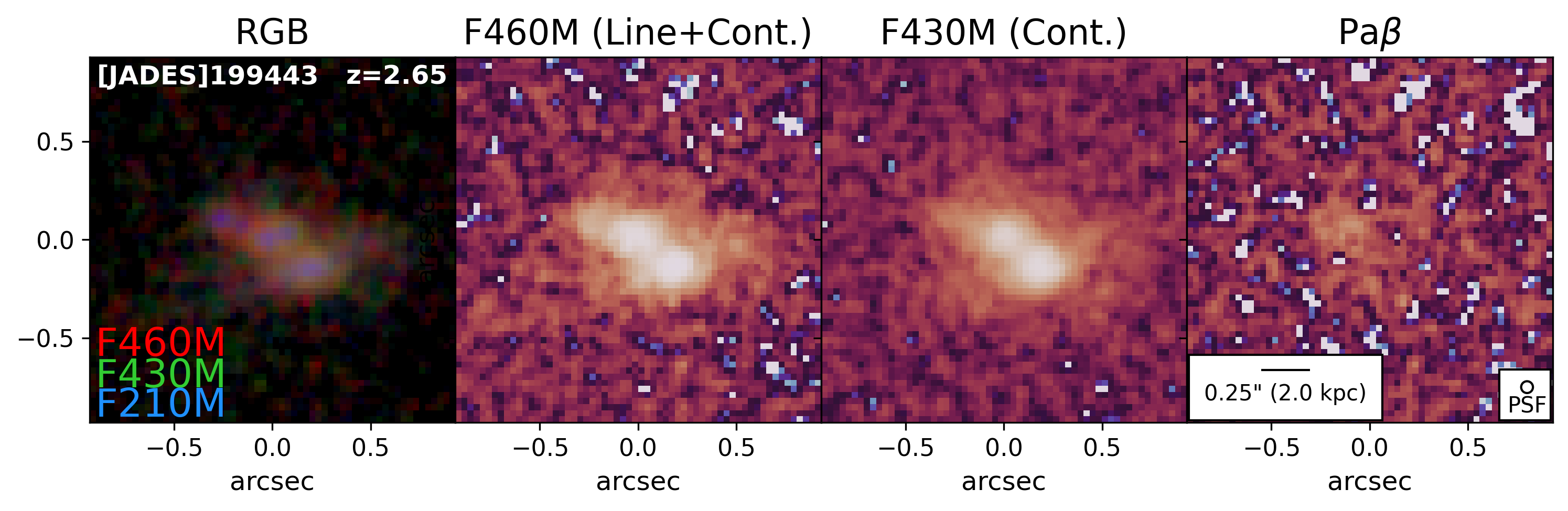"}
        \caption{JWST/NIRCam observations of [JADES]199443, an AGN at $\mathrm{z}=2.65$. Top Panel: Shows the [\ion{O}{3}]$+\mathrm{H}\beta$ emission. First column: RGB image with [\ion{O}{3}]$+\mathrm{H}\beta$ emission represented by the green color. Second column: F182M image showing both emission-line $+$ continuum emission. Third column: F210M continuum image. Fourth column: Continuum subtracted [\ion{O}{3}]$+\mathrm{H}\beta$ map. Bottom Panel: Shows the Pa$\beta$ emission. First column: RGB image with Pa$\beta$ emission represented by the red color. Second column: F460M image showing both emission-line $+$ continuum emission. Third column: F430M continuum image. Fourth column: Continuum subtracted Pa$\beta$ map.}
        \label{fig:Emission-line_maps}
\end{figure*}

For the [\ion{O}{3}]$+\mathrm{H}\beta$ maps, we subtract the F210M image from F182M. At $z=2.5-2.9$, F182M encompasses both the [\ion{O}{3}] and H$\beta$ emission lines, which are not spectrally resolved at the width of the medium-band filter. We therefore treat the combined flux as a single [\ion{O}{3}]$+\mathrm{H}\beta$ component. We adopt an analogous procedure for Pa$\beta$. To demonstrate that these filter pairs effectively sample the target emission lines and adjacent continuum without contamination from other lines, we show in Figure \ref{fig:spec_w_filter} example \texttt{EAZY} SEDs for an AGN (top panel) and a control sample galaxy (bottom panel), with the NIRCam F182M, F210M, F430M, F460M, and F480M filter curves overlaid. For galaxies with $z<2.67$, Pa$\beta$ falls in F460M, while for galaxies with $z>2.67$, the line shifts into F480M. In both cases, we subtract F430M to remove the continuum. The paired filters are sufficiently close in wavelength that their NIRCam PSFs are effectively identical, and we therefore do not apply PSF matching prior to subtraction. For visualization purposes, we also constructed RGB images using F210M/F182M/F150W for [\ion{O}{3}]$+\mathrm{H}\beta$ and F460M/F430M/F210M or F480M/F460M/F430M for Pa$\beta$, depending on redshift. In Figure \ref{fig:Emission-line_maps}, we display an example of the RGB images, filter images, and emission-line maps for an AGN in our sample (ID $=199443$), showing the [\ion{O}{3}]$+\mathrm{H}\beta$ emission in the top panel and Pa$\beta$ emission in the bottom panel.

To ensure accurate continuum subtractions, we also include a scaling factor, $f$, to account for the slope of the continuum. For the continuum normalizations applied to F210M, the $f$-factor is derived from the best-fitting EAZY template photometry for each source \citep{Hainline2023}. However, we find that EAZY does not robustly model the rest-frame near-infrared regime, and in particular does not accurately capture the Pa$\beta$ emission. We instead derive $f$ for the F430M continuum normalizations from the CIGALE (Code Investigating GALaxy Emission; \citealt{Boquien}) SED fits, which include a more complete treatment of nebular emission, specifically for Pa$\beta$-detected objects. Details of the CIGALE models and parameter choices will be presented in Section \ref{subsec:SED_modeling}. These scaling factors are calculated by taking the ratio of the estimated average continuum flux through the filter targeting the emission line, measured in narrow wavelength windows on either side of the emission line, to the average flux measured in the filter sampling the continuum emission. This scaling factor is then applied across the pixel fluxes of the continuum image during the subtraction step (e.g. $\mathrm{F}_{\lambda}(\mathrm{F182M})-f \times \mathrm{F}_{\lambda}(\mathrm{F210M})$). Across our sample, the median scaling factor applied to F210M is $f=1.03$ ($16^{th}–84^{th}$ percentile: $0.95$–$1.05$) and $f=1.05$ ($16^{th}–84^{th}$ percentile: $1.02$–$1.11$) for F430M, indicating only modest continuum slope corrections between the paired filters. The sensitivity of measured ionized gas extents to variations in $f$ was previously evaluated in \LAGN{}, where a $10\%$ change in the scaling factor altered the median radius by only $\sim5\%$, demonstrating that the inferred extents are weakly sensitive to uncertainties in the continuum correction. Next, we present our methods for measuring the spatial extents of the resulting emission-line maps.

\subsection{Size Measurements of Ionized Gas} \label{subsec:size_measurements}
We measure the radial extent of ionized gas using a fixed surface brightness threshold applied to the continuum-subtracted emission-line maps. While many low-redshift studies have defined NLR sizes using the $10^{-15}/(1+z)^4$ erg s$^{-1}$ cm$^{-2}$ arcsec$^{-2}$ isophote \citep{Liu2014, Hainline2013, Haineline2014, Sun2017}, \LAGN{} showed that at $z\sim3$ this threshold approaches the residual background uncertainty in their medium-band NIRCam maps (constructed from the same data and reductions used in this work), leading to false detections of extended emission. Instead, a limiting surface brightness corresponding to the $3\sigma$ median noise level of the sample was adopted, calculated from the standard deviation of $15 \times 15$ pixel corner regions in each emission-line map. In this work, we tested the same approach but found that a $3\sigma$ threshold was not sufficiently conservative for the largest objects, resulting in false-positive detections of background fluctuations. Conversely, a $5\sigma$ threshold was too restrictive to robustly recover extended emission in our more compact sources. To balance the detection of both extended and compact sources, we adopt a uniform $4\sigma$ threshold across our sample. For the [\ion{O}{3}]$+\mathrm{H}\beta$ maps, this corresponds to a limiting surface brightness of SB$_{\mathrm{limit}} = 8.7\times10^{-17}$ erg s$^{-1}$ cm$^{-2}$, and SB$_{\mathrm{limit}} = 2.0\times10^{-17}$ erg s$^{-1}$ cm$^{-2}$ for the Pa$\beta$ maps. 

To measure the spatial extents of the [\ion{O}{3}]$+\mathrm{H}\beta$ and Pa$\beta$ emission, we convert the emission line maps to surface brightness units and mask any pixel below the adopted surface brightness thresholds. We note that many of our sources have discrete regions of ionized emission. To accurately quantify their spatial distribution, we measure the physical properties of each discrete ionized feature per object, defined such that one feature must be separated from another by at least one pixel (i.e., no edge- or corner-touching pixels). We note that since features are defined purely by pixel connectivity above the adopted surface brightness threshold, emission regions that exhibit visual substructure or clumpiness are treated as a single feature if they remain connected by at least one above-threshold pixel. Applying the $4\sigma$ surface brightness limits and removing obvious noise detections, we obtain 367 distinct [\ion{O}{3}]$+\mathrm{H}\beta$ features across our full JEMS sample ($n=208$), and 63 Pa$\beta$ features across the Pa$\beta$-detected sample ($n=32$). 

We sought to calculate sizes for each source so that we could compare ionized gas extents across each sample, taking into account that some targets possess more than one discrete feature. All sizes reported in this work are PSF-convolved (i.e., no PSF correction is applied). For each discrete ionized feature, we compute the following quantities:
\begin{enumerate}
    \item \textbf{Average radial size}, $R_{\mathrm{avg}}$: the mean distance from the feature centroid to its edge pixels.
    \item \textbf{Maximum radial size}, $R_{\mathrm{max}}$: the maximum distance from the centroid to the furthest edge pixel.
    \item \textbf{Maximum radial extent}, $D_{\mathrm{max}}$: the maximum distance from the galaxy center to the furthest edge pixel.
\end{enumerate}

\begin{figure*}
        \centering
        \includegraphics[width=\textwidth]{"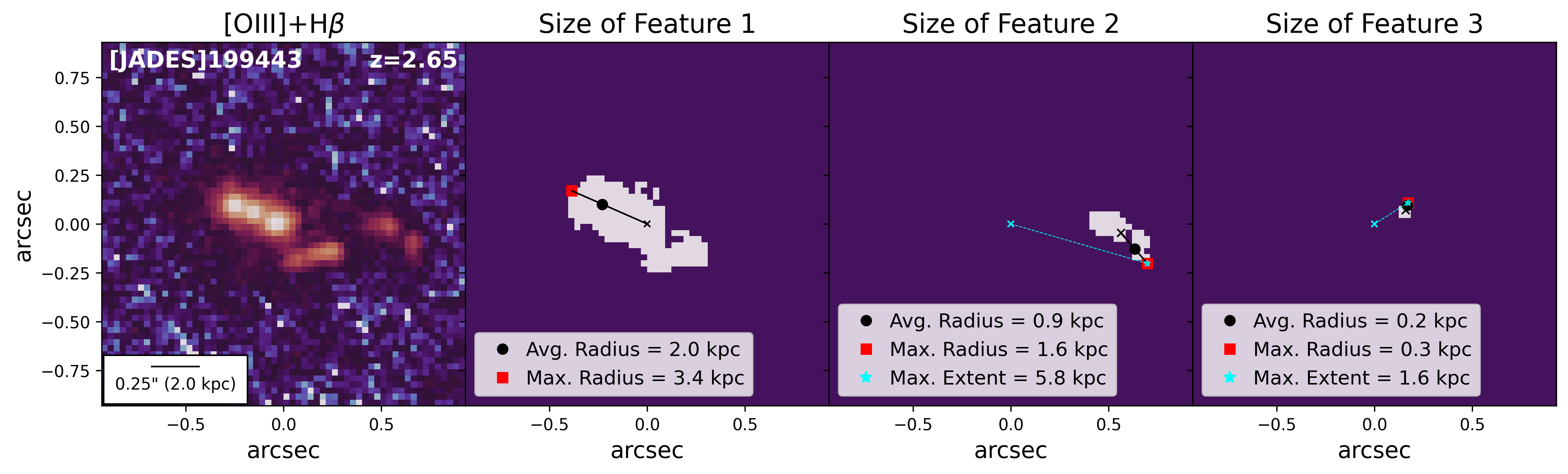"}
        \caption{Example of size calculations for [JADES]199443 ($z=2.65$), an AGN with three discrete [\ion{O}{3}]$+\mathrm{H}\beta$ features at a surface brightness limit of $8.7\times10^{-17}$ erg s$^{-1}$ cm$^{-2}$. The [\ion{O}{3}]$+\mathrm{H}\beta$ emission-line map is shown in the first column and each of the three features are shown in columns 2-4 in order of decreasing area. The average radial size of each features is shown by a black circle and the maximum radial size is shown by a red square. For features that are offset from the galaxy's center, we plot a light blue 'x' at the center and a black 'x' at the feature's centroid. For these offset features, we also plot the maximum extent from the galaxy's center to the furthest edge pixel of the feature represented by a light blue star.}
        \label{fig:Multi_feature_sizes}
\end{figure*}

Throughout this work, size–luminosity relations are constructed using three spatial metrics. For each galaxy, $R_{\mathrm{avg}}$ and $R_{\mathrm{max}}$ are measured from the largest detected emission feature, with $R_{\mathrm{max}}$ capturing potential asymmetric morphology. In contrast, $D_{\mathrm{max}}$ is defined as the maximum distance from the galaxy center to the furthest edge pixel across all detected features, tracing the full spatial extent of the ionized gas. These complementary measurements allow us to compare how the characteristic size of the dominant ionized region and the total extent of the ionized gas evolve with AGN luminosity and host galaxy properties (i.e. galaxy luminosity, SFR, and stellar mass). In Figure \ref{fig:Multi_feature_sizes}, we show an example of these size metrics for an AGN with multiple discrete features. We next describe the SED-fitting methodology used to derive AGN and host-galaxy properties for the full sample.

\subsection{SED Modeling} \label{subsec:SED_modeling}
To model the best-fit SEDs of our JEMS sample we used the SED modeling code, AGNfitter-rx \citep{Rivera2016, Martinez-Ramirez2024}. AGNfitter-rx is a Bayesian MCMC-based code built upon AGNfitter that models the AGN and host galaxy emission from the radio to the X-ray regime. This code uses a combination of theoretical and semi-empirical models to model four physical components of AGN emission: the accretion disk, a dust-obscuring torus, relativistic jets/core, and the X-ray-emitting corona; and three host galaxy components: stellar populations, cold dust emission from starbursts, and stellar radio emission. By using MCMC sampling to explore the full parameter space, AGNfitter fits each emission component independently using a "flexible" energy balance, enabling more effective disentanglement of the nuclear and host contributions.

To model the accretion disk component (i.e. the Big Blue Bump (BBB); \citealt{Malkan1982}), we use the AGNfitter \texttt{THB21} model based on the semi-empirical template from \citet{Temple2021}, which represents the accretion disk emission with a broken power law, models the hot-dust emission at $1-3 \mu m$, and includes both broad and narrow emission lines. To fit the torus component, we use the \texttt{SKIRTOR} template from \citet{Stalevski2016}, which models the torus as a two-phase dust distribution, consisting of a high-density clumpy structure embedded within a smoother, low-density medium. We model the stellar emission from the host galaxy using the \texttt{BC03\_metal} template, which implements the simple stellar population library from \citet{Bruzual2003} and allows for variations in stellar age, metallicity, star formation history (SFH) decay timescales, and line-of-sight dust attenuation according to the reddening law from \citet{Calzetti2000}. We fit the cold-dust component associated with starburst activity using the \texttt{S17} template from \citet{Schreiber2018}, which describes the far-IR emission with a dust temperature that evolves with redshift calibrated on main sequence galaxies. 

As allowed by the data, we also fit the radio and X-ray components for galaxies with VLA detections \citep[SNR~$>5$ at 3 or 6 GHz,][]{Alberts2020} and Chandra 7Ms detections (SNR$>5$ in the soft ($0.5-2.0$ keV) or hard ($2.0-7.0$ keV) bands) \citep{Luo2017}. To model the radio contribution from AGN synchrotron emission, AGNfitter applies a simple power law, $L \propto \nu^{-\alpha}$, when one or two radio bands are provided. For detections in two bands, AGNfitter tests $\alpha$ values over the range $[-2.0, 1.0]$; while for detections in only one band, it assumes a fixed slope of $-0.75$ for non-thermal emission \citep{Baan2006}. The X-ray–emitting corona is modeled using the empirical $\alpha_{\mathrm{ox}}$–$L_{2500\text{\AA}}$ relation from \citet{Lusso2017}, which connects the coronal 2 keV emission to the AGN ultraviolet emission at 2500 $\text{\AA}$. Throughout this analysis, we adopt the default AGNfitter input parameter priors, with the exception of the BBB reddening parameter, which is expanded from [0, 1] to [0, 10] to allow for increased nuclear dust attenuation, consistent with the elevated incidence of dust-obscured AGN at these redshifts. The default model parameters are listed in Table 1 of \citet{Martinez-Ramirez2024}. All AGNfitter runs employ 100 walkers and two burn-in phases, each with a chain length of 25,000 using the \texttt{ultranest} MCMC algorithm with a random mixture of oriented walkers. 

As a cross-check, we compare the parameters derived from AGNfitter’s best-fit SEDs with those obtained using an independent SED-fitting code, the Code Investigating GALaxy Emission (CIGALE, \cite{Boquien}). CIGALE is a galaxy SED-fitting code that applies energy balance principles to derive global physical properties from multiwavelength photometry. CIGALE explores a pre-computed grid of models from user-defined parameter values, selecting the model with the best $\chi^{2}$ value. For this comparison, we adopt a standard set of modules, including a delayed star-formation history (\texttt{sfhdelayed}), stellar population synthesis (\texttt{bc03}), nebular emission (\texttt{nebular}), dust attenuation (\texttt{dustatt\_modified\_starburst}), dust emission (\texttt{dale2014}), and an optional AGN component (\texttt{skirtor2016}). Unlike AGNfitter, CIGALE does not assume the presence of an AGN and will only include an AGN component if AGN models produce a lower $\chi^{2}$ than purely galactic ones, which may cause weak or subtle AGN signatures to be missed.

\begin{figure*}
        \centering
        \includegraphics[width=\textwidth]{"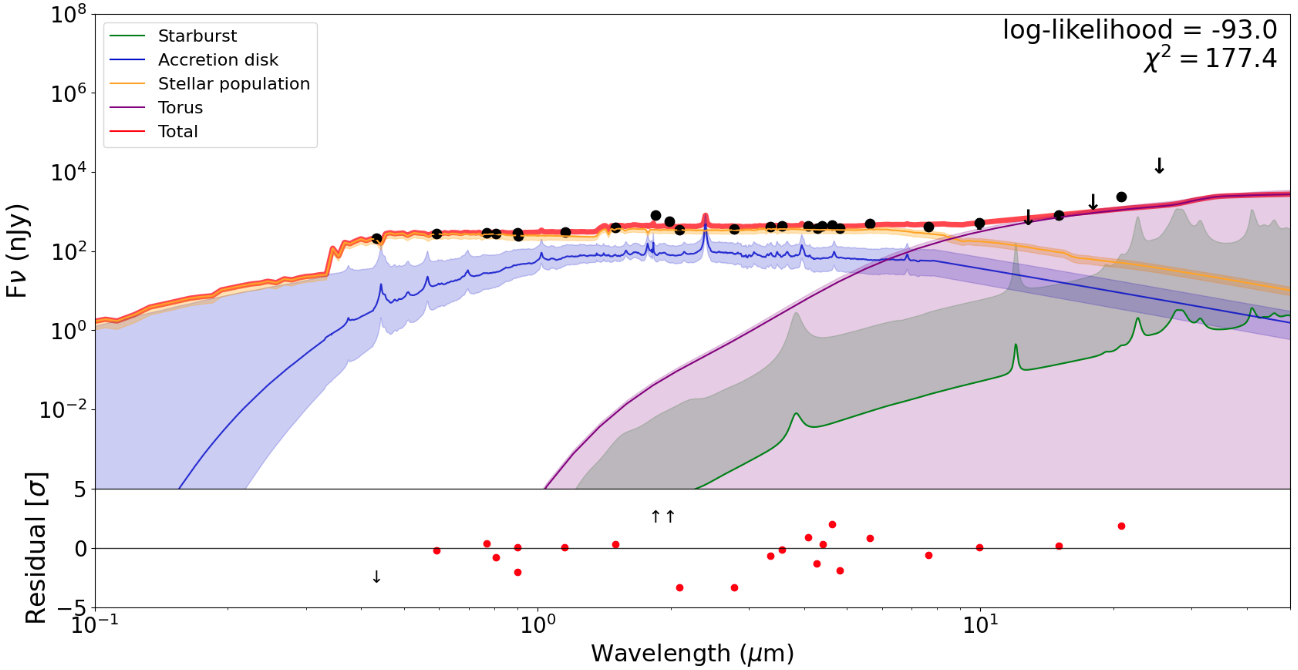"}
        \caption{AGNfitter’s best-fit SED in the observed frame for [JADES]199443 using the AGN and galaxy components described in Section \ref{subsec:SED_modeling}. The cold-dust emission from star formation is shown in green, and the stellar emission is shown in yellow. The accretion-disk component is modeled in blue, and the hot-dust torus emission is shown in purple. The width of each curve reflects the uncertainty in that component. The red line shows the total model fit, with residuals displayed in the bottom panel as red points. The total model is computed by summing the four emission components at each wavelength. Observed photometry is plotted as black points, including medium- and wide-band measurements from HST/ACS, JWST/NIRCam, and JWST/MIRI (downward arrows indicate MIRI upper limits at $\mathrm{SNR}<3$). This object does not have Chandra or VLA observations. For this AGN, the AGNfitter solution favors strong stellar emission and a weak accretion-disk component in the rest-frame UV–optical, while at mid-IR wavelengths the model finds that a prominent torus component best reproduces the MIRI data. We note that an independent SED fit with \texttt{CIGALE} yields consistent results for this object, including a similarly strong torus component at mid-IR wavelengths.}
        \label{fig:Example_SED}
\end{figure*}

To compare the two codes, we performed SED modeling on the full galaxy sample using both AGNfitter-rx and CIGALE, running each with and without an AGN component. We compared the resulting best-fit estimates of the star-formation rate (SFR), stellar mass, galaxy luminosity (attenuated stellar emission integrated over $0.1-1\mu$m), and AGN luminosity (integrated over $0.1-30\mu$m). We find that AGNfitter and CIGALE yield closely consistent estimates of SFR, stellar mass, and galaxy luminosity across the sample, with median offsets of 0.1, -0.1, and 0.1 dex, respectively. For sources where an AGN component is favored, the inferred AGN luminosities show a somewhat larger systematic offset, with AGNfitter predicting a higher median value by 0.4 dex. Despite including AGN templates in its model grid, CIGALE frequently assigns negligible AGN fractions to sources independently identified as AGN through X-ray diagnostics (see Section \ref{subsec:AGN_selection}). Given this behavior, and AGNfitter’s explicit treatment of the radio and X-ray AGN components, we adopt AGNfitter output parameters for deriving the AGN and host-galaxy properties used in this work. We show an example of the AGNfitter best-fit SED for [JADES]199443 in Figure \ref{fig:Example_SED}. 

The SED-derived parameters are then used differently depending on AGN classification (described in the following section). For galaxies classified as non-AGN (the control sample), we adopt SFR, stellar mass, and galaxy luminosity estimates from fits in which the AGN component is forcibly turned off, thereby avoiding potential contamination from nuclear emission. For galaxies identified as AGN, we run the fit with the AGN component enabled and use the derived $0.1-30\mu$m AGN luminosity as a proxy for the AGN bolometric luminosity. Through this methodology, we are able to use the same SED-fitting code for the control and AGN samples, ensuring consistency across the full galaxy sample. We now describe the criteria used to identify AGN within the sample and define the corresponding control population.

\subsection{AGN Selection} \label{subsec:AGN_selection}
To identify AGN within our sample, we combined previously established AGN classifications from the literature with additional multiwavelength diagnostics applied directly to our data. An overview of the selection criteria used to classify our AGN sample is provided in Table \ref{tab:AGN_selection}. We required at least one criterion to be met to classify a source as an AGN. Galaxies that did not satisfy any AGN selection criterion are hereafter referred to as the control sample. This classification is used throughout the remainder of the analysis to compare ionized gas properties between AGN hosts and likely non-AGN systems.

First, we cross-matched our galaxy sample with the GOODS-S pre-JWST and post-JWST AGN catalogs presented by \cite{Lyu2022} and \cite{Lyu2024}. These two catalogs jointly contain $\sim 900$ AGN in the 3D-HST GOODS-S footprint. The post-JWST catalog expands on the AGN selection from the \cite{Lyu2022} by performing SED analysis of MIRI-selected AGN from the SMILES to identify dust-obscured AGN \citep{Alberts2024}. AGN are classified according to nine selection techniques, including: mid-IR SED, mid-IR color type, X-ray luminosity, X-ray-radio luminosity relation, radio loudness, radio slope index, optical spectrum, optical SED, and variability \citep{Lyu2022}. We refer the reader to \citet{Lyu2022} and \citet{Lyu2024} for more details on the AGN selection. Cross-matching our galaxy sample with these AGN catalogs identified 19 AGN. 

We then applied X-ray diagnostics to select AGN missed by these catalogs. We cross-matched our JEMS sample with the Chandra 7Ms GOODS-S catalog from \citet{Luo2017}, identifying 27 X-ray detected objects (SNR$>5$ in the soft ($0.5-2.0$ keV) or hard ($2.0-7.0$ keV) bands). We examined the Chandra $0.5-2/2-7$ keV images to confirm that the x-ray emission is within the positional uncertainty of each source. After completing this check, we then accepted any objects classified as an X-ray AGN from \citet{Luo2017}, which identified AGN according to the following standard X-ray thresholds: X-ray luminosity ($\mathrm{L}_\mathrm{X-ray} \geq 3\times10^{42}$erg s$^{-1}$), X-ray hardness, X-ray-to-optical flux ratio, X-ray-to-radio luminosity ratio, and/or X-ray-to-NIR flux ratio. We identify fifteen AGN based on the X-ray criteria of \citet{Luo2017}, including three sources that were not flagged as AGN in \citet{Lyu2022} or \citet{Lyu2024}, likely reflecting differences in catalog construction and selection criteria, as some AGN may not be identified depending on the adopted methodology.

To identify any additional missed AGN within the sample, we applied a combination of color–color and SED-based diagnostics. We visually inspected the SEDs of all galaxies with an estimated AGN fraction $> 0.5$ from our SED modeling (see Section \ref{subsec:SED_modeling}), as well as those that showed a clear improvement in the CIGALE reduced $\chi^{2}$ when an AGN component was included. We also flagged galaxies with $m(\mathrm{F277W}) - m(\mathrm{F356W}) > 0.15$. At $z=2.5-2.9$, this color probes the rest-frame optical continuum slope and is sensitive to excess red emission that can arise from hot dust associated with AGN activity. We find that many of the confirmed AGN in our sample occupy this region of color–color space. These same sources consistently fall in the upper-right regions of other NIRCam color–color diagrams ($m(\mathrm{F150W}) - m(\mathrm{F200W}) > 0.3$, $m(\mathrm{F200W}) - m(\mathrm{F277W}) > 0.1$, $m(\mathrm{F356W}) - m(\mathrm{F444W}) > 0.1$), reinforcing the interpretation that they exhibit reddened continuum slopes indicative of nuclear activity. Using this combined approach, we identified a subset of 22 possible AGN.

We then compared the best-fit SEDs from both \texttt{CIGALE} and \texttt{AGNfitter-rx}, testing models with and without an AGN component to assess the consistency of the fits. We classified as AGN any sources with a mid-IR SED strongly indicative of torus dust-reprocessing and/or a radio contribution consistent with AGN synchrotron emission. Using these criteria, we identified three new AGN. During this process, we found that ten control-sample galaxies failed to converge when \texttt{AGNfitter-rx} was run without an AGN component. Visual inspection of their positions on the $m(\mathrm{F200W}) - m(\mathrm{F277W})$ versus $m(\mathrm{F277W}) - m(\mathrm{F356W})$ color-color plot showed that eight of these ten objects clustered with the confirmed AGN population, and their [\ion{O}{3}]$+\mathrm{H}\beta$ morphologies were consistent with the presence of an NLR or ionization cone. We therefore classified these eight sources as AGN as well, bringing the final AGN sample to 33 AGN ($16\%$ of the full galaxy sample). The full list of AGN and their derived properties is presented in Table \ref{tab:AGN_props} in the Appendix. We note, however, that AGN identification remains subject to significant limitations even when multiple diagnostics are combined. Consequently, our AGN sample is likely incomplete, and additional AGN may remain undetected within the control population. We now present our results exploring the spatial extent of the [\ion{O}{3}]$+\mathrm{H}\beta$ and Pa$\beta$ emission as a function of AGN and galaxy properties.

\begin{deluxetable*}{c c c c c c c}
\tablecolumns{7}
\tablecaption{AGN selection methods used to identify AGN in this work. An ``x" indicates that a source is classified as an AGN by the corresponding method.} \label{tab:AGN_selection}
\tablehead{JADES ID & Pre-JWST Cat. & Post-JWST Cat. & X-ray Diagnostics & MIR SED & Radio SED & AGNfitter/color-color} 
\startdata
193332 &   &   &   &   &   & x \\
193915 &   & x &   &   &   &   \\
194269 & x &   &   &   &   &   \\
194373 & x &   & x &   &   &   \\
194952 &   &   &   &   &   & x \\
195412 &   &   & x & x &   &   \\
196134 &   &   &   &   &   & x \\
196184 & x & x & x & x & x &   \\
196187 &   &   & x &   &   &   \\
196290 & x & x & x & x & x &   \\
197581 & x &   & x &   &   &   \\
198790 &   &   &   &   &   & x \\
199443 &   & x &   &   &   &   \\
199494 &   & x &   &   &   &   \\
199996 &   &   &   & x & x &   \\
200800 &   &   &   &   &   & x \\
201584 &   &   &   &   &   & x \\
202378 &   & x & x &   & x &   \\
202380 & x &   & x &   &   &   \\
202484 &   & x & x &   &   &   \\
202597 &   &   &   & x &   &   \\
204449 &   &   &   &   &   & x \\
207277 &   & x &   &   &   &   \\
207592 &   & x &   &   &   &   \\
208000 & x & x & x &   &   & x \\
208176 &   &   &   &   &   & x \\
208820 &   &   &   &   & x &   \\
209026 & x & x & x & x & x &   \\
209027 &   & x & x &   &   &   \\
209116 &   &   & x &   &   &   \\
209117 & x &   & x & x & x &   \\
209617 & x &   & x &   &   &   \\
419920 &   &   &   & x &   &   \\
\enddata
\end{deluxetable*}

\section{Results} \label{sec:results}
We present our results on the [\ion{O}{3}]$+\mathrm{H}\beta$ and Pa$\beta$ properties of our JEMS sample at Cosmic Noon. By dividing the full galaxy sample into AGN hosts ($16\%$), control galaxies ($84\%$), and Pa$\beta$-detected systems ($15\%$) (the latter including both AGN and control galaxies), we examine how the morphology and spatial extent of ionized gas vary with nuclear and stellar activity. Differences in spatial structure can provide insight into the dominant ionization mechanisms and the impact of AGN or stellar-driven radiation on their host galaxies. In Section \ref{subsec:OIII}, we summarize the morphologies and spatial extents of the [\ion{O}{3}]$+\mathrm{H}\beta$ emission in each sample. In Section \ref{subsec:Pbeta}, we describe the Pa$\beta$ morphologies and spatial extents of the Pa$\beta$-detected galaxies. Finally, in Section \ref{subsec:Galaxy_props}, we present the observed and derived galaxy properties and compare trends across each sample.

\subsection{[\ion{O}{3}]$+\mathrm{H}\beta$ Morphologies and Spatial Extents} \label{subsec:OIII}
\begin{figure}
        \centering
        \includegraphics[width=0.49\textwidth, trim={0.3cm 0.2cm 0cm 0cm}, clip]{"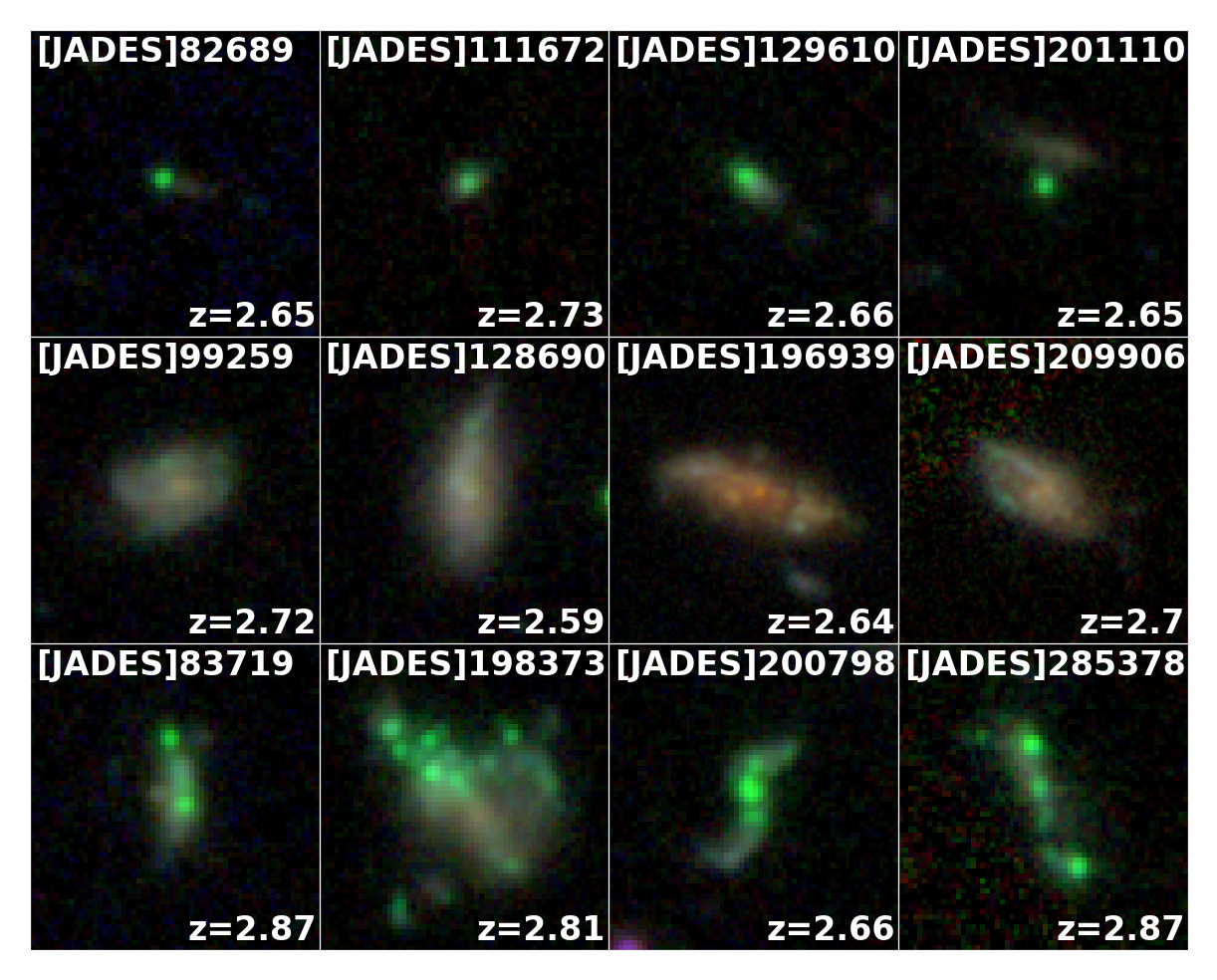"}
        \caption{Montage of RGB (F182M/F210M/F150W) thumbnails showing the [\ion{O}{3}]$+\mathrm{H}\beta$ morphologies (green) for the control sample. The top row show objects with compact, centrally concentrated [\ion{O}{3}]$+\mathrm{H}\beta$ emission. The middle row shows sources with extended [\ion{O}{3}]$+\mathrm{H}\beta$ emission. The bottom row shows sources with [\ion{O}{3}]$+\mathrm{H}\beta$ emission resembling bright knots.}
        \label{fig:Control_thumbnails}
\end{figure}

Both AGN and non-AGN galaxies at Cosmic Noon exhibit a wide range of [\ion{O}{3}]$+\mathrm{H}\beta$ morphologies and spatial extents. In this section, we summarize the observed properties of the [\ion{O}{3}]$+\mathrm{H}\beta$ emission in the AGN, control, and Pa$\beta$-detected samples. Through visual inspection of the continuum-subtracted NIRCam medium-band images, we classify the [\ion{O}{3}]$+\mathrm{H}\beta$ morphologies into four categories: compact, extended, knotted, and (among AGN) conical. These categories are defined qualitatively and are intended to provide interpretive context for the spatial extent measurements, rather than serve as strict quantitative classifications. Representative examples of compact (top row), extended (middle row), and knotted (bottom row) [\ion{O}{3}]$+\mathrm{H}\beta$ morphologies observed in the control sample are shown in Figure \ref{fig:Control_thumbnails}.

Galaxies with compact morphologies exhibit centrally concentrated [\ion{O}{3}]$+\mathrm{H}\beta$ emission. This morphology is observed in $\sim35\%$ of the full sample and is more prevalent among the control galaxies ($\sim40\%$) than the AGN ($\sim9\%$). The lower incidence of compact morphologies among AGN suggests that AGN hosts more frequently exhibit extended or asymmetric ionized gas distributions, consistent with AGN ionizing radiation influencing gas on larger spatial scales. Approximately $20\%$ of the full sample shows multiple bright [\ion{O}{3}]$+\mathrm{H}\beta$ clumps, which we classify as knotted morphologies. This morphology comprises $\sim20\%$ of the control sample and $\sim20\%$ of the AGN. Galaxies that exhibit irregular or asymmetric [\ion{O}{3}]$+\mathrm{H}\beta$ emission extended over multiple pixels, but lacking the distinct bright clumps characteristic of the knotted category are classified as extended. This category encompasses morphologies with elongated emission as well as those with spiral-arm or ring-like features. Extended morphologies occur with an incidence comparable to compact systems ($\sim40\%$ overall), but are somewhat more common among AGN ($\sim50\%$) than in the control sample ($\sim35\%$). 

\begin{figure*}[t]
        \centering
        \includegraphics[width=1\textwidth]{"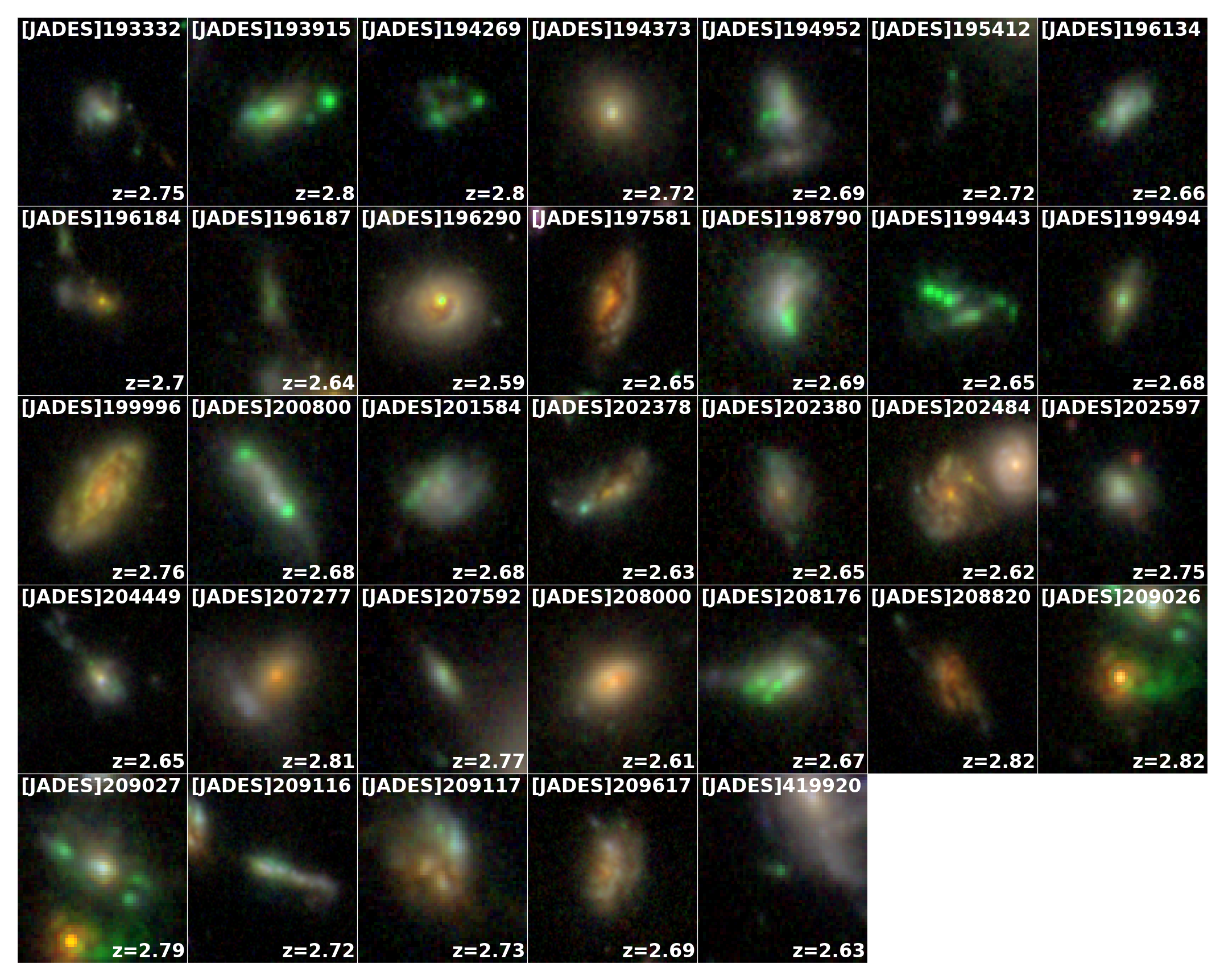"}
        \caption{Montage of RGB (F182M/F210M/F150W) thumbnails showing [\ion{O}{3}]$+\mathrm{H}\beta$ (green) morphologies for the 33 AGN in our JEMS sample. The AGN demonstrate a range of [\ion{O}{3}]$+\mathrm{H}\beta$ morphologies including compact, extended, conical, and knots.}
        \label{fig:AGN_thumbnails}
\end{figure*}

In Figure \ref{fig:AGN_thumbnails}, we present the [\ion{O}{3}]$+\mathrm{H}\beta$ morphologies of the AGN sample ($n=33$). AGN comprise $16\%$ of the full sample, with $52\%$ of AGN ($n=17$) detected in Pa$\beta$. In addition to compact, extended, and knotted morphologies, we introduce a conical class for the AGN sample to account for asymmetric, extended emission that is inconsistent with spiral-arm or ring-like structures and may instead trace ionization cones or a bicone. Approximately $25\%$ of the AGN display conical or biconical [\ion{O}{3}]$+\mathrm{H}\beta$ morphologies. 

\begin{figure*}
        \centering
        \includegraphics[width=1\textwidth]{"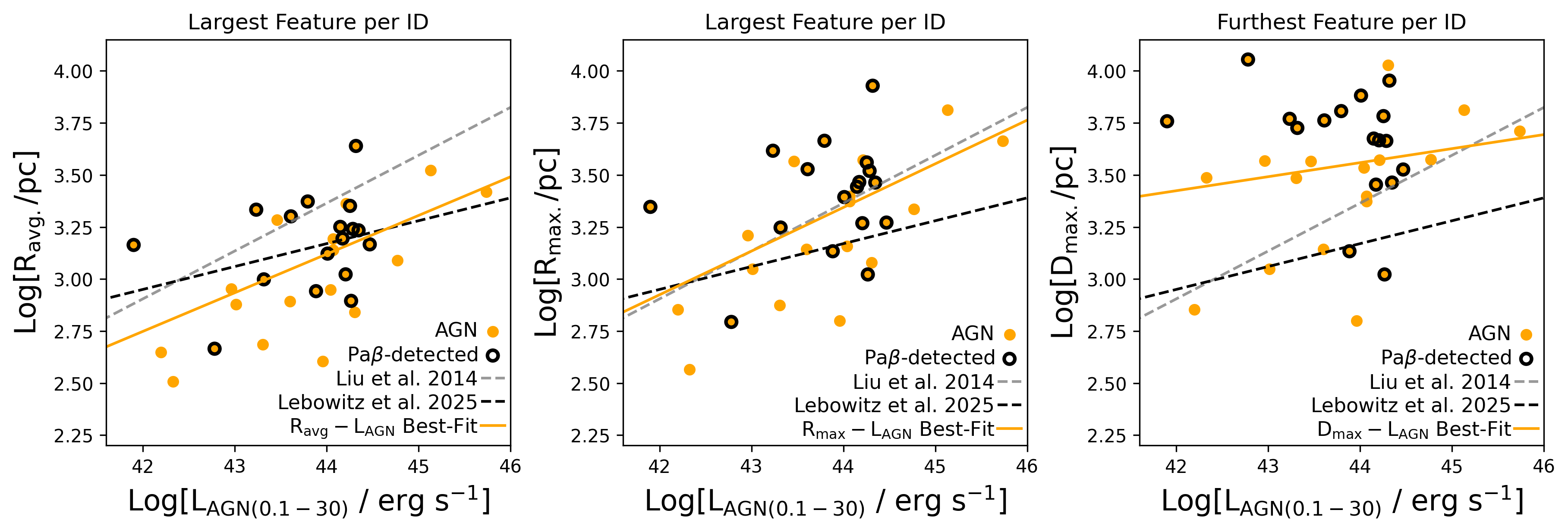"}
        \caption{Plots showing the [\ion{O}{3}]$+\mathrm{H}\beta$ average radial sizes (left), maximum radial sizes (middle), and maximum extents (right) versus the $0.1-30 \mu m$ integrated AGN luminosity for the AGN sample. Pa$\beta$-detected AGN are outlined in black. The gray dashed line shows the observed low-redshift best-fit line from \cite{Liu2014} The black dashed line represents the high-redshift trend derived from simulated observations of AGN NLRs at $z=2.7$ from \cite{Lebowitz2025}. The solid yellow line shows the derived best-fit line for our AGN sample.}
        \label{fig:Size_AGN_lum}
\end{figure*}

\begin{figure*}
        \centering
        \includegraphics[width=1\textwidth]{"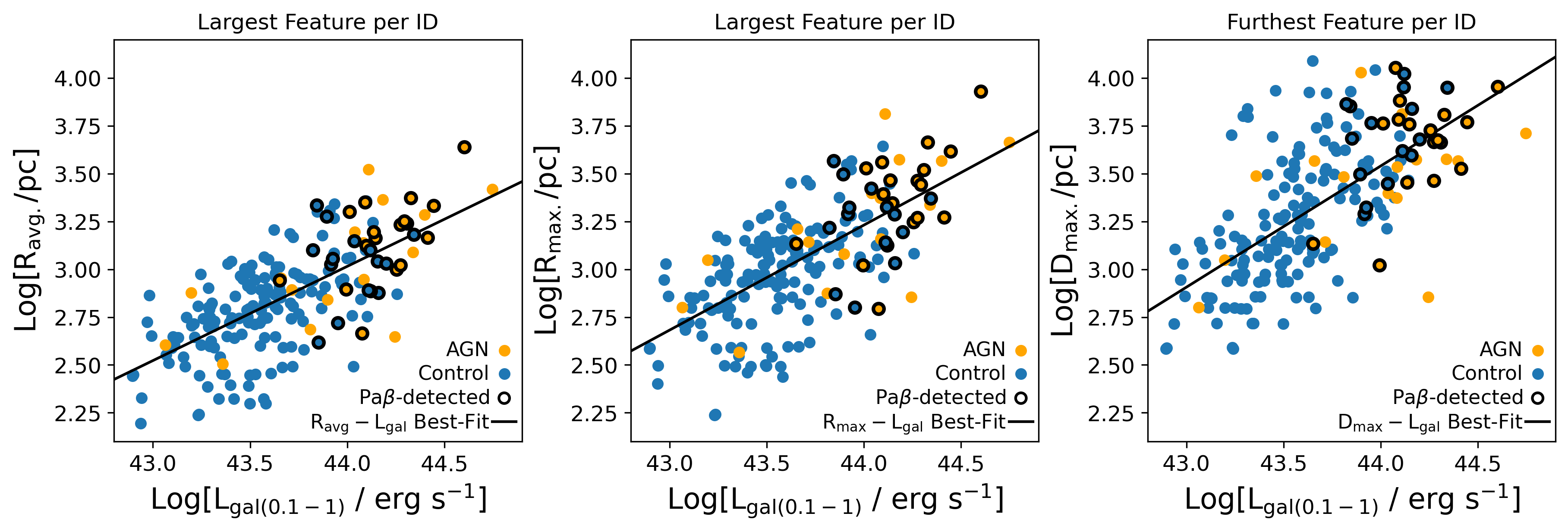"}
        \caption{Plots showing the [\ion{O}{3}]$+\mathrm{H}\beta$ average radial sizes (left), maximum radial sizes (middle), and maximum extents (right) versus the $0.1-1 \mu m$ integrated galaxy luminosity for the AGN (yellow points) and control (blue points) samples. Pa$\beta$-detected control and AGN are outlined in black. We overlay the derived best-fit line for the full galaxy sample as the solid black line.}
        \label{fig:Size_Gal_lum}
\end{figure*}

To gain a more complete picture of how oxygen is ionized, we examine the full spatial extent of the [\ion{O}{3}]$+\mathrm{H}\beta$ emission across different morphologies as a function of both AGN (Figure \ref{fig:Size_AGN_lum}) and galaxy luminosities (Figure \ref{fig:Size_Gal_lum}). We report three measurements of the [\ion{O}{3}]$+\mathrm{H}\beta$ spatial extents: average radial size ($\mathrm{R}_{\mathrm{avg}}$) of the largest feature per object (left column), maximum radial size ($\mathrm{R}_{\mathrm{max}}$) of the largest feature per object (middle column), and the maximum extent ($\mathrm{D}_{\mathrm{max}}$) of the the furthest feature per object (right column). For details on the calculation of these spatial extent measurements, we refer the reader back to Section \ref{subsec:size_measurements}. We find that both the AGN (median R$_{\mathrm{max}}=2.2$ kpc) and Pa$\beta$-detected (median R$_{\mathrm{max}}=2.1$ kpc) samples exhibit systematically larger [\ion{O}{3}]$+\mathrm{H}\beta$ regions than the control sample (median R$_{\mathrm{max}}=1.1$ kpc), with the Pa$\beta$-detected galaxies preferentially being associated with higher AGN and galaxy luminosities. For clarity, we summarize the properties of $R_{\max}$ for the control, AGN, and Pa$\beta$-detected samples in Table \ref{tab:rmax_stats}. This metric captures potential asymmetric morphologies for the dominant emission feature and is the primary size metric used in our analysis. Summary trends for $R_{\mathrm{avg}}$ and $D_{\max}$ are described in the text.

\setlength{\tabcolsep}{3pt}
\begin{deluxetable}{l c c c c}
\tablecolumns{5}
\tablecaption{Summary statistics for the maximum [\ion{O}{3}]$+\mathrm{H}\beta$ radial sizes ($\mathrm{R}_{\max}$) of each sample. $\sigma$ is the standard deviation of the size distribution.} \label{tab:rmax_stats} 
\tablehead{
\colhead{Sample} &
\colhead{Min. (kpc)} &
\colhead{Max. (kpc)} &
\colhead{Median (kpc)} &
\colhead{$\sigma$ (kpc)} }
\startdata
Control          & 0.2 & 7.2 & 1.1 & 0.9 \\
AGN              & 0.4 & 8.5 & 2.2 & 1.7 \\
Pa$\beta$-detected & 0.6 & 8.5 & 2.1 & 1.5 \\
\enddata
\end{deluxetable}

In Figure \ref{fig:Size_AGN_lum}, we plot the [\ion{O}{3}]$+\mathrm{H}\beta$ spatial extents of the AGN sample as a function of AGN luminosity, using the AGNfitter-derived luminosity integrated over $0.1-30 \mu m$ as a proxy for the bolometric luminosity. We similarly plot the [\ion{O}{3}]$+\mathrm{H}\beta$ spatial extents of the full JEMS sample as a function of galaxy luminosity, using the AGNfitter-derived galaxy luminosity integrated over $0.1-1 \mu m$. In both panels, AGN are shown as yellow points and the control sample as blue points, with Pa$\beta$-detected AGN highlighted by black outlines. For the average [\ion{O}{3}]$+\mathrm{H}\beta$ radial extent of the largest feature (left panel), we find a median size of 1.4 kpc ($\sigma=0.9$ kpc). Using the maximum radial extent of that same feature (middle panel), the median increases to 2.2 kpc ($\sigma = 1.7$ kpc). When considering the maximum radial distance to the furthest discrete feature (right panel), we obtain a larger median size of 3.4 kpc ($\sigma=2.5$ kpc).

For comparison with one NLR size-AGN luminosity trend measured at low-redshift, we overlay the best-fit relationship from \cite{Liu2014} (gray dashed line) in terms of the bolometric luminosity in Figure \ref{fig:Size_AGN_lum}. This best-fit line (Equation \ref{eq:R_NLR}) is derived from a sample of $z<0.6$ AGN that include obscured quasars \citep{Liu2013a, Hainline2013, Greene2011, Humphrey2010}, unobscured quasars \citep{Liu2014,Husemann2013}, and Seyfert 2 galaxies \citep{Bennert2006, Fraquelli2003}, where the NLR radii are measured above a limiting surface brightness of $10^{-15}/(1+z)^{4}$ erg s$^{-1}$ cm$^{-2}$ arcsec$^{-2}$. To enable comparison with previous work at $z\sim3$, we also overlay the best-fit relation from \cite{Lebowitz2025} (black dashed line; Equation \ref{eq:R_NLR_sim}), which was derived by fitting [\ion{O}{3}]$+\mathrm{H}\beta$ spatial extents above a limiting surface brightness of $1.4\times10^{-16}$ erg s$^{-1}$ cm$^{-2}$ arcsec$^{-2}$ using mock $z=2.7$ NIRCam images of nearby AGN NLRs.

\begin{equation}
\begin{aligned}
&\log(\frac{\mathrm{R}_{\mathrm{NLR}}}{\mathrm{pc}}) = 0.23 \times \log(\frac{\mathrm{L}_{[\mathrm{OIII}]}}{10^{42} \mathrm{erg s}^{-1}}) + 3.72
\label{eq:R_NLR}
\end{aligned}
\end{equation} 

where $\mathrm{L}_{\mathrm{bol}}$, from \cite{Lamastra2009}, is given by:

\begin{equation}
\begin{aligned}
&\mathrm{L}_{[\mathrm{OIII}]} \simeq \frac{\mathrm{L}_{\mathrm{bol}}}{3500}
\label{eq:L_bol}
\end{aligned}
\end{equation} 

\begin{equation}
\begin{aligned}
&\log(\frac{\mathrm{R}_{\mathrm{NLR, noise}}}{\mathrm{pc}}) = 0.11 \times \log(\frac{\mathrm{L}_{[\mathrm{OIII}]}}{10^{42} \mathrm{erg s}^{-1}}) + 3.34
\label{eq:R_NLR_sim}
\end{aligned}
\end{equation} 

To compare with previous work from the literature, we perform linear fits to the [\ion{O}{3}]$+\mathrm{H}\beta$ spatial extents as a function of AGN luminosity for the AGN sample (yellow line in Figure \ref{fig:Size_AGN_lum}) and as a function of galaxy luminosity for the full JEMS sample (black line in Figure \ref{fig:Size_Gal_lum}). The slopes and intercepts are derived using Equation \ref{eq:R[OIII]_LAGN} for AGN luminosity and Equation \ref{eq:R[OIII]_Lgal} for galaxy luminosity, and are reported in Table \ref{tab:slopes}. For the average radial extent of the largest [\ion{O}{3}]$+\mathrm{H}\beta$ feature versus AGN luminosity (left panel), we derive a slope of 0.19. Using the maximum radial extent of that same feature (middle panel), we obtain a slope of 0.21. In contrast, when considering the maximum radial distance to the furthest discrete feature (right panel), the slope decreases to 0.07. We will discuss the uncertainties on these slopes further in Section \ref{subsec:size-lum_discussion}.

\begin{equation}
\begin{aligned}
&\log(\frac{\mathrm{R}_{\mathrm{[OIII]}+\mathrm{H}\beta}}{\mathrm{pc}}) = m \times \log[\frac{\mathrm{L}_{\mathrm{AGN(0.1-30)}\mu m}}{10^{42} \mathrm{erg s}^{-1}(3500)}] + b
\label{eq:R[OIII]_LAGN}
\end{aligned}
\end{equation} 

\begin{equation}
\begin{aligned}
&\log(\frac{\mathrm{R}_{\mathrm{[OIII]}+\mathrm{H}\beta}}{\mathrm{pc}}) = m \times \log[\frac{\mathrm{L}_{\mathrm{gal(0.1-1)}\mu m}}{10^{42} \mathrm{erg s}^{-1}}] + b
\label{eq:R[OIII]_Lgal}
\end{aligned}
\end{equation} 

\setlength{\tabcolsep}{6pt}
\begin{deluxetable}{l l c c}
\tablecolumns{4}
\tablecaption{Best-fit slopes and intercepts for [\ion{O}{3}]$+\mathrm{H}\beta$ size--luminosity relations\label{tab:slopes}}
\tablehead{
\colhead{Size Metric} & \colhead{Luminosity} & \colhead{$m$} & \colhead{$b$}}
\startdata
$\mathrm{R}_{\mathrm{avg}}\mathrm{[OIII]}+\mathrm{H}\beta$ & L$_{\rm AGN}$ & 0.19 & 3.41 \\
                   & L$_{\rm gal}$ & 0.49 & 2.03 \\
\tableline
$\mathrm{R}_{\mathrm{max}}\mathrm{[OIII]}+\mathrm{H}\beta$ & L$_{\rm AGN}$ & 0.21 & 3.67 \\
                   & L$_{\rm gal}$ & 0.55 & 2.13 \\
\tableline
$\mathrm{D}_{\mathrm{max}}\mathrm{[OIII]}+\mathrm{H}\beta$ & L$_{\rm AGN}$ & 0.07 & 3.66 \\
                   & L$_{\rm gal}$ & 0.63 & 2.27 \\
\enddata
\end{deluxetable}

We find that the best-fit relation between maximum [\ion{O}{3}]$+\mathrm{H}\beta$ radial size and AGN luminosity yields a slope and intercept ($m=0.21$, $b=3.67$) that closely match the low-redshift relation from \cite{Liu2014} ($m=0.23$, $b=3.72$). In contrast, we derive steeper slopes than those measured at $z\sim3$ in our previous work ($m=0.11$; \citealt{Lebowitz2025}) across all three R$_{\mathrm{[OIII]}+\mathrm{H}\beta}$–L$_{\mathrm{AGN}}$ relations. This difference is likely driven by the deeper surface-brightness limit adopted in this work (by a factor of $\sim1.6$) and the larger AGN sample size ($n=33$ versus $n=9$), discussed further in Section \ref{subsec:size-lum_discussion}. When plotted against galaxy luminosity, the [\ion{O}{3}]$+\mathrm{H}\beta$ size metrics yield steeper slopes ($m\sim0.5$–$0.6$) than those found for the AGN luminosity relations ($m\sim0.1$–$0.2$). However, this tighter correlation may partly reflect the fact that both the continuum-subtracted emission maps and the SED-derived galaxy luminosities are based on the same underlying photometric measurements. Although the continuum flux is removed in constructing the emission-line maps, the strength and spatial extent of the line emission remain linked to the rest-frame optical flux, which also contributes to the derived galaxy luminosity. We further discuss these size–luminosity trends in Section \ref{subsec:size-lum_discussion}. Next, we present our measurements of the Pa$\beta$ spatial extents for the Pa$\beta$-detected sample.

\subsection{Pa$\beta$ Morphologies and Spatial Extents} \label{subsec:Pbeta}
\begin{figure*}[t]
        \centering
        \includegraphics[width=1\textwidth]{"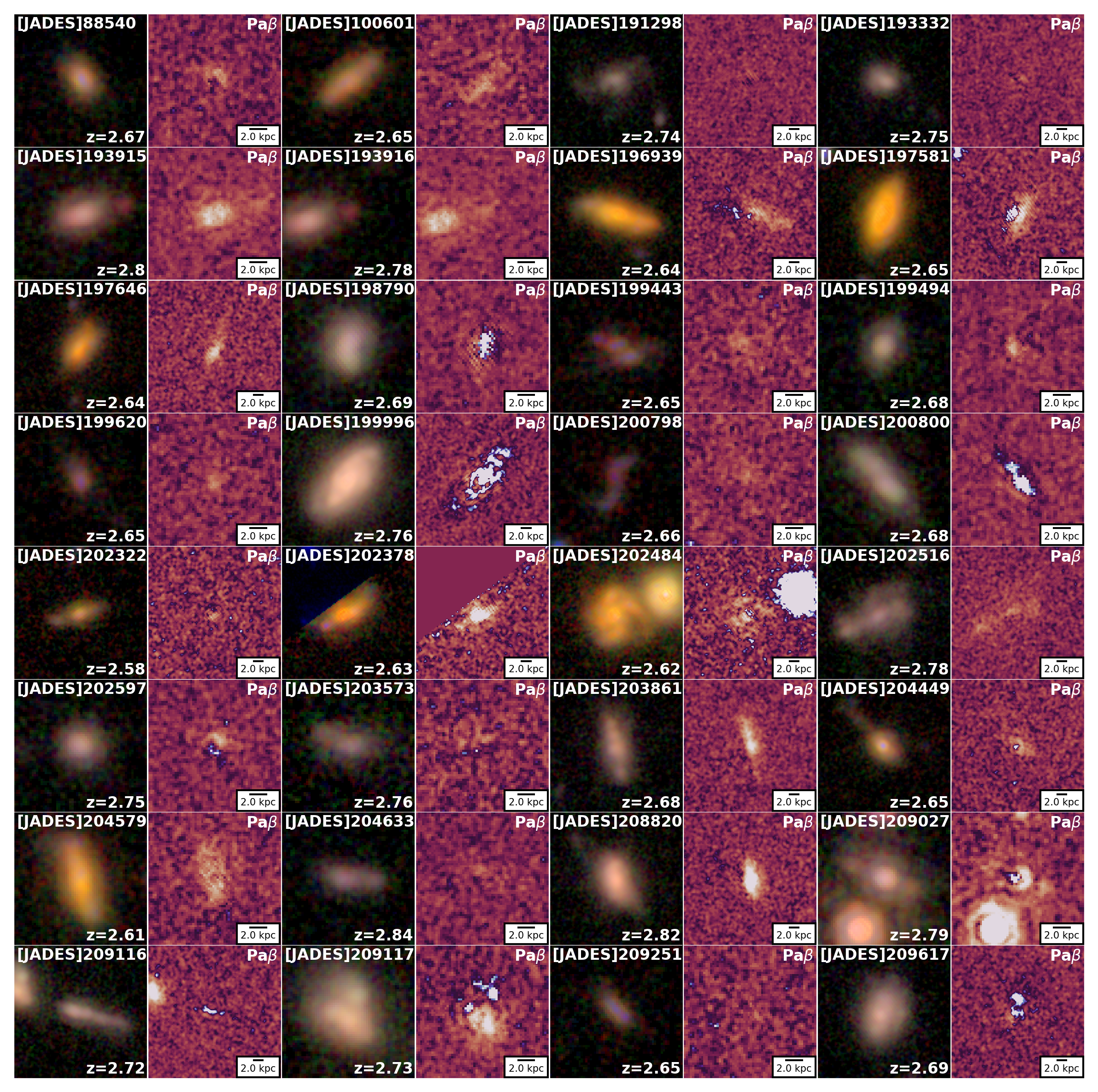"}
        \caption{Montage of RGB thumbnails (left panel) and emission-line maps (right) panel showing Pa$\beta$ morphologies for the Pa$\beta$-detected sample. RGB images are constructed using F460M/F430M/F210M for $z<2.7$ and F480M/F460M/F430M for $z>2.7$, to highlight the Pa$\beta$ emission in red. All emission-line maps are confirmed Pa$\beta$ detections according to our requirement that at least four continuous pixels must possess emission above the adopted surface brightness limit.} 
        \label{fig:Pbeta_thumbnails}
\end{figure*}

In this section, we summarize the observed morphologies and spatial extents of the Pa$\beta$-detected sample. Through visual inspection of the continuum-subtracted Pa$\beta$ emission-line maps, we identify 32 galaxies with Pa$\beta$ detections above a limiting surface brightness of $2.0\times10^{-17}$ erg s$^{-1}$ cm$^{-2}$ (see Section \ref{subsec:size_measurements} for details on the construction of the emission-line maps and the adopted surface-brightness limit). The Pa$\beta$-detected sample comprises approximately $15\%$ of the full JEMS galaxy sample, and includes 17 sources classified as AGN based on our selection criteria. The detection fraction of Pa$\beta$ among AGN is $52\%$ compared to $9\%$ among control galaxies. In Figure \ref{fig:Pbeta_thumbnails}, we present a montage of the continuum-subtracted Pa$\beta$ maps alongside RGB thumbnails of the Pa$\beta$-detected galaxies, with the Pa$\beta$ emission highlighted in red. Some objects show over-subtraction features (e.g., ID=199996), likely due to our use of a constant continuum scaling factor. We retain these objects in the sample, as over-subtraction makes the Pa$\beta$ detections more conservative. We find that, unlike the [\ion{O}{3}]$+\mathrm{H}\beta$ maps, the Pa$\beta$ maps do not display a wide range of morphologies; instead, they appear either compact or broadly extended. The reduced level of detail in Pa$\beta$ relative to [\ion{O}{3}]$+\mathrm{H}\beta$ is likely driven by both intrinsically weaker line emission and reduced spatial resolution at longer wavelengths. 

Next, we examine the Pa$\beta$ spatial extents as a function of both AGN and galaxy luminosity and compare them to the corresponding [\ion{O}{3}]$+\mathrm{H}\beta$ spatial extents. As in our [\ion{O}{3}]$+\mathrm{H}\beta$ analysis, we measure three size metrics for each source: the average and maximum radial sizes of the largest Pa$\beta$ feature, and the maximum extent of the most distant Pa$\beta$ feature. We find a median size of 1.0 kpc ($\sigma=1.2$ kpc) when measuring the average Pa$\beta$ radial extent of the largest feature. Using the maximum radial extent of that same feature, the median increases to 1.8 kpc ($\sigma = 2.0$ kpc). When considering the maximum radial distance to the furthest discrete feature, we obtain a median size of 2.6 kpc ($\sigma=2.0$ kpc). Given the relatively small number of Pa$\beta$-detected systems and to maintain consistency with our [\ion{O}{3}]$+\mathrm{H}\beta$ analysis, we focus on the maximum radial size of the largest Pa$\beta$ feature per object in the discussion that follows. Summary statistics for the maximum Pa$\beta$ radial sizes of the AGN and control subsets of the Pa$\beta$-detected sample are reported in Table \ref{tab:rmax_stats_Pbeta}. 

\setlength{\tabcolsep}{3pt}
\begin{deluxetable}{l c c c c}
\tablecolumns{5}
\tablecaption{Summary statistics for the maximum Pa$\beta$ radial sizes ($\mathrm{R}_{\max}$) of each Pa$\beta$ subset. $\sigma$ is the standard deviation of the size distribution.}\label{tab:rmax_stats_Pbeta}
\tablehead{
\colhead{Pa$\beta$ Subset} &
\colhead{Min. (kpc)} &
\colhead{Max. (kpc)} &
\colhead{Median (kpc)} &
\colhead{$\sigma$ (kpc)}}
\startdata
Control & 0.3 & 8.0 & 1.1 & 2.1 \\
AGN & 0.5 & 7.1 & 2.3 & 1.8 \\
\enddata
\end{deluxetable}

\begin{figure}
        \centering
        \includegraphics[width=0.48\textwidth, trim={0.3cm 0.2cm 0cm 0cm}, clip]{"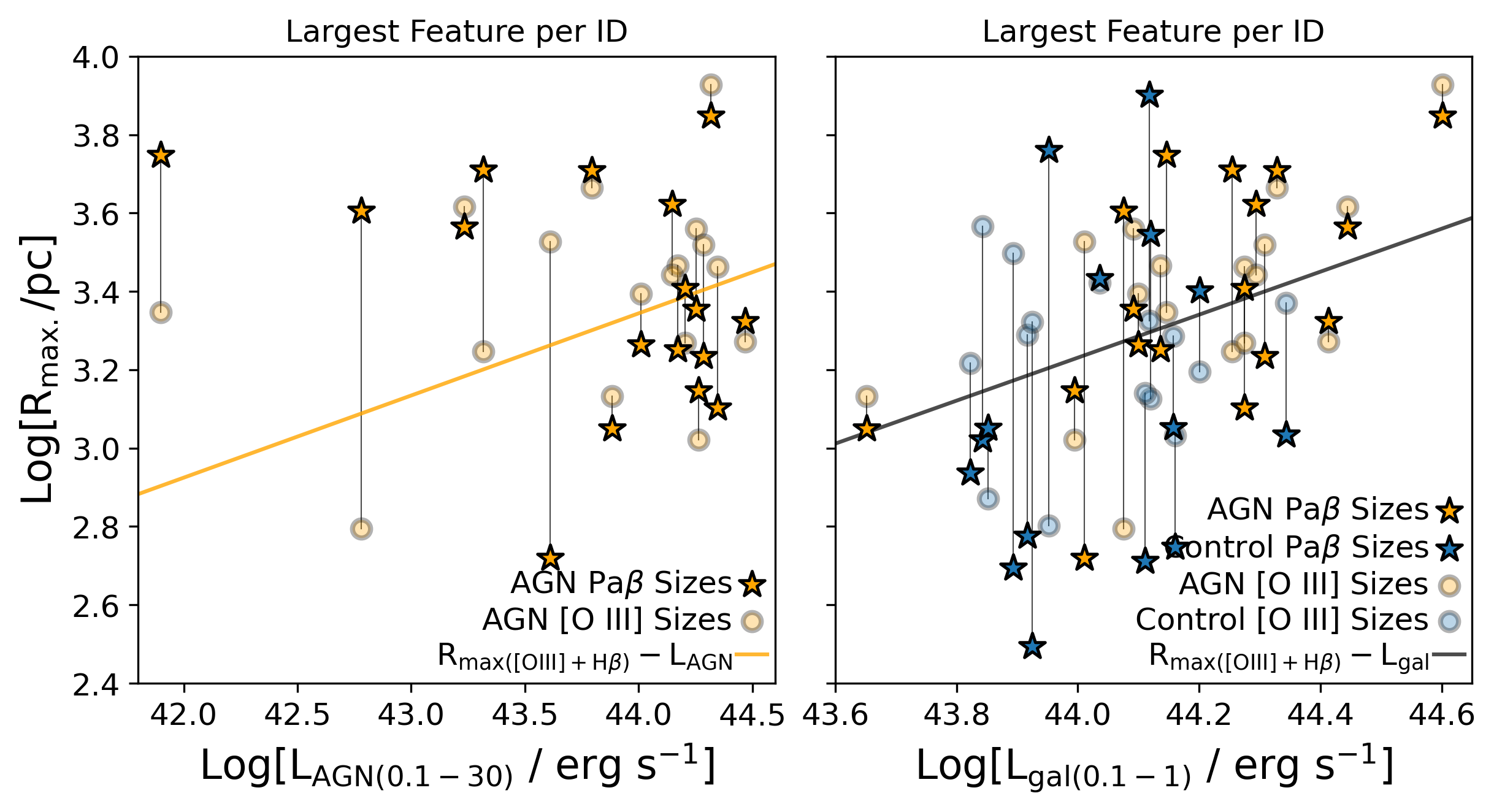"}
        \caption{Plots showing the Pa$\beta$ maximum radial sizes (star points) versus the $0.1-30 \mu m$ integrated AGN luminosity for Pa$\beta$-detected AGN (left) and the $0.1-1 \mu m$ integrated galaxy luminosity for Pa$\beta$-detected AGN and control galaxies (right). For comparison purposes, the [\ion{O}{3}]$+\mathrm{H}\beta$ maximum radial sizes are overlaid on both plots as faint circular points, with black lines linking them to their corresponding Pa$\beta$ sizes. We also include the R$_{\mathrm{max [O III]}+\mathrm{H}\beta}$-L$_{\mathrm{AGN}}$ and R$_{\mathrm{max [O III]}+\mathrm{H}\beta}$-L$_{\mathrm{gal}}$ best-fit lines on their respective plots.}
        \label{fig:Pbeta_size_lum}
\end{figure}

In Figure \ref{fig:Pbeta_size_lum}, we show the maximum Pa$\beta$ radial sizes (star symbols) plotted as a function of AGN luminosity for Pa$\beta$-detected AGN (left panel) and as a function of galaxy luminosity for both Pa$\beta$-detected AGN and control galaxies (right panel), adopting the same AGNfitter-derived luminosities used in the [\ion{O}{3}]$+\mathrm{H}\beta$ analysis. For comparison, we overlay the [\ion{O}{3}]$+\mathrm{H}\beta$ maximum radial sizes as faint outlined circles, along with the best-fit relations derived from the [\ion{O}{3}]$+\mathrm{H}\beta$ R$_{\mathrm{max}}$–L$_{\mathrm{AGN}}$ (yellow line) and R$_{\mathrm{max}}$–L$_{\mathrm{gal}}$ (black line) plots. We follow the same plotting convention as in the previous section, with AGN shown in yellow and control galaxies in blue.

Our Pa$\beta$-detected AGN exhibit systematically larger Pa$\beta$ regions (median R$_{\mathrm{max},\mathrm{Pa}\beta}=2.3$ kpc) than the Pa$\beta$-detected control galaxies (median R$_{\mathrm{max},\mathrm{Pa}\beta}=1.1$ kpc). Following the procedure described in Section \ref{subsec:OIII}, we also perform linear fits to the Pa$\beta$ size–luminosity relations. Given the small number of Pa$\beta$-detected systems and the substantial scatter in the measurements, we do not display these fits in Figure \ref{fig:Pbeta_size_lum}. We find a weak negative correlation between the Pa$\beta$ radial sizes and AGN luminosities ($m\sim-0.16$), and a steep positive correlation with galaxy luminosity ($m\sim0.92$). However, the limited sample size prevents robust constraints on these slopes, so we interpret these trends as qualitative and revisit them in the discussion.

To place these Pa$\beta$ sizes in context, we compare their spatial extents to those of the [\ion{O}{3}]$+\mathrm{H}\beta$ regions, finding that the Pa$\beta$ radial sizes are a median $0.05-0.08$ dex ($\sim 0.2-0.5$ kpc) smaller across AGN and control subsets, respectively. Eight AGN (ID = 197581, 198790, 202378, 202484, 202597, 208820, 209027, and 209117) and six control galaxies (ID = 100601, 196939, 197646, 202322, 203861, 204579) exhibit Pa$\beta$ regions that are more extended than their [\ion{O}{3}]$+\mathrm{H}\beta$ emission, suggesting that stellar ionization may dominate over the AGN in these objects. We will return to these objects in Section \ref{subsec:color-color}. Next, we explore the general galaxy properties (i.e., color, [\ion{O}{3}]$+\mathrm{H}\beta$ equivalent width, SFR and stellar mass) of each sample.

\subsection{Galaxy Properties} \label{subsec:Galaxy_props}
\begin{figure}
        \centering
        \includegraphics[width=0.47\textwidth, trim={0.3cm 0.2cm 0cm 0cm}, clip]{"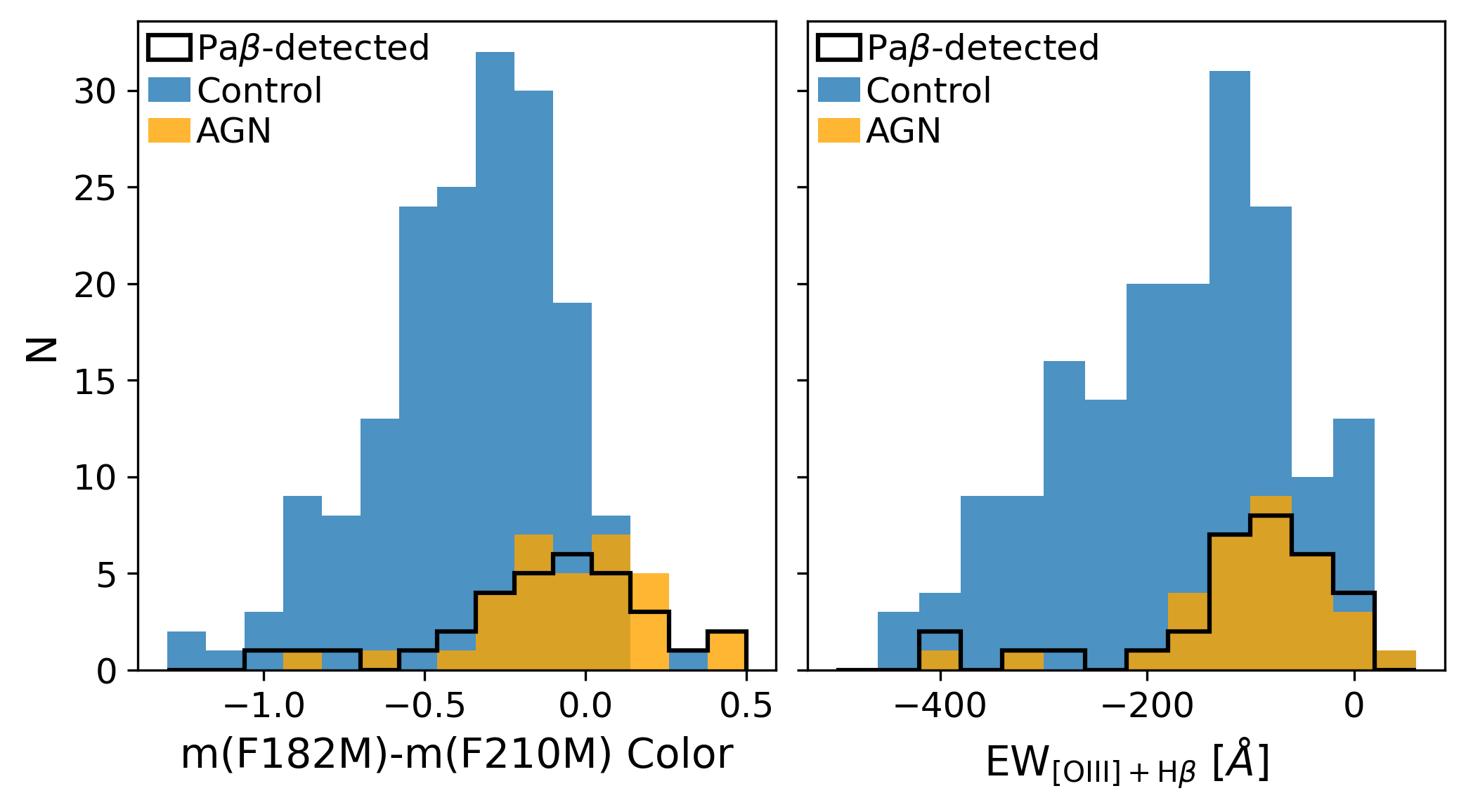"}
        \includegraphics[width=0.47\textwidth, trim={0.3cm 0.2cm 0cm 0cm}, clip]{"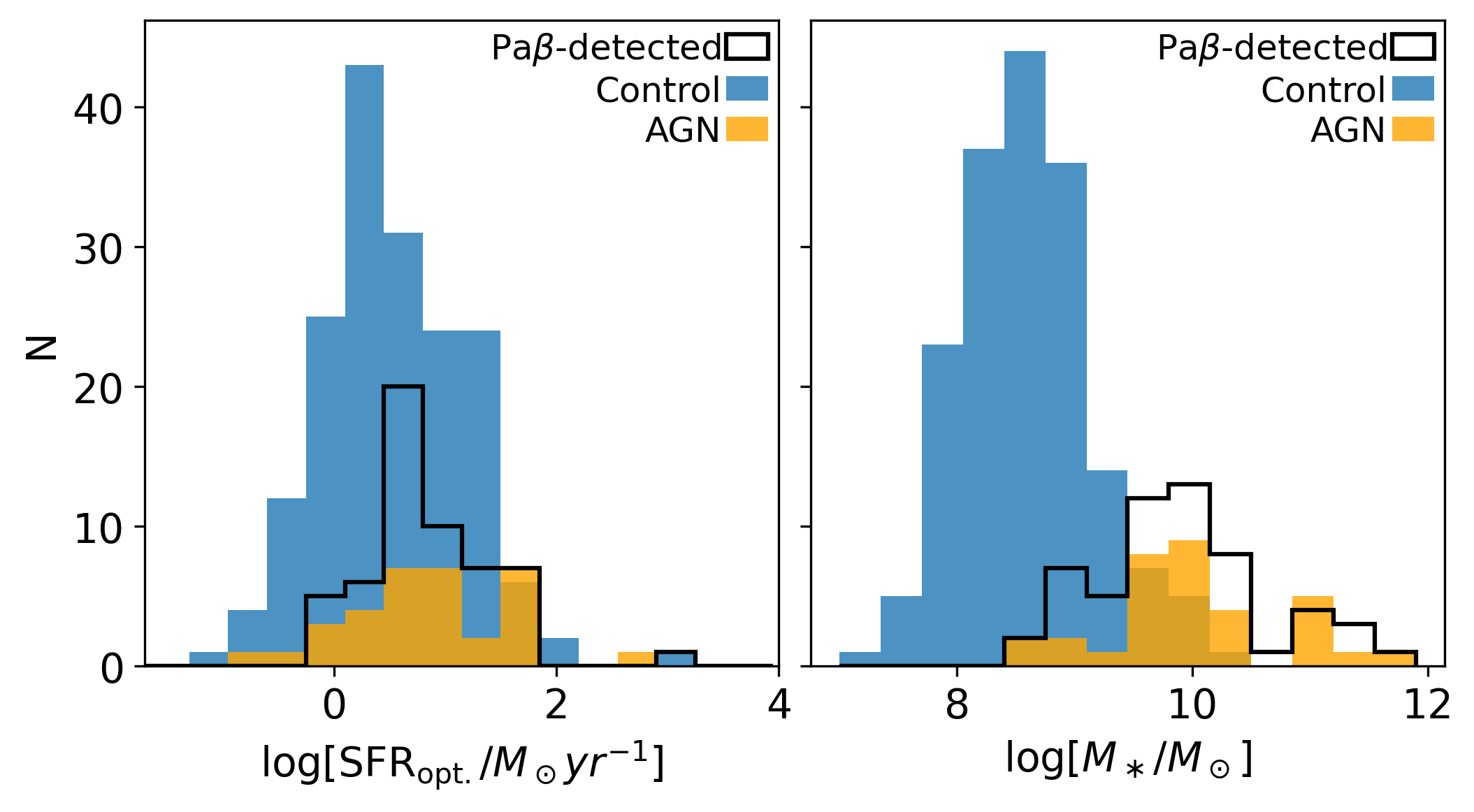"}
        \caption{Histograms showing the $m(\mathrm{F182M})-m(\mathrm{F210M})$ color (top left), [\ion{O}{3}]$+\mathrm{H}\beta$ equivalent width (top right), star-formation rate (bottom left), and stellar mass (bottom right) distributions, with the AGN sample portrayed as the shaded yellow region, control represented by the blue shaded region, and Pa$\beta$-detected objects as the black outlined region.}
        \label{fig:gal_props_hists}
\end{figure}

To assess whether differences in ionized gas morphology and extent are linked to underlying host galaxy properties, we compare the $m(\mathrm{F182M})-m(\mathrm{F210M})$ colors, [\ion{O}{3}]$+\mathrm{H}\beta$ equivalent widths, star-formation rates, and stellar masses of the AGN, Pa$\beta$-detected, and control galaxies. Median values and standard deviations for each quantity are listed in Table \ref{tab:median_props}. In Figure \ref{fig:gal_props_hists}, we present the distributions of $m(\mathrm{F182M})-m(\mathrm{F210M})$ color (top left), [\ion{O}{3}]$+\mathrm{H}\beta$ equivalent width (top right), star-formation rate (bottom left), and stellar mass (bottom right). We adopt a consistent plotting convention throughout, showing AGN in yellow, control galaxies in blue, and Pa$\beta$-detected sources outlined in black.

We compute the $m(\mathrm{F182M})-m(\mathrm{F210M})$ colors using the F182M and F210M JADES ``CIRC6'' fluxes, corresponding to fixed circular apertures with radii of $0.5''$. The [\ion{O}{3}]$+\mathrm{H}\beta$ equivalent widths are estimated following a prescription similar to \cite{Hainline2012}, in which the observed color excess is converted to a line-to-continuum flux ratio, multiplied by the effective width of the F182M filter, and corrected to the rest frame (Equation \ref{eq:EW}). However, if the continuum slope is not flat; for example, due to dust reddening or older stellar populations, this approach overestimates the equivalent widths. To mitigate this, we apply a color correction using the same scaling factor, $f$, discussed in Section \ref{subsec:emission_maps}, which accounts for non-zero galaxy continuum slope at $\sim 2$ $\mu$m (Equation \ref{eq:m_corr}). Following convention, we define negative equivalent widths as arising from line emission rather than absorption. Star-formation rates and stellar masses are derived from the AGNfitter best-fit SEDs according to the methodology detailed in Section \ref{subsec:SED_modeling}.

\begin{equation}
\begin{aligned}
&\mathrm{EW}_{0}=\frac{\Delta\lambda_{\mathrm{F182M}}}{1+\mathrm{z}}[10^{\Delta m_{\mathrm{corrected}}/2.5}-1] 
\label{eq:EW}
\end{aligned}
\end{equation}

\begin{equation}
\begin{aligned}
&\Delta m_{\mathrm{corrected}}=\Delta m_{\mathrm{observed}}+2.5\log(f)
\label{eq:m_corr}
\end{aligned}
\end{equation}

We find that both the AGN and Pa$\beta$-detected samples show less color excess (redder $m(\mathrm{F182M})-m(\mathrm{F210M})$ colors) and smaller [\ion{O}{3}]$+\mathrm{H}\beta$ equivalent widths compared to the control sample. Although there is substantial overlap across the three samples, the control galaxies dominate the bluer end of the $m(\mathrm{F182M})-m(\mathrm{F210M})$ color distribution and exhibit the largest [\ion{O}{3}]$+\mathrm{H}\beta$ equivalent widths. Specifically, $37\%$ of the control sample exhibit [\ion{O}{3}]$+\mathrm{H}\beta$ equivalent widths between $-450$ and $-200$~\AA, compared to only $9\%$ of AGN and $16\%$ of Pa$\beta$-detected galaxies in this range. The Pa$\beta$-detected sample closely overlaps the AGN population in both color and equivalent-width space.

The AGN and Pa$\beta$-detected samples also extend to higher star-formation rates and stellar masses relative to the control sample. As seen in the color and equivalent width distributions, the Pa$\beta$-detected sample occupies similar regions of parameter space as the AGN in both star-formation rate and stellar mass distributions. This behavior is broadly expected, as more massive galaxies host more massive black holes and larger gas reservoirs, supporting both elevated star formation and AGN activity \citep{Hickox&Alexander2018}. This trend is also consistent with our finding in Section \ref{subsec:OIII} that AGN and Pa$\beta$-detected galaxies preferentially occupy the high-luminosity end of the galaxy population in the rest-frame optical. Collectively, these investigations of the [\ion{O}{3}]$+\mathrm{H}\beta$, Pa$\beta$, and global galaxy properties in our JEMS sample point to a strong link between AGN activity and stellar processes at Cosmic Noon, which we explore further in the following section.

\setlength{\tabcolsep}{6pt}
\begin{deluxetable*}{l c c c c}
\tablecolumns{5}
\tablecaption{Median galaxy properties of the control, AGN, and Pa$\beta$-detected samples \label{tab:median_props}}
\tablehead{
\colhead{Sample} &
\colhead{$m(\mathrm{F182M})-m(\mathrm{F210M})$} &
\colhead{[O III]+H$\beta$ EW (\AA)} &
\colhead{$\log[\mathrm{SFR}/M_\odot\,\mathrm{yr}^{-1}]$} &
\colhead{$\log[M_\star/M_\odot]$}}
\startdata
Control & $-0.33 ~(\sigma=0.28)$ & $-154 ~(\sigma=111)$ & $0.5 ~(\sigma=0.6)$ & $8.6 ~(\sigma=0.7)$ \\
AGN & $-0.04 ~(\sigma=0.28)$ & $-95 ~(\sigma=82)$ & $0.9 ~(\sigma=0.7)$ & $10.0 ~(\sigma=0.7)$ \\
Pa$\beta$-detected & $-0.05 ~(\sigma=0.34)$ & $-96 ~(\sigma=100)$ & $0.8 ~(\sigma=0.6)$ & $9.7 ~(\sigma=1.7)$ \\
\enddata
\end{deluxetable*}

\section{Discussion} \label{sec:discussion}
At Cosmic Noon, galaxies experience peak star formation and black hole accretion rates, with both processes contributing to the ionization of the interstellar and circumgalactic mediums \citep{Madau2014}. Therefore, probing the properties and spatial extents of ionized gas is critical for investigating the relative roles of stellar and AGN-driven excitation in shaping the environments of galaxies at this epoch. The [\ion{O}{3}] emission line is a particularly powerful probe of the physical processes that excite gas within galaxies, tracing the most extreme, low-density environments such as in the vicinity of massive, short-lived O stars or in the extended narrow line regions of AGN \citep{Heckman&Best2014}. While many studies at low and intermediate redshift ($z\sim0.5-3$) have used the radial extent of [\ion{O}{3}] emission to evaluate the AGN's influence on its host \citep[][\LAGN{}]{Liu2014, Haineline2014}, comparatively little work has been done exploring the stellar contribution to spatially extended ionized emission, specifically using hydrogen recombination-line tracers of stellar activity, such as Pa$\beta$, to disentangle ionization sources. 

In Section \ref{subsec:size-lum_discussion}, we first place the AGN-host [\ion{O}{3}]$+\mathrm{H}\beta$ extent–luminosity relation measured in this work in the context of low- and high-redshift studies, and assess the dominant systematic uncertainties that affect its interpretation. In Section \ref{subsec:Pabeta_size-lum_discussion}, we then examine Pa$\beta$ spatial extents across AGN and non-AGN hosts, using hydrogen recombination emission as a complementary tracer of ionized gas that is differentially sensitive to dust and excitation conditions. Finally, in Section \ref{subsec:color-color}, we interpret the overlap in photometric and host-galaxy properties between the Pa$\beta$-detected and AGN samples, discussing the relative roles of selection effects and physical coupling between star formation and AGN activity at Cosmic Noon.

\subsection{[\ion{O}{3}] extent–AGN luminosity across redshifts}\label{subsec:size-lum_discussion}
The [\ion{O}{3}] extent–AGN luminosity relation provides an empirical link between the spatial distribution of AGN-ionized gas in the narrow-line region and the radiative output of the central engine. At Cosmic Noon ($z\sim2$–3), however, quantifying this relation for AGN NLRs is complicated by the frequent coexistence of vigorous star formation and AGN activity, as well as by strong observational biases, including cosmological surface-brightness dimming and instrumental limitations. Building on the methodology established in \LAGN{}, we use a larger sample and deeper surface brightness limits to measure the [\ion{O}{3}]$+\mathrm{H}\beta$ extent–AGN luminosity relation at this epoch.

At low redshift ($z<0.5$), NLR sizes are reported to range from $2-20$ kpc, increasing linearly with AGN luminosity \citep{Liu2013a, Liu2014, Hainline2013, Bennert2006, Haineline2014, Sun2017}. Many studies report a flattening of the NLR size-AGN luminosity slope at the high-luminosity end, suggesting an upper limit to the NLR extent set by the host galaxy \citep{Hainline2013, Bergmann2018}. Quantitatively, the slope of the low-redshift NLR size–AGN luminosity relation is reported to range from $m \sim 0.2-0.5$, with variations attributed to differences in quasar sample selection and in the luminosity tracers used to estimate AGN power \citep{Greene2011, Liu2014, Husemann2014, Hainline2013, Bergmann2018}. At high redshift ($z \sim 2-3$), however, measurements of the [\ion{O}{3}] extents have yielded mixed results, with reported sizes varying widely across studies. Many ground-based IFU observations of luminous quasars and extremely red quasars (ERQs) report modest [\ion{O}{3}] extents of $\sim0.5-3$ kpc, often associated with the narrow component of ionized outflows rather than the full quiescent NLR \citep{Harrison2012, Carniani2015, Vayner2021, Lau2024}. In contrast, deeper observations of select systems with JWST have revealed substantially more extended ionized gas, reaching $\sim10-20$ kpc in some ERQs \citep{Wylezalek2022, Vayner2024}, highlighting the strong sensitivity of inferred extents to surface-brightness limits and instrumental depth. As a result, high-redshift studies have struggled to robustly constrain the slope of the $\mathrm{R}_{\mathrm{[O III]}}-\mathrm{L}_{\mathrm{AGN}}$ relation.

\LAGN{} attempted to measure this relation from NIRCam medium-band observations of nine AGN at $z\sim3$ with evidence for NLRs. Their reported average (R${\mathrm{avg.}}=1.5$ kpc) and maximum (R${\mathrm{max.}}=2.4$ kpc) [\ion{O}{3}]$+\mathrm{H}\beta$ extents were broadly consistent with previous high-redshift measurements, if slightly lower, though this result may reflect the modest AGN luminosities of their sample. However, given the significant scatter introduced by the small sample size, a tight correlation was difficult to discern. These authors instead used a simulation framework to predict the $\mathrm{R}_{\mathrm{[O III]}}-\mathrm{L}_{\mathrm{AGN}}$ relation at $z=2.7$ by constructing mock high-redshift NIRCam medium-band images based on low-redshift AGN with prominent ionization cones observed by MUSE. Adopting the same noise properties and limiting surface brightness as the real NIRCam images, they found that the linear correlation between $\mathrm{R}_{\mathrm{[O III]}}-\mathrm{L}_{\mathrm{AGN}}$ holds, but with a shallower slope ($m=0.11$) than those reported in low-redshift studies. \LAGN{} interpreted this shallower slope as a consequence of observational biases arising from higher instrumental noise at this redshift and cosmological surface-brightness dimming, which together limit the depth to which reliable measurements can be made.  

In the current study, two improvements have been made to our procedure for measuring spatial extents that likely contribute to more accurate [\ion{O}{3}]$+\mathrm{H}\beta$ measurements: (1) we allowed noncontinuous [\ion{O}{3}]$+\mathrm{H}\beta$ features to be accounted for across a wide range of galaxy sizes and morphologies, and (2) we adopted a uniform surface brightness limit ($\mathrm{SB}_{\mathrm{limit}}=8.7\times10^{-17}$ erg s$^{-1}$ cm$^{-2}$) across our full sample that was a factor of two deeper than the limit ($\mathrm{SB}_{\mathrm{limit}}=1.4\times10^{-16}$ erg s$^{-1}$ cm$^{-2}$) used in \LAGN{}. In our extended AGN sample, we find characteristic [\ion{O}{3}]$+\mathrm{H}\beta$ extents ranging from $0.3-4.4$ kpc when using a conservative, average size metric. Using maximum radial extents, the measured sizes span $0.4$–$8.5$ kpc. The increased sample size of 33 AGN now allows us to measure a slope for the $\mathrm{R}_{\mathrm{[O III]+H}\beta}-\mathrm{L}_{\mathrm{AGN}}$ relation. Plotting the average and maximum [\ion{O}{3}]$+\mathrm{H}\beta$ radial extents against the derived AGN luminosities, we measure slopes of 0.19 and 0.21, respectively, consistent with the low-redshift slopes reported by ($m=0.22$; \citealt{Greene2011}) and ($m=0.24$; \citealt{Liu2014}).

When placed in the context of low-redshift scaling relations, the slopes we recover for the $\mathrm{R}_{\mathrm{[O III]+H}\beta}-\mathrm{L}_{\mathrm{AGN}}$ relation at $z\sim3$ are consistent with the shallower end of the range reported at low redshift. In the local universe, such shallow slopes have often been interpreted as evidence for a matter-bounded regime, in which the radial extent of the NLR is regulated primarily by the availability and spatial distribution of ionized gas rather than by the intensity of the AGN radiation field \citep{Greene2011, Liu2014}. In contrast, several low-redshift studies report steeper slopes \citep{Hainline2013, Husemann2014, Bergmann2018}, which are commonly interpreted as reflecting ionization-bounded conditions and a stronger coupling between the AGN radiation field and the spatial extent of the NLR. Importantly, these studies also emphasize that the inferred slope depends sensitively on the choice of bolometric luminosity proxy, with shallower relations typically obtained when using $L_{\mathrm{[O III]}}$ and steeper slopes recovered when adopting optical continuum ($L_{5100}$; \citealt{Husemann2014}) or mid-infrared ($L_{8\mu\mathrm{m}}$; \citealt{Hainline2013}) tracers of AGN power. 

In this context, it is notable that our measurements yield a comparatively shallow slope despite relying on independently derived AGN luminosities, suggesting that gas availability may continue to play an important role in setting the observable extent of AGN-ionized gas at Cosmic Noon. We note, however, that galaxies at $z\sim3$ are systematically more compact at fixed stellar mass compared to their low-redshift counterparts \citep[e.g.,][]{Martorano2024, Lyu2025}, such that a given ionized gas extent may represent a larger fraction of the host galaxy, potentially implying a greater relative impact from AGN ionization. Nevertheless, considering the observational challenges inherent to high-redshift measurements and the sensitivity of inferred slopes to analysis choices, any physical interpretation of the measured relation must be treated with caution. In particular, several sources of uncertainty are expected to systematically bias the inferred [\ion{O}{3}] extent–AGN luminosity relation toward shallower slopes, potentially obscuring an intrinsically steeper correlation. These include uncertainties inherent to the adoption of a uniform surface brightness limit despite variations in noise properties among the NIRCam images (particularly between large and small targets), contamination from H$\beta$ emission within the medium-band filters, and our use of the $0.1-30 \mu m$ AGNfitter-derived AGN luminosities as a proxy for the bolometric luminosity. 

For a detailed discussion of the uncertainties associated with adopting a uniform surface-brightness limit and with contamination in the medium-band filters, we refer the reader to \LAGN{}. Briefly, based on simulations, \LAGN{} found that (1) increasing the surface-brightness depth by a factor of four would nearly recover the intrinsic NLR extents of their low-redshift AGN sample, and (2) contamination from H$\beta$ emission and stellar continuum within the medium-band filters leads to systematically overestimated radial extents and a shallower inferred slope by a factor of $\sim1.5$. In this work, we adopt a surface-brightness limit that is a factor of two deeper than that used in \LAGN{}, enabling more substantial recovery of the [\ion{O}{3}]$+\mathrm{H}\beta$ emission across our sample. Consequently, the remaining uncertainty in the measured [\ion{O}{3}] extent–AGN luminosity relation is likely dominated by stellar contamination, which cannot be mitigated without spectroscopic observations, as well as by uncertainties in the AGNfitter-derived $0.1$–$30 \mu$m AGN luminosities, which may underestimate the true bolometric luminosity for radio- and/or X-ray–detected AGN. 

To assess this uncertainty, we recomputed the slope of the $\mathrm{R}_{\mathrm{[O III]+H}\beta}$–$\mathrm{L}_{\mathrm{AGN}}$ relation using only the fifteen X-ray–detected AGN and estimating bolometric luminosities via the X-ray bolometric correction (Equation 2) from \citet{Brown2019}. Using this approach, the median AGN luminosity decreases by 0.1 dex, leading to a modest change in the inferred slope: from 0.19 to 0.16 when adopting the average size metric, and from 0.21 to 0.19 when using the maximum size metric. Given the small magnitude of this shift compared to the systematic uncertainties discussed above and in \LAGN{}, uncertainties in the bolometric correction are unlikely to drive the observed slope. Rather, contamination in the medium-band measurements and the adopted surface-brightness limit likely represent the dominant systematics and may bias the inferred relation toward shallower slopes, suggesting that the intrinsic correlation could be steeper.

Beyond these uncertainties, further caution is advised when comparing the $\mathrm{R}_{\mathrm{[O III]+H}\beta}-\mathrm{L}_{\mathrm{AGN}}$ relation derived in this work to previous measurements of the $\mathrm{R}_{\mathrm{NLR}}-\mathrm{L}_{\mathrm{AGN}}$ relation. In particular, we do not restrict our sample to include only AGN with classical conical NLR morphologies; as a result, contamination from stellar photoionization in [\ion{O}{3}]$+\mathrm{H}\beta$ cannot be fully excluded in our AGN-classified hosts, especially at this epoch. In the following section, we therefore examine [\ion{O}{3}]$+\mathrm{H}\beta$ and Pa$\beta$ emission together in AGN and non-AGN hosts to assess the potential stellar contribution to gas ionization at Cosmic Noon.

\subsection{Ionized gas extents in Pa$\beta$-detected galaxies}\label{subsec:Pabeta_size-lum_discussion}
As a near-IR, hydrogen recombination line, Pa$\beta$ is a relatively dust-insensitive tracer of ionizing radiation produced by both massive stars and AGN \citep{Cleri2022, Lamperti2017}. In star-forming regions, Pa$\beta$ arises in the H II regions surrounding OB stars. In AGN, the same recombination physics applies, but the ionizing continuum is supplied by the accretion disk, with Pa$\beta$ arising in the BLR and/or NLR \citep{Osterbrock2006}. Pa$\beta$ is an intrinsically weaker line compared to [\ion{O}{3}] since hydrogen recombination emission is distributed over several transitions; therefore, a statistically significant detection of Pa$\beta$ generally signals a substantial ionized gas reservoir. However, its presence does not uniquely distinguish between stellar and AGN ionization in composite systems \citep{Larkin1998}; therefore, interpreting its origin requires careful treatment in galaxies hosting both AGN and star formation. In this regard, our Pa$\beta$-detected sample provides a natural framework to explore the relative contributions of AGN and stellar-driven ionization by comparing the Pa$\beta$ and [\ion{O}{3}]$+\mathrm{H}\beta$ extents---which are differentially sensitive to dust and ionization conditions---among AGN hosts and and control galaxies.

Across our Pa$\beta$-detected sample, we detect Pa$\beta$ emission on kiloparsec scales, with median maximum radial extents of 1.8 kpc. These extents are only slightly smaller than those measured for [\ion{O}{3}]$+\mathrm{H}\beta$, which exhibit a median radial extent of 2.1 kpc, corresponding to an offset of $\sim$0.1 dex. The discrepancy is reduced among Pa$\beta$-detected AGN, which show median maximum [\ion{O}{3}]$+\mathrm{H}\beta$ and Pa$\beta$ extents of 2.8 kpc and 2.3 kpc, respectively. Notably, eight AGN (ID = 197581, 198790, 202378, 202484, 202597, 208820, 209027, and 209117) and six control galaxies (ID = 100601, 196939, 197646, 202322, 203861, 204579) exhibit larger Pa$\beta$ nebulae than [\ion{O}{3}]$+\mathrm{H}\beta$. For these objects, we find their rest-frame optical SED slopes are suggestive of possible dust reddening, while their NIRCam F210M/F182M/F150W RGB images reveal morphological features consistent with dust lanes. Since [\ion{O}{3}]$+\mathrm{H}\beta$ is sensitive to dust extinction, we surmise that attenuation likely suppresses the observed [\ion{O}{3}]$+\mathrm{H}\beta$ emission in these systems, causing the Pa$\beta$ emission to appear comparatively more extended.

The smallest Pa$\beta$ extents are found among our 15 Pa$\beta$-detected control galaxies. Since these systems do not meet any of our AGN selection criteria, the Pa$\beta$ emission in these galaxies likely traces star formation alone. The discrepancy between the [\ion{O}{3}]$+\mathrm{H}\beta$ and Pa$\beta$ extents is statistically significant in these systems, with median maximum Pa$\beta$ extents of 1.1 kpc compared to 1.9 kpc for [\ion{O}{3}]$+\mathrm{H}\beta$. If these systems are correctly identified as non-AGN hosts, then both the [\ion{O}{3}]$+\mathrm{H}\beta$ and Pa$\beta$ emission must be arising from stellar ionization. Without the presence of hard ionizing radiation from an AGN, recombination emission is expected to play a relatively larger role in producing extended ionized gas nebulae \citep{Osterbrock2006}. In such systems, H$\beta$ may contribute more significantly to the total [\ion{O}{3}]$+\mathrm{H}\beta$ emission compare to AGN hosts \citep{Kewley2013}. 

Given that Pa$\beta$ is intrinsically weaker than both [\ion{O}{3}] and H$\beta$, we expect [\ion{O}{3}]$+\mathrm{H}\beta$ will remain detectable at significantly lower surface brightness, allowing it to trace the full spatial extent of ionized gas, except in systems subject to heavy dust attenuation, while Pa$\beta$ is confined to compact, high-surface-brightness regions associated with active star formation \citep{Haffner2009}. This interpretation is supported by our analysis of the [\ion{O}{3}]$+\mathrm{H}\beta$ and Pa$\beta$ emission-line maps, which reveal Pa$\beta$ emission to be centralized for objects with extended [\ion{O}{3}]$+\mathrm{H}\beta$ morphologies, whereas galaxies hosting bright [\ion{O}{3}]$+\mathrm{H}\beta$ clumps exhibit Pa$\beta$ emission that is spatially coincident with these knots (see for example ID = 199443 and 202378 in Figures \ref{fig:AGN_thumbnails} and \ref{fig:Pbeta_thumbnails}). Since [\ion{O}{3}]$+\mathrm{H}\beta$ knots are present in $\sim20\%$ of our control sample, the observed spatial correlation between the two tracers in these systems suggests a shared origin in dense, compact, dust-enshrouded star-forming regions, in agreement with a recent JWST study showing that Paschen recombination emission traces such features at Cosmic Noon \citep{Liu2024}. Our reported Pa$\beta$ spatial extents ($\sim1.1$ kpc in control galaxies and $\sim2.3$ kpc in AGN hosts) are also broadly consistent with a $z\sim4-6$ JWST grism study of H$\alpha$, which traces more extended ionized gas due to its sensitivity to low surface-brightness recombination emission. These authors found typical ionized gas sizes of $\sim1.2$ kpc for galaxies of similar stellar mass \citep{Danhaive2026}.

While these spatial comparisons provide insight into the physical regions traced by Pa$\beta$, they do not uniquely determine the mechanisms governing the extent of the Pa$\beta$ emission. We turn to the Pa$\beta$ extent–luminosity relations to assess whether the size of the Pa$\beta$ nebulae scales with AGN luminosity or host-galaxy luminosity, noting that the limited sample size of Pa$\beta$-detected AGN introduces substantial uncertainty in the inferred AGN luminosity trend. We find a negative, although weak correlation ($m\sim-0.16$) between the Pa$\beta$ radial extents and AGN luminosities. This relation may suggest that increased AGN activity is associated with a more centrally concentrated distribution of recombination emission, possibly reflecting a reduced contribution from extended star-forming regions at higher AGN luminosities. By contrast, the Pa$\beta$ extent–galaxy luminosity relation appears steeper (with a best-fit slope of $m\sim0.9$) across AGN and non-AGN hosts. This behavior is likely driven by the well-established correlation between star formation and rest-frame optical galaxy luminosity. Indeed, we find that Pa$\beta$ extent scales similarly with the derived SFR estimates and stellar masses, consistent with previous studies that report strong trends between the spatial distribution of hydrogen recombination emission and global galaxy properties \citep{Nelson2016, Liu2024}. In the next section, we discuss the observed overlap in the photometric and galaxy properties of our AGN and Pa$\beta$-detected samples.

\subsection{Pa$\beta$-detected galaxies overlap AGN across photometric and galaxy properties} \label{subsec:color-color}
The correlation between the Pa$\beta$ spatial extents with host galaxy properties across both AGN and Pa$\beta$-detected samples suggest underlying similarities in the host environments of both populations. As shown in Section \ref{subsec:Galaxy_props}, we found that Pa$\beta$-detected galaxies indeed occupied similar regions of photometric and galaxy parameter space as AGN, including in color–color space, [\ion{O}{3}]$+\mathrm{H}\beta$ equivalent width, star-formation rate, and stellar mass. This overlap may reflect a combination of selection effects inherent to the detection of Pa$\beta$ and a physical connection between star formation and AGN activity in massive galaxies at Cosmic Noon. We discuss both interpretations and their implications for AGN-host galaxy evolution at this epoch.

As discussed in the previous section, the inherent weakness of the Pa$\beta$ emission line may naturally bias its detection towards galaxies with significant ionized gas reservoirs and/or high-surface-brightness features associated with strong and potentially dust-obscured star formation. Both of these properties are expected to preferentially favor higher-stellar-mass galaxies, consistent with a recent JWST study that measured the star-forming sequence at Cosmic Noon using Pa$\alpha$ as a SFR indicator \citep{Neufeld2024}, and another that found a positive correlation between Pa$\alpha$ extent and stellar mass \citep{Liu2024}. In fact, from the AGNfitter results, $38\%$ of the Pa$\beta$-detected sample possess stellar masses above $10^{10} M\odot$, compared to just $3\%$ of the control. Likewise, our AGN sample also favors higher stellar masses with $46\%$ lying above $10^{10} M\odot$. This result is consistent with the widely accepted trend that AGN detection rates rise toward higher stellar masses \citep{Kauffmann2003, Mainieri2011, Juneau2013, Bongiorno2016}.

\begin{figure}
        \centering
        \includegraphics[width=0.48\textwidth]{"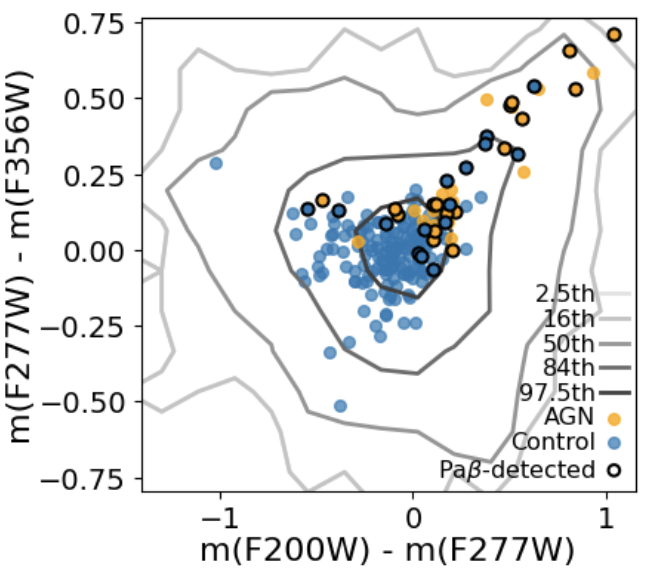"}
        \caption{$m(\mathrm{F200W}) - m(\mathrm{F277W})$ versus $m(\mathrm{F277W}) - m(\mathrm{F356W})$ color-color plot. The AGN sample is represented by the orange points and the control sample is represented by the blue points. The gray scale contours show the full JEMS sample in the same redshift range at 2.5, 16, 50, 84, and 97.5 percentiles.}
        \label{fig:color-color_plot}
\end{figure}

In addition to AGN and Pa$\beta$ detections favoring more massive galaxies, our selection methods may also bias both samples toward dust-rich host environments. The use of a dust-insensitive, near-IR recombination line, along with AGN diagnostics based on mid-infrared and X-ray emission, naturally increases sensitivity to systems with substantial dust attenuation. Consistent with this picture, Pa$\beta$-detected galaxies appear to be color outliers in the rest-frame optical, occupying the same region of color–color space as the AGN sample. In Figure \ref{fig:color-color_plot}, we show the $m(\mathrm{F200W}) - m(\mathrm{F277W})$ versus $m(\mathrm{F277W}) - m(\mathrm{F356W})$ colors of the AGN, control, and Pa$\beta$-detected samples, illustrating how AGN (yellow points) and Pa$\beta$-detected galaxies (black outlined points) cluster in the far upper right-hand quadrant. Notably, four Pa$\beta$-detected galaxies (ID = 196939, 197646, 203861, 204579) occupy this region despite showing no discernible evidence for an AGN. These systems exhibit rest-frame optical SED slopes and continuum morphologies suggestive of significant dust reddening. These objects also have large discrepancies between their optical ($\sim 5-20 \mathrm{M}_{\odot}/\mathrm{yr}$) and infrared-derived ($\sim 90-150 \mathrm{M}_{\odot}/\mathrm{yr}$) star-formation rates, indicating substantial dust-enshrouded star formation that can drive photometric properties similar to those of obscured AGN \citep{Donley2012}.

However, selection effects and dust-obscuration alone are unlikely to fully explain the observed photometric, stellar mass, and star-formation rate overlap. In particular, we find that our Pa$\beta$-detected sample possesses an AGN detection fraction ($f_{\mathrm{AGN}}=52\%$) that is a factor of six higher than the control ($f_{\mathrm{AGN}}=9\%$). If the Pa$\beta$ emission in these systems traces primarily stellar activity, then this high AGN incidence suggests a physical connection between AGN activity and star formation at this Cosmic Noon that extends beyond simple selection biases. At this epoch, galaxies are characterized by high gas fractions and turbulent disks which can drive efficient gas inflows toward galactic centers \citep{Schreiber2020}. Such inflows can simultaneously fuel rapid black hole growth and ignite star formation, leading to concurrent AGN and stellar activity in massive systems \citep{Hickox&Alexander2018}, with both processes contributing to the ionization of the ISM. Ultimately, spatially resolved spectroscopy and line-ratio diagnostics are necessary to robustly determine the relative contribution of stellar and AGN ionization in these systems.

Regardless of the dominant ionizing source, the overlap in rest-frame optical colors and [\ion{O}{3}]$+\mathrm{H}\beta$ equivalent widths between the Pa$\beta$-detected and AGN samples suggests that both populations occupy a similar region of parameter space in massive galaxies that is systematically offset from the control sample. Both Pa$\beta$-detected galaxies and AGN exhibit redder $m(\mathrm{F182M})-m(\mathrm{F210M})$ colors and smaller [\ion{O}{3}]$+\mathrm{H}\beta$ equivalent widths than the control galaxies. This is notable given that the median [\ion{O}{3}]$+\mathrm{H}\beta$ radial extents are approximately a factor of $\sim2$ larger in the AGN and Pa$\beta$-detected systems compared to the control sample, indicating more spatially extended ionized gas even as the line emission is weaker relative to the underlying continuum. Because equivalent width measures the strength of the line relative to the underlying stellar continuum, massive galaxies with with low specific star-formation rates and strong continuum emission can exhibit small equivalent widths even when the line-emitting nebulae are spatially extended as a result of continuum dilution. This interpretation is consistent with the results of \citet{Reddy2018}, who showed that [\ion{O}{3}]$+\mathrm{H}\beta$ equivalent width increases systematically toward lower stellar masses and higher specific star-formation rates, with the highest equivalent widths most often found in low-stellar-mass galaxies with high specific star-formation rates and low metallicities. 

The redder $m(\mathrm{F182M})-m(\mathrm{F210M})$ colors and smaller [\ion{O}{3}]$+\mathrm{H}\beta$ equivalent widths in the AGN and Pa$\beta$-detected samples may also arise from a combination of higher dust content, more evolved stellar populations, and differences in the physical conditions of the ionized gas. Although redder colors can be produced by older stellar populations, the robust detection of [\ion{O}{3}]$+\mathrm{H}\beta$ emission among Pa$\beta$ control galaxies suggests the presence of young, massive stars in these systems, favoring an interpretation in which wavelength-dependent dust extinction preferentially suppresses the observed [\ion{O}{3}]$+\mathrm{H}\beta$ emission. Additional contributions from differences in interstellar medium conditions, such as higher metallicities and lower electron temperatures, may further reduce [\ion{O}{3}]$+\mathrm{H}\beta$ emissivity in these systems \citep{Maiolino2008, Reddy2018}. For the AGN sample, the lack of clear Type I signatures in their SEDs as well as the mid-IR emission traced by MIRI in several sources suggest that many of these systems are likely to be dust-obscured, in which case dust attenuation provides a natural explanation for both their redder rest-frame optical colors and the reduced [\ion{O}{3}]$+\mathrm{H}\beta$ equivalent widths. 

Our findings are also broadly consistent with the medium-band–selected extended emission-line galaxy sample presented by \citet{zhu2025}, who also report elevated dust attenuation and enhanced star formation among systems exhibiting extended emission features. Although their selection is based on morphological excess in medium-band imaging rather than explicit AGN diagnostics, the similarity in star-formation rate and dust properties suggests that spatially extended ionized gas at Cosmic Noon is frequently associated with dusty, actively growing systems. In this context, it is possible that AGN and stellar processes may drive the ionization of the gas at this epoch, while host-galaxy properties such as stellar mass, gas content, and dust attenuation influence the visibility and morphology of the resulting emission. Further work will be required to disentangle the relative roles of these processes and host-galaxy properties in shaping the observed ionized gas structures.

\section{Conclusions} \label{sec:conclusion}
In this paper, we presented the first statistically robust, spatially resolved measurements of [\ion{O}{3}]$+\mathrm{H}\beta$ and Pa$\beta$ emission in $\sim200$ galaxies at $2.5<z<2.9$, and investigated the relative roles of AGN and stellar processes in producing these features at Cosmic Noon. Using JWST/NIRCam medium-band imaging from JEMS, we measure ionized gas extents across AGN and control populations and relate these to AGN and host-galaxy properties derived from spectral energy distribution modeling. Our main results are summarized as follows:

\begin{enumerate}
    \item Applying multiple multiwavelength AGN selection techniques, we identify 33 AGN ($16\%$) within our full galaxy sample ($n=208$), with 32 galaxies ($15\%$) detected in Pa$\beta$.    
    \item The characteristic [\ion{O}{3}]$+\mathrm{H}\beta$ extents measured for our AGN are broadly consistent with previous studies at $z\sim3$, spanning $0.3$–$4.4$ kpc using average size metrics and $0.4$–$8.5$ kpc when adopting maximum radial extents. AGN more frequently exhibit conical [\ion{O}{3}]$+\mathrm{H}\beta$ morphologies than control galaxies, but many also show knotted or irregular structure, indicating that nuclear ionization often coexists with clumpy star-forming regions at Cosmic Noon.
    \item We derive slopes of $m=0.19$ (average size) and $m=0.21$ (maximum size) for the $\mathrm{R}_{\mathrm{[O III]+H}\beta}$–$\mathrm{L}_{\mathrm{AGN}}$ relation at $z\sim3$. These values lie at the shallow end of those reported at low redshift, supporting a matter-bounded interpretation in which gas availability limits the observable extent of AGN-ionized gas at $z\sim3$.
    \item We compare the radial extents of the [\ion{O}{3}]$+\mathrm{H}\beta$ and Pa$\beta$ emission across AGN and control galaxies. In both samples, [\ion{O}{3}]$+\mathrm{H}\beta$ emission extends to slightly larger radii than Pa$\beta$, by $\sim0.1$ dex. Both tracers also show systematically larger maximum radial extents in AGN hosts than in control galaxies, with median sizes of 2.2 kpc ([\ion{O}{3}]$+\mathrm{H}\beta$) and 2.3 kpc (Pa$\beta$) for AGN, compared to 1.1 kpc for both tracers in the control sample. These results indicate more extended ionized gas reservoirs in AGN hosts.
    \item AGN and Pa$\beta$-detected galaxies occupy similar regions of photometric and host-galaxy parameter space, including color–color space, [\ion{O}{3}]$+\mathrm{H}\beta$ equivalent width, star-formation rate, and stellar mass. This overlap suggests that elevated star formation and AGN activity frequently coexist in massive, gas-rich galaxies at Cosmic Noon.
\end{enumerate}

\section{Acknowledgments} 
NIRCam was built by a team at the University of Arizona (UofA) and Lockheed Martin's Advanced Technology Center, led by Prof. Marcia Rieke at UoA. JADES data taken under the JWST/NIRCam contract to the University of Arizona NAS5-02105. This research uses services or data provided by the Astro Data Lab, which is part of the Community Science and Data Center (CSDC) Program of NSF NOIRLab. NOIRLab is operated by the Association of Universities for Research in Astronomy (AURA), Inc. under a cooperative agreement with the U.S. National Science Foundation. The material contained in this document is based upon work supported by a National Aeronautics and Space Administration (NASA) cooperative agreement 80NSSC25M7084. Any opinions, findings, conclusions or recommendations expressed in this material are those of the author and do not necessarily reflect the views of NASA. The work of SL was supported through a NASA grant awarded to the Arizona/NASA Space Grant Consortium and [in part] by NASA program JWST-GO-01837.068-A .

This work is based [in part] on observations made with the NASA/ESA/CSA James Webb Space Telescope. The data were obtained from the Mikulski Archive for Space Telescopes at the Space Telescope Science Institute, which is operated by the Association of Universities for Research in Astronomy, Inc., under NASA contract NAS 5-03127 for JWST. These observations are associated with programs: JADES DOI: http://dx.doi.org/10.17909/8tdj-8n28 \citep{JADES_doi}, JADES DR2 DOI: http://dx.doi.org/10.17909/z2gw-mk31 \citep{JADES_DR2_doi}, JEMS DOI: https://dx.doi.org/10.17909/fsc4-dt61 \citep{JEMS_doi}, and FRESCO DOI: http://dx.doi.org/10.17909/gdyc-7g80 \citep{FRESCO_doi}. SL and CCW gratefully acknowledge support for program JWST-GO-1963 provided by NASA through a grant from the Space Telescope Science Institute, which is operated by the Association of Universities for Research in Astronomy, Inc., under NASA contract NAS 5-03127. The work of SJ and CCW is supported by NOIRLab, which is managed by the Association of Universities for Research in Astronomy (AURA) under a cooperative agreement with the National Science Foundation.  SP is supported by the international Gemini Observatory, a program of NSF NOIRLab, which is managed by the Association of Universities for Research in Astronomy (AURA) under a cooperative agreement with the U.S. National Science Foundation, on behalf of the Gemini partnership of Argentina, Brazil, Canada, Chile, the Republic of Korea, and the United States of America.

We respectfully acknowledge the University of Arizona is on the land and territories of Indigenous peoples. Today, Arizona is home to 22 federally recognized tribes, with Tucson being home to the O’odham and the Yaqui. Committed to diversity and inclusion, the University strives to build sustainable relationships with sovereign Native Nations and Indigenous communities through education offerings, partnerships, and community service.

\vspace{3mm}
\facilities{JWST (NIRCam), VLT (MUSE), Astro Data Lab}
\software{\texttt{numpy} \citep{harris2020}, \texttt{matplotlib} \citep{Hunter2007}, \texttt{astropy} \citep{astropy2013, astropy2018, astropy2022}, \texttt{WebbPSF} \citep{Perrin2014}, \texttt{astro-datalab} \citep{Fitzpatrick2014,Nikutta2020, Juneau2021_jupyter}}

\appendix
\label{Appendix}
\vspace{-0.75\baselineskip}
\enlargethispage{2\baselineskip}
\noindent
In this appendix, we present the full table of AGN identified in our JEMS sample at $z\sim2.5-2.9$. This table includes all 33 AGN used in the analysis and lists their coordinates, redshifts, galaxy properties, and [\ion{O}{3}]$+\mathrm{H}\beta$ and Pa$\beta$ spatial extent measurements. The complete catalog, including both the AGN and control samples, is available in a machine-readable format.

\begin{deluxetable}{ccccclcccc
@{\hspace{1pt}}c
@{\hspace{1pt}}c
@{\hspace{1pt}}c
@{\hspace{1pt}}c
@{\hspace{1pt}}c
@{\hspace{1pt}}c
@{\hspace{1pt}}c
@{\hspace{1pt}}c
@{\hspace{1pt}}c
@{\hspace{1pt}}c}
\setlength{\tabcolsep}{3pt}
\rotate
\tablecaption{Properties of JEMS AGN sample. Columns include source identifiers, coordinates, redshifts, Pa$\beta$ detection flags, AGN and galaxy properties, and spatial extent measurements of [\ion{O}{3}]$+\mathrm{H}\beta$ and Pa$\beta$. The full table, which includes the control sample, is available in machine-readable format. \label{tab:AGN_props}}
\tablehead{\colhead{ID} & \colhead{RA} & \colhead{DEC} & \colhead{z$_{\mathrm{phot}}$} & \colhead{z$_{\mathrm{spec}}$} &
\colhead{Pa$\beta$} & \colhead{f$_{\mathrm{F210M}}$} & \colhead{f$_{\mathrm{F430M}}$} &
\colhead{m$_{\mathrm{(F182M-F210M)}}$} & \colhead{EW$_{\mathrm{[OIII]}+\mathrm{H}\beta}$} &
\colhead{L$_{\mathrm{AGN},~0.1-30}$} & 
\colhead{L$_{\mathrm{ga},~0.1-1}$} &
\colhead{logMstar} & \colhead{SFR$_{\mathrm{opt}}$} &
\colhead{$\mathrm{R}_{\mathrm{avg, [OIII]}}$} & \colhead{$\mathrm{R}_{\mathrm{max, [OIII]}}$} & \colhead{$\mathrm{D}_{\mathrm{max, [OIII]}}$} &
\colhead{$\mathrm{R}_{\mathrm{avg,Pa}\beta}$} & \colhead{$\mathrm{R}_{\mathrm{max,Pa}\beta}$} & \colhead{$\mathrm{D}_{\mathrm{max,Pa}\beta}$}}
\startdata
193332 & 53.127 & -27.830 & 2.75 & \nodata & True & 1.01 & 1.01 & -0.14 & -76.7 & 44.01 & 44.10 & 9.84 & 3.36 & 1.32 & 2.48 & 7.61 & 1.01 & 1.84 & 1.11 \\
193915 & 53.144 & -27.828 & 2.80 & \nodata & True & 0.97 & 1.08 & -0.39 & -209.5 & 44.25 & 44.09 & 9.61 & 8.99 & 2.25 & 3.63 & 6.06 & 1.56 & 2.26 & 2.26 \\
194269 & 53.128 & -27.827 & 2.80 & \nodata & False & 0.98 & \nodata & -0.69 & -313.2 & 42.96 & 43.66 & 8.68 & 2.01 & 0.90 & 1.62 & 3.70 & \nodata & \nodata & \nodata \\
194373 & 53.140 & -27.827 & 2.72 & 2.670 & False & 0.95 & \nodata & 0.11 & 32.5 & 42.20 & 44.24 & 10.14 & 2.19 & 0.44 & 0.72 & 0.72 & \nodata & \nodata & \nodata \\
194952 & 53.138 & -27.825 & 2.69 & \nodata & False & 0.98 & \nodata & -0.17 & -103.1 & 44.07 & 44.08 & 9.72 & 4.19 & 1.38 & 2.37 & 2.37 & \nodata & \nodata & \nodata \\
195412 & 53.131 & -27.824 & 2.72 & \nodata & False & 1.05 & \nodata & -0.33 & -153.2 & 42.33 & 43.36 & 8.63 & 1.01 & 0.32 & 0.37 & 3.07 & \nodata & \nodata & \nodata \\
196134 & 53.129 & -27.822 & 2.66 & \nodata & False & 1.00 & \nodata & -0.21 & -116.1 & 44.07 & 44.04 & 9.49 & 7.80 & 1.56 & 2.50 & 2.50 & \nodata & \nodata & \nodata \\
196184 & 53.101 & -27.822 & 2.70 & \nodata & False & 0.79 & \nodata & 0.14 & -70.3 & 44.30 & 43.90 & 10.19 & 41.98 & 0.69 & 1.20 & 10.67 & \nodata & \nodata & \nodata \\
196187 & 53.102 & -27.821 & 2.64 & \nodata & False & 0.88 & \nodata & -0.02 & -94.7 & 43.01 & 43.20 & 8.92 & 0.22 & 0.75 & 1.12 & 1.12 & \nodata & \nodata & \nodata \\
196290 & 53.149 & -27.821 & 2.59 & \nodata & False & 0.77 & \nodata & 0.10 & -105.0 & 45.73 & 44.75 & 11.07 & 63.14 & 2.62 & 4.61 & 5.14 & \nodata & \nodata & \nodata \\
197581 & 53.142 & -27.817 & 2.65 & \nodata & True & 0.71 & 1.11 & 0.39 & 9.5 & 43.32 & 44.25 & 11.34 & 20.34 & 1.00 & 1.77 & 5.32 & 2.65 & 5.13 & 4.73 \\
198790 & 53.114 & -27.813 & 2.69 & \nodata & True & 1.04 & 1.02 & -0.22 & -100.8 & 44.47 & 44.41 & 10.03 & 10.85 & 1.47 & 1.87 & 3.36 & 1.12 & 2.11 & 2.65 \\
199443 & 53.128 & -27.811 & 2.65 & 2.617 & True & 1.01 & 1.01 & -0.91 & -381.3 & 43.61 & 44.01 & 8.94 & 4.70 & 2.00 & 3.37 & 5.78 & 0.35 & 0.52 & 2.13 \\
199494 & 53.115 & -27.810 & 2.68 & \nodata & True & 0.91 & 1.07 & -0.11 & -118.5 & 43.88 & 43.65 & 9.54 & 3.32 & 0.87 & 1.36 & 1.36 & 0.65 & 1.12 & 1.22 \\
199996 & 53.120 & -27.808 & 2.76 & \nodata & True & 0.68 & 1.11 & 0.20 & -118.9 & 44.32 & 44.60 & 11.76 & 3.83 & 4.36 & 8.48 & 8.98 & 5.09 & 7.06 & 6.00 \\
200800 & 53.118 & -27.805 & 2.68 & \nodata & True & 1.04 & 1.02 & -0.30 & -142.1 & 44.28 & 44.31 & 9.85 & 12.45 & 1.74 & 3.31 & 4.61 & 0.93 & 1.72 & 2.63 \\
201584 & 53.164 & -27.803 & 2.68 & \nodata & False & 1.02 & \nodata & -0.12 & -61.7 & 44.04 & 44.09 & 9.77 & 3.17 & 0.89 & 1.44 & 3.43 & \nodata & \nodata & \nodata \\
202378 & 53.130 & -27.800 & 2.63 & \nodata & True & 0.88 & 1.10 & 0.10 & -24.1 & 41.90 & 44.15 & 10.25 & 63.95 & 1.46 & 2.22 & 5.73 & 2.99 & 5.59 & 3.82 \\
202380 & 53.130 & -27.799 & 2.65 & \nodata & False & 0.93 & \nodata & 0.02 & -36.3 & 43.31 & 43.81 & 9.58 & 2.19 & 0.49 & 0.75 & 3.05 & \nodata & \nodata & \nodata \\
202484 & 53.120 & -27.799 & 2.62 & \nodata & True & 0.68 & 1.09 & 0.25 & -99.2 & 43.79 & 44.33 & 10.93 & 1.03 & 2.36 & 4.62 & 6.41 & 2.60 & 5.12 & 5.15 \\
202597 & 53.156 & -27.799 & 2.75 & \nodata & True & 1.02 & 1.06 & -0.05 & -15.5 & 44.26 & 43.99 & 9.79 & 3.27 & 0.79 & 1.05 & 1.05 & 0.95 & 1.40 & 1.93 \\
204449 & 53.161 & -27.792 & 2.65 & \nodata & True & 0.95 & 1.02 & -0.05 & -58.1 & 44.17 & 44.14 & 10.11 & 0.84 & 1.57 & 2.92 & 2.84 & 0.97 & 1.78 & 2.60 \\
207277 & 53.176 & -27.783 & 2.81 & \nodata & False & 0.74 & \nodata & 0.19 & -74.1 & 44.77 & 44.34 & 9.84 & 718.46 & 1.23 & 2.17 & 3.76 & \nodata & \nodata & \nodata \\
207592 & 53.173 & -27.781 & 2.77 & \nodata & False & 0.99 & \nodata & -0.00 & -8.8 & 43.60 & 43.71 & 9.26 & 2.80 & 0.78 & 1.39 & 1.39 & \nodata & \nodata & \nodata \\
208000 & 53.146 & -27.780 & 2.61 & 2.582 & False & 0.68 & \nodata & 0.24 & -103.0 & 43.46 & 44.40 & 10.87 & 19.13 & 1.93 & 3.68 & 3.68 & \nodata & \nodata & \nodata \\
208176 & 53.182 & -27.780 & 2.67 & \nodata & False & 1.00 & \nodata & -0.29 & -158.0 & 44.21 & 44.18 & 9.97 & 11.50 & 2.31 & 3.74 & 3.74 & \nodata & \nodata & \nodata \\
208820 & 53.181 & -27.778 & 2.82 & \nodata & True & 0.61 & 1.23 & 0.40 & -80.6 & 42.78 & 44.08 & 10.99 & 43.34 & 0.46 & 0.62 & 11.34 & 2.48 & 4.03 & 4.03 \\
209026 & 53.183 & -27.777 & 2.82 & \nodata & False & 0.71 & \nodata & 0.09 & -150.0 & 45.13 & 44.11 & 11.01 & 48.34 & 3.33 & 6.49 & 6.49 & \nodata & \nodata & \nodata \\
209027 & 53.183 & -27.776 & 2.79 & \nodata & True & 1.01 & 1.07 & -0.19 & -97.6 & 44.21 & 44.27 & 10.05 & 11.95 & 1.05 & 1.85 & 4.64 & 1.42 & 2.56 & 4.24 \\
209116 & 53.160 & -27.776 & 2.72 & \nodata & True & 0.99 & 1.02 & -0.16 & -95.1 & 44.35 & 44.27 & 9.97 & 7.77 & 1.71 & 2.91 & 2.91 & 0.79 & 1.27 & 2.58 \\
209117 & 53.161 & -27.776 & 2.73 & 2.542 & True & 0.88 & 1.12 & 0.03 & -59.0 & 44.15 & 44.29 & 10.50 & 56.55 & 1.78 & 2.77 & 4.73 & 2.41 & 4.19 & 2.90 \\
209617 & 53.151 & -27.774 & 2.69 & \nodata & True & 0.78 & 1.06 & 0.17 & -54.4 & 43.23 & 44.44 & 10.35 & 50.49 & 2.15 & 4.13 & 5.89 & 1.92 & 3.67 & 4.35 \\
419920 & 53.136 & -27.814 & 2.63 & \nodata & False & 0.99 & \nodata & -0.04 & -35.1 & 43.96 & 43.06 & 9.50 & 0.51 & 0.40 & 0.63 & 0.63 & \nodata & \nodata & \nodata
\enddata
\end{deluxetable}

\clearpage

\bibliography{ms}{}

@ARTICLE{Juneau2013,
       author = {{Juneau}, St{\'e}phanie and {Dickinson}, Mark and {Bournaud}, Fr{\'e}d{\'e}ric and {Alexander}, David M. and {Daddi}, Emanuele and {Mullaney}, James R. and {Magnelli}, Benjamin and {Kartaltepe}, Jeyhan S. and {Hwang}, Ho Seong and {Willner}, S.~P. and {Coil}, Alison L. and {Rosario}, David J. and {Trump}, Jonathan R. and {Weiner}, Benjamin J. and {Willmer}, Christopher N.~A. and {Cooper}, Michael C. and {Elbaz}, David and {Faber}, S.~M. and {Frayer}, David T. and {Kocevski}, Dale D. and {Laird}, Elise S. and {Monkiewicz}, Jacqueline A. and {Nandra}, Kirpal and {Newman}, Jeffrey A. and {Salim}, Samir and {Symeonidis}, Myrto},
        title = "{Widespread and Hidden Active Galactic Nuclei in Star-forming Galaxies at Redshift >0.3}",
      journal = {\apj},
         year = 2013,
        month = feb,
       volume = {764},
       number = {2},
          eid = {176},
        pages = {176},
          doi = {10.1088/0004-637X/764/2/176},
archivePrefix = {arXiv},
       eprint = {1211.6436},
 primaryClass = {astro-ph.CO},
       adsurl = {https://ui.adsabs.harvard.edu/abs/2013ApJ...764..176J}
}

@ARTICLE{Juneau2022,
       author = {{Juneau}, St{\'e}phanie and {Goulding}, Andy D. and {Banfield}, Julie and {Bianchi}, Stefano and {Duc}, Pierre-Alain and {Ho}, I. -Ting and {Dopita}, Michael A. and {Scharw{\"a}chter}, Julia and {Bauer}, Franz E. and {Groves}, Brent and {Alexander}, David M. and {Davies}, Rebecca L. and {Elbaz}, David and {Freeland}, Emily and {Hampton}, Elise and {Kewley}, Lisa J. and {Nikutta}, Robert and {Shastri}, Prajval and {Shu}, Xinwen and {Vogt}, Fr{\'e}d{\'e}ric P.~A. and {Wang}, Tao and {Wong}, O. Ivy and {Woo}, Jong-Hak},
        title = "{The Black Hole-Galaxy Connection: Interplay between Feedback, Obscuration, and Host Galaxy Substructure}",
      journal = {\apj},
         year = 2022,
        month = feb,
       volume = {925},
       number = {2},
          eid = {203},
        pages = {203},
          doi = {10.3847/1538-4357/ac425f},
archivePrefix = {arXiv},
       eprint = {2112.08380},
 primaryClass = {astro-ph.GA},
       adsurl = {https://ui.adsabs.harvard.edu/abs/2022ApJ...925..203J}
}

@ARTICLE{Lopez-coba,
       author = {{L{\'o}pez-Cob{\'a}}, Carlos and {S{\'a}nchez}, Sebasti{\'a}n F. and {Anderson}, Joseph P. and {Cruz-Gonz{\'a}lez}, Irene and {Galbany}, Llu{\'\i}s and {Ruiz-Lara}, Tom{\'a}s and {Barrera-Ballesteros}, Jorge K. and {Prieto}, Jos{\'e} L. and {Kuncarayakti}, Hanindyo},
        title = "{The AMUSING++ Nearby Galaxy Compilation. I. Full Sample Characterization and Galactic-scale Outflow Selection}",
      journal = {\aj},
         year = 2020,
        month = apr,
       volume = {159},
       number = {4},
          eid = {167},
        pages = {167},
          doi = {10.3847/1538-3881/ab7848},
archivePrefix = {arXiv},
       eprint = {2002.09328},
 primaryClass = {astro-ph.GA},
       adsurl = {https://ui.adsabs.harvard.edu/abs/2020AJ....159..167L}
}

@ARTICLE{Fiore2017,
       author = {{Fiore}, F. and {Feruglio}, C. and {Shankar}, F. and {Bischetti}, M. and {Bongiorno}, A. and {Brusa}, M. and {Carniani}, S. and {Cicone}, C. and {Duras}, F. and {Lamastra}, A. and {Mainieri}, V. and {Marconi}, A. and {Menci}, N. and {Maiolino}, R. and {Piconcelli}, E. and {Vietri}, G. and {Zappacosta}, L.},
        title = "{AGN wind scaling relations and the co-evolution of black holes and galaxies}",
      journal = {\aap},
         year = 2017,
        month = may,
       volume = {601},
          eid = {A143},
        pages = {A143},
          doi = {10.1051/0004-6361/201629478},
archivePrefix = {arXiv},
       eprint = {1702.04507},
 primaryClass = {astro-ph.GA},
       adsurl = {https://ui.adsabs.harvard.edu/abs/2017A&A...601A.143F}
}

@ARTICLE{Luo2017,
       author = {{Luo}, B. and {Brandt}, W.~N. and {Xue}, Y.~Q. and {Lehmer}, B. and {Alexander}, D.~M. and {Bauer}, F.~E. and {Vito}, F. and {Yang}, G. and {Basu-Zych}, A.~R. and {Comastri}, A. and {Gilli}, R. and {Gu}, Q. -S. and {Hornschemeier}, A.~E. and {Koekemoer}, A. and {Liu}, T. and {Mainieri}, V. and {Paolillo}, M. and {Ranalli}, P. and {Rosati}, P. and {Schneider}, D.~P. and {Shemmer}, O. and {Smail}, I. and {Sun}, M. and {Tozzi}, P. and {Vignali}, C. and {Wang}, J. -X.},
        title = "{The Chandra Deep Field-South Survey: 7 Ms Source Catalogs}",
      journal = {\apjs},
         year = 2017,
        month = jan,
       volume = {228},
       number = {1},
          eid = {2},
        pages = {2},
          doi = {10.3847/1538-4365/228/1/2},
archivePrefix = {arXiv},
       eprint = {1611.03501},
 primaryClass = {astro-ph.GA},
       adsurl = {https://ui.adsabs.harvard.edu/abs/2017ApJS..228....2L}
}

@ARTICLE{Rieke2023b,
       author = {{Rieke}, Marcia J. and {Robertson}, Brant and {Tacchella}, Sandro and {Hainline}, Kevin and {Johnson}, Benjamin D. and {Hausen}, Ryan and {Ji}, Zhiyuan and {Willmer}, Christopher N.~A. and {Eisenstein}, Daniel J. and {Pusk{\'a}s}, D{\'a}vid and {Alberts}, Stacey and {Arribas}, Santiago and {Baker}, William M. and {Baum}, Stefi and {Bhatawdekar}, Rachana and {Bonaventura}, Nina and {Boyett}, Kristan and {Bunker}, Andrew J. and {Cameron}, Alex J. and {Carniani}, Stefano and {Charlot}, Stephane and {Chevallard}, Jacopo and {Chen}, Zuyi and {Curti}, Mirko and {Curtis-Lake}, Emma and {Danhaive}, A. Lola and {DeCoursey}, Christa and {Dressler}, Alan and {Egami}, Eiichi and {Endsley}, Ryan and {Helton}, Jakob M. and {Hviding}, Raphael E. and {Kumari}, Nimisha and {Looser}, Tobias J. and {Lyu}, Jianwei and {Maiolino}, Roberto and {Maseda}, Michael V. and {Nelson}, Erica J. and {Rieke}, George and {Rix}, Hans-Walter and {Sandles}, Lester and {Saxena}, Aayush and {Sharpe}, Katherine and {Shivaei}, Irene and {Skarbinski}, Maya and {Smit}, Renske and {Stark}, Daniel P. and {Stone}, Meredith and {Suess}, Katherine A. and {Sun}, Fengwu and {Topping}, Michael and {{\"U}bler}, Hannah and {Villanueva}, Natalia C. and {Wallace}, Imaan E.~B. and {Williams}, Christina C. and {Willott}, Chris and {Whitler}, Lily and {Witstok}, Joris and {Woodrum}, Charity},
        title = "{JADES Initial Data Release for the Hubble Ultra Deep Field: Revealing the Faint Infrared Sky with Deep JWST NIRCam Imaging}",
      journal = {\apjs},
         year = 2023,
        month = nov,
       volume = {269},
       number = {1},
          eid = {16},
        pages = {16},
          doi = {10.3847/1538-4365/acf44d},
archivePrefix = {arXiv},
       eprint = {2306.02466},
 primaryClass = {astro-ph.GA},
       adsurl = {https://ui.adsabs.harvard.edu/abs/2023ApJS..269...16R}
}

@ARTICLE{Alberts2020,
       author = {{Alberts}, Stacey and {Rujopakarn}, Wiphu and {Rieke}, George H. and {Jagannathan}, Preshanth and {Nyland}, Kristina},
        title = "{Completing the Census of AGN in GOODS-S/HUDF: New Ultradeep Radio Imaging and Predictions for JWST}",
      journal = {\apj},
         year = 2020,
        month = oct,
       volume = {901},
       number = {2},
          eid = {168},
        pages = {168},
          doi = {10.3847/1538-4357/abb1a0},
archivePrefix = {arXiv},
       eprint = {2008.11208},
 primaryClass = {astro-ph.GA},
       adsurl = {https://ui.adsabs.harvard.edu/abs/2020ApJ...901..168A}
}

@ARTICLE{Hainline2023,
       author = {{Hainline}, Kevin N. and {Johnson}, Benjamin D. and {Robertson}, Brant and {Tacchella}, Sandro and {Helton}, Jakob M. and {Sun}, Fengwu and {Eisenstein}, Daniel J. and {Simmonds}, Charlotte and {Topping}, Michael W. and {Whitler}, Lily and {Willmer}, Christopher N.~A. and {Rieke}, Marcia and {Suess}, Katherine A. and {Hviding}, Raphael E. and {Cameron}, Alex J. and {Alberts}, Stacey and {Baker}, William M. and {Bhatawdekar}, Rachana and {Boyett}, Kristan and {Bunker}, Andrew J. and {Carniani}, Stefano and {Charlot}, Stephane and {Chen}, Zuyi and {Curti}, Mirko and {Curtis-Lake}, Emma and {D'Eugenio}, Francesco and {Egami}, Eiichi and {Endsley}, Ryan and {Hausen}, Ryan and {Ji}, Zhiyuan and {Looser}, Tobias J. and {Lyu}, Jianwei and {Maiolino}, Roberto and {Nelson}, Erica and {Puskas}, David and {Rawle}, Tim and {Sandles}, Lester and {Saxena}, Aayush and {Smit}, Renske and {Stark}, Daniel P. and {Williams}, Christina C. and {Willott}, Chris and {Witstok}, Joris},
        title = "{The Cosmos in its Infancy: JADES Galaxy Candidates at z > 8 in GOODS-S and GOODS-N}",
      journal = {arXiv e-prints},
         year = 2023,
        month = jun,
          eid = {arXiv:2306.02468},
        pages = {arXiv:2306.02468},
          doi = {10.48550/arXiv.2306.02468},
archivePrefix = {arXiv},
       eprint = {2306.02468},
 primaryClass = {astro-ph.GA},
       adsurl = {https://ui.adsabs.harvard.edu/abs/2023arXiv230602468H}
}

@ARTICLE{Eisenstein2023a,
       author = {{Eisenstein}, Daniel J. and {Willott}, Chris and {Alberts}, Stacey and {Arribas}, Santiago and {Bonaventura}, Nina and {Bunker}, Andrew J. and {Cameron}, Alex J. and {Carniani}, Stefano and {Charlot}, Stephane and {Curtis-Lake}, Emma and {D'Eugenio}, Francesco and {Endsley}, Ryan and {Ferruit}, Pierre and {Giardino}, Giovanna and {Hainline}, Kevin and {Hausen}, Ryan and {Jakobsen}, Peter and {Johnson}, Benjamin D. and {Maiolino}, Roberto and {Rieke}, Marcia and {Rieke}, George and {Rix}, Hans-Walter and {Robertson}, Brant and {Stark}, Daniel P. and {Tacchella}, Sandro and {Williams}, Christina C. and {Willmer}, Christopher N.~A. and {Baker}, William M. and {Baum}, Stefi and {Bhatawdekar}, Rachana and {Boyett}, Kristan and {Chen}, Zuyi and {Chevallard}, Jacopo and {Circosta}, Chiara and {Curti}, Mirko and {Danhaive}, A. Lola and {DeCoursey}, Christa and {de Graaff}, Anna and {Dressler}, Alan and {Egami}, Eiichi and {Helton}, Jakob M. and {Hviding}, Raphael E. and {Ji}, Zhiyuan and {Jones}, Gareth C. and {Kumari}, Nimisha and {L{\"u}tzgendorf}, Nora and {Laseter}, Isaac and {Looser}, Tobias J. and {Lyu}, Jianwei and {Maseda}, Michael V. and {Nelson}, Erica and {Parlanti}, Eleonora and {Perna}, Michele and {Pusk{\'a}s}, D{\'a}vid and {Rawle}, Tim and {Rodr{\'\i}guez Del Pino}, Bruno and {Sandles}, Lester and {Saxena}, Aayush and {Scholtz}, Jan and {Sharpe}, Katherine and {Shivaei}, Irene and {Silcock}, Maddie S. and {Simmonds}, Charlotte and {Skarbinski}, Maya and {Smit}, Renske and {Stone}, Meredith and {Suess}, Katherine A. and {Sun}, Fengwu and {Tang}, Mengtao and {Topping}, Michael W. and {{\"U}bler}, Hannah and {Villanueva}, Natalia C. and {Wallace}, Imaan E.~B. and {Whitler}, Lily and {Witstok}, Joris and {Woodrum}, Charity},
        title = "{Overview of the JWST Advanced Deep Extragalactic Survey (JADES)}",
      journal = {arXiv e-prints},
         year = 2023,
        month = jun,
          eid = {arXiv:2306.02465},
        pages = {arXiv:2306.02465},
          doi = {10.48550/arXiv.2306.02465},
archivePrefix = {arXiv},
       eprint = {2306.02465},
 primaryClass = {astro-ph.GA},
       adsurl = {https://ui.adsabs.harvard.edu/abs/2023arXiv230602465E}
}

@ARTICLE{Eisenstein2023b,
       author = {{Eisenstein}, Daniel J. and {Johnson}, Benjamin D. and {Robertson}, Brant and {Tacchella}, Sandro and {Hainline}, Kevin and {Jakobsen}, Peter and {Maiolino}, Roberto and {Bonaventura}, Nina and {Bunker}, Andrew J. and {Cameron}, Alex J. and {Cargile}, Phillip A. and {Curtis-Lake}, Emma and {Hausen}, Ryan and {Pusk{\'a}s}, D{\'a}vid and {Rieke}, Marcia and {Sun}, Fengwu and {Willmer}, Christopher N.~A. and {Willott}, Chris and {Alberts}, Stacey and {Arribas}, Santiago and {Baker}, William M. and {Baum}, Stefi and {Bhatawdekar}, Rachana and {Carniani}, Stefano and {Charlot}, Stephane and {Chen}, Zuyi and {Chevallard}, Jacopo and {Curti}, Mirko and {DeCoursey}, Christa and {D'Eugenio}, Francesco and {de Graaff}, Anna and {Egami}, Eiichi and {Helton}, Jakob M. and {Ji}, Zhiyuan and {Jones}, Gareth C. and {Kumari}, Nimisha and {L{\"u}tzgendorf}, Nora and {Laseter}, Isaac and {Looser}, Tobias J. and {Lyu}, Jianwei and {Maseda}, Michael V. and {Nelson}, Erica and {Parlanti}, Eleonora and {Rauscher}, Bernard J. and {Rawle}, Tim and {Rieke}, George and {Rix}, Hans-Walter and {Rujopakarn}, Wiphu and {Sandles}, Lester and {Saxena}, Aayush and {Scholtz}, Jan and {Sharpe}, Katherine and {Shivaei}, Irene and {Simmonds}, Charlotte and {Smit}, Renske and {Topping}, Michael W. and {{\"U}bler}, Hannah and {Venturi}, Giacomo and {Williams}, Christina C. and {Witstok}, Joris and {Woodrum}, Charity},
        title = "{The JADES Origins Field: A New JWST Deep Field in the JADES Second NIRCam Data Release}",
      journal = {arXiv e-prints},
         year = 2023,
        month = oct,
          eid = {arXiv:2310.12340},
        pages = {arXiv:2310.12340},
          doi = {10.48550/arXiv.2310.12340},
archivePrefix = {arXiv},
       eprint = {2310.12340},
 primaryClass = {astro-ph.GA},
       adsurl = {https://ui.adsabs.harvard.edu/abs/2023arXiv231012340E}
}

@ARTICLE{Hunter2007,
       author = {{Hunter}, John D.},
        title = "{Matplotlib: A 2D Graphics Environment}",
      journal = {Computing in Science and Engineering},
         year = 2007,
        month = may,
       volume = {9},
       number = {3},
        pages = {90-95},
          doi = {10.1109/MCSE.2007.55},
       adsurl = {https://ui.adsabs.harvard.edu/abs/2007CSE.....9...90H}
}

@ARTICLE{Lyu2022,
       author = {{Lyu}, Jianwei and {Alberts}, Stacey and {Rieke}, George H. and {Rujopakarn}, Wiphu},
        title = "{AGN Selection and Demographics in GOODS-S/HUDF from X-Ray to Radio}",
      journal = {\apj},
         year = 2022,
        month = dec,
       volume = {941},
       number = {2},
          eid = {191},
        pages = {191},
          doi = {10.3847/1538-4357/ac9e5d},
archivePrefix = {arXiv},
       eprint = {2209.06219},
 primaryClass = {astro-ph.GA},
       adsurl = {https://ui.adsabs.harvard.edu/abs/2022ApJ...941..191L}
}

@ARTICLE{Oesch2023,
       author = {{Oesch}, P.~A. and {Brammer}, G. and {Naidu}, R.~P. and {Bouwens}, R.~J. and {Chisholm}, J. and {Illingworth}, G.~D. and {Matthee}, J. and {Nelson}, E. and {Qin}, Y. and {Reddy}, N. and {Shapley}, A. and {Shivaei}, I. and {van Dokkum}, P. and {Weibel}, A. and {Whitaker}, K. and {Wuyts}, S. and {Covelo-Paz}, A. and {Endsley}, R. and {Fudamoto}, Y. and {Giovinazzo}, E. and {Herard-Demanche}, T. and {Kerutt}, J. and {Kramarenko}, I. and {Labbe}, I. and {Leonova}, E. and {Lin}, J. and {Magee}, D. and {Marchesini}, D. and {Maseda}, M. and {Mason}, C. and {Matharu}, J. and {Meyer}, R.~A. and {Neufeld}, C. and {Prieto Lyon}, G. and {Schaerer}, D. and {Sharma}, R. and {Shuntov}, M. and {Smit}, R. and {Stefanon}, M. and {Wyithe}, J.~S.~B. and {Xiao}, M.},
        title = "{The JWST FRESCO survey: legacy NIRCam/grism spectroscopy and imaging in the two GOODS fields}",
      journal = {\mnras},
         year = 2023,
        month = oct,
       volume = {525},
       number = {2},
        pages = {2864-2874},
          doi = {10.1093/mnras/stad2411},
archivePrefix = {arXiv},
       eprint = {2304.02026},
 primaryClass = {astro-ph.GA},
       adsurl = {https://ui.adsabs.harvard.edu/abs/2023MNRAS.525.2864O}
}

@ARTICLE{Williams2023,
       author = {{Williams}, Christina C. and {Tacchella}, Sandro and {Maseda}, Michael V. and {Robertson}, Brant E. and {Johnson}, Benjamin D. and {Willott}, Chris J. and {Eisenstein}, Daniel J. and {Willmer}, Christopher N.~A. and {Ji}, Zhiyuan and {Hainline}, Kevin N. and {Helton}, Jakob M. and {Alberts}, Stacey and {Baum}, Stefi and {Bhatawdekar}, Rachana and {Boyett}, Kristan and {Bunker}, Andrew J. and {Carniani}, Stefano and {Charlot}, Stephane and {Chevallard}, Jacopo and {Curtis-Lake}, Emma and {de Graaff}, Anna and {Egami}, Eiichi and {Franx}, Marijn and {Kumari}, Nimisha and {Maiolino}, Roberto and {Nelson}, Erica J. and {Rieke}, Marcia J. and {Sandles}, Lester and {Shivaei}, Irene and {Simmonds}, Charlotte and {Smit}, Renske and {Suess}, Katherine A. and {Sun}, Fengwu and {{\"U}bler}, Hannah and {Witstok}, Joris},
        title = "{JEMS: A Deep Medium-band Imaging Survey in the Hubble Ultra Deep Field with JWST NIRCam and NIRISS}",
      journal = {\apjs},
         year = 2023,
        month = oct,
       volume = {268},
       number = {2},
          eid = {64},
        pages = {64},
          doi = {10.3847/1538-4365/acf130},
archivePrefix = {arXiv},
       eprint = {2301.09780},
 primaryClass = {astro-ph.GA},
       adsurl = {https://ui.adsabs.harvard.edu/abs/2023ApJS..268...64W}
}

@ARTICLE{Durre&Mould2018,
       author = {{Durr{\'e}}, Mark and {Mould}, Jeremy},
        title = "{The AGN Ionization Cones of NGC 5728. I. Excitation and Nuclear Structure}",
      journal = {\apj},
         year = 2018,
        month = nov,
       volume = {867},
       number = {2},
          eid = {149},
        pages = {149},
          doi = {10.3847/1538-4357/aae68e},
archivePrefix = {arXiv},
       eprint = {1810.03258},
 primaryClass = {astro-ph.GA},
       adsurl = {https://ui.adsabs.harvard.edu/abs/2018ApJ...867..149D}
}

@ARTICLE{Sun2017,
       author = {{Sun}, Ai-Lei and {Greene}, Jenny E. and {Zakamska}, Nadia L.},
        title = "{Sizes and Kinematics of Extended Narrow-line Regions in Luminous Obscured AGN Selected by Broadband Images}",
      journal = {\apj},
         year = 2017,
        month = feb,
       volume = {835},
       number = {2},
          eid = {222},
        pages = {222},
          doi = {10.3847/1538-4357/835/2/222},
archivePrefix = {arXiv},
       eprint = {1611.04469},
 primaryClass = {astro-ph.GA},
       adsurl = {https://ui.adsabs.harvard.edu/abs/2017ApJ...835..222S}
}

@ARTICLE{Hainline2013,
       author = {{Hainline}, Kevin N. and {Hickox}, Ryan and {Greene}, Jenny E. and {Myers}, Adam D. and {Zakamska}, Nadia L.},
        title = "{SALT Long-slit Spectroscopy of Luminous Obscured Quasars: An Upper Limit on the Size of the Narrow-line Region?}",
      journal = {\apj},
         year = 2013,
        month = sep,
       volume = {774},
       number = {2},
          eid = {145},
        pages = {145},
          doi = {10.1088/0004-637X/774/2/145},
archivePrefix = {arXiv},
       eprint = {1307.5852},
 primaryClass = {astro-ph.CO},
       adsurl = {https://ui.adsabs.harvard.edu/abs/2013ApJ...774..145H}
}

@ARTICLE{Haineline2014,
       author = {{Hainline}, Kevin N. and {Hickox}, Ryan C. and {Greene}, Jenny E. and {Myers}, Adam D. and {Zakamska}, Nadia L. and {Liu}, Guilin and {Liu}, Xin},
        title = "{Gemini Long-slit Observations of Luminous Obscured Quasars: Further Evidence for an Upper Limit on the Size of the Narrow-line Region}",
      journal = {\apj},
         year = 2014,
        month = may,
       volume = {787},
       number = {1},
          eid = {65},
        pages = {65},
          doi = {10.1088/0004-637X/787/1/65},
archivePrefix = {arXiv},
       eprint = {1404.1921},
 primaryClass = {astro-ph.GA},
       adsurl = {https://ui.adsabs.harvard.edu/abs/2014ApJ...787...65H}
}

@ARTICLE{Bennert2002,
       author = {{Bennert}, Nicola and {Falcke}, Heino and {Schulz}, Hartmut and {Wilson}, Andrew S. and {Wills}, Beverley J.},
        title = "{Size and Structure of the Narrow-Line Region of Quasars}",
      journal = {\apjl},
         year = 2002,
        month = aug,
       volume = {574},
       number = {2},
        pages = {L105-L109},
          doi = {10.1086/342420},
archivePrefix = {arXiv},
       eprint = {astro-ph/0206334},
 primaryClass = {astro-ph},
       adsurl = {https://ui.adsabs.harvard.edu/abs/2002ApJ...574L.105B}
}

@ARTICLE{Schreiber2020,
       author = {{F{\"o}rster Schreiber}, Natascha M. and {Wuyts}, Stijn},
        title = "{Star-Forming Galaxies at Cosmic Noon}",
      journal = {\araa},
         year = 2020,
        month = aug,
       volume = {58},
        pages = {661-725},
          doi = {10.1146/annurev-astro-032620-021910},
archivePrefix = {arXiv},
       eprint = {2010.10171},
 primaryClass = {astro-ph.GA},
       adsurl = {https://ui.adsabs.harvard.edu/abs/2020ARA&A..58..661F}
}

@ARTICLE{Heckman&Best2014,
       author = {{Heckman}, Timothy M. and {Best}, Philip N.},
        title = "{The Coevolution of Galaxies and Supermassive Black Holes: Insights from Surveys of the Contemporary Universe}",
      journal = {\araa},
         year = 2014,
        month = aug,
       volume = {52},
        pages = {589-660},
          doi = {10.1146/annurev-astro-081913-035722},
archivePrefix = {arXiv},
       eprint = {1403.4620},
 primaryClass = {astro-ph.GA},
       adsurl = {https://ui.adsabs.harvard.edu/abs/2014ARA&A..52..589H}
}

@ARTICLE{Hickox&Alexander2018,
       author = {{Hickox}, Ryan C. and {Alexander}, David M.},
        title = "{Obscured Active Galactic Nuclei}",
      journal = {\araa},
         year = 2018,
        month = sep,
       volume = {56},
        pages = {625-671},
          doi = {10.1146/annurev-astro-081817-051803},
archivePrefix = {arXiv},
       eprint = {1806.04680},
 primaryClass = {astro-ph.GA},
       adsurl = {https://ui.adsabs.harvard.edu/abs/2018ARA&A..56..625H}
}

@ARTICLE{Antonucci1993,
       author = {{Antonucci}, Robert},
        title = "{Unified models for active galactic nuclei and quasars.}",
      journal = {\araa},
         year = 1993,
        month = jan,
       volume = {31},
        pages = {473-521},
          doi = {10.1146/annurev.aa.31.090193.002353},
       adsurl = {https://ui.adsabs.harvard.edu/abs/1993ARA&A..31..473A}
}

@ARTICLE{Netzer2015,
       author = {{Netzer}, Hagai},
        title = "{Revisiting the Unified Model of Active Galactic Nuclei}",
      journal = {\araa},
         year = 2015,
        month = aug,
       volume = {53},
        pages = {365-408},
          doi = {10.1146/annurev-astro-082214-122302},
archivePrefix = {arXiv},
       eprint = {1505.00811},
 primaryClass = {astro-ph.GA},
       adsurl = {https://ui.adsabs.harvard.edu/abs/2015ARA&A..53..365N}
}

@ARTICLE{Lamastra2009,
       author = {{Lamastra}, A. and {Bianchi}, S. and {Matt}, G. and {Perola}, G.~C. and {Barcons}, X. and {Carrera}, F.~J.},
        title = "{The bolometric luminosity of type 2 AGN from extinction-corrected [OIII]. No evidence of Eddington-limited sources}",
      journal = {\aap},
         year = 2009,
        month = sep,
       volume = {504},
       number = {1},
        pages = {73-79},
          doi = {10.1051/0004-6361/200912023},
archivePrefix = {arXiv},
       eprint = {0905.4439},
 primaryClass = {astro-ph.CO},
       adsurl = {https://ui.adsabs.harvard.edu/abs/2009A&A...504...73L}
}

@ARTICLE{DiMatteo2005,
       author = {{Di Matteo}, Tiziana and {Springel}, Volker and {Hernquist}, Lars},
        title = "{Energy input from quasars regulates the growth and activity of black holes and their host galaxies}",
      journal = {\nat},
         year = 2005,
        month = feb,
       volume = {433},
       number = {7026},
        pages = {604-607},
          doi = {10.1038/nature03335},
archivePrefix = {arXiv},
       eprint = {astro-ph/0502199},
 primaryClass = {astro-ph},
       adsurl = {https://ui.adsabs.harvard.edu/abs/2005Natur.433..604D}
}

@ARTICLE{Baldwin1981,
       author = {{Baldwin}, J.~A. and {Phillips}, M.~M. and {Terlevich}, R.},
        title = "{Classification parameters for the emission-line spectra of extragalactic objects.}",
      journal = {\pasp},
         year = 1981,
        month = feb,
       volume = {93},
        pages = {5-19},
          doi = {10.1086/130766},
       adsurl = {https://ui.adsabs.harvard.edu/abs/1981PASP...93....5B}
}

@ARTICLE{Madau2014,
       author = {{Madau}, Piero and {Dickinson}, Mark},
        title = "{Cosmic Star-Formation History}",
      journal = {\araa},
         year = 2014,
        month = aug,
       volume = {52},
        pages = {415-486},
          doi = {10.1146/annurev-astro-081811-125615},
archivePrefix = {arXiv},
       eprint = {1403.0007},
 primaryClass = {astro-ph.CO},
       adsurl = {https://ui.adsabs.harvard.edu/abs/2014ARA&A..52..415M}
}

@BOOK{Peterson1997,
       author = {{Peterson}, Bradley M.},
        title = "{An Introduction to Active Galactic Nuclei}",
         year = 1997,
       adsurl = {https://ui.adsabs.harvard.edu/abs/1997iagn.book.....P}
}

@ARTICLE{Liu2014,
       author = {{Liu}, Guilin and {Zakamska}, Nadia L. and {Greene}, Jenny E.},
        title = "{Similarity of ionized gas nebulae around unobscured and obscured quasars}",
      journal = {\mnras},
         year = 2014,
        month = aug,
       volume = {442},
       number = {2},
        pages = {1303-1318},
          doi = {10.1093/mnras/stu974},
archivePrefix = {arXiv},
       eprint = {1401.0536},
 primaryClass = {astro-ph.GA},
       adsurl = {https://ui.adsabs.harvard.edu/abs/2014MNRAS.442.1303L}
}

@ARTICLE{Brammer2008,
       author = {{Brammer}, Gabriel B. and {van Dokkum}, Pieter G. and {Coppi}, Paolo},
        title = "{EAZY: A Fast, Public Photometric Redshift Code}",
      journal = {\apj},
         year = 2008,
        month = oct,
       volume = {686},
       number = {2},
        pages = {1503-1513},
          doi = {10.1086/591786},
archivePrefix = {arXiv},
       eprint = {0807.1533},
 primaryClass = {astro-ph},
       adsurl = {https://ui.adsabs.harvard.edu/abs/2008ApJ...686.1503B}
}

@ARTICLE{Maiolino2008,
       author = {{Maiolino}, R. and {Nagao}, T. and {Grazian}, A. and {Cocchia}, F. and {Marconi}, A. and {Mannucci}, F. and {Cimatti}, A. and {Pipino}, A. and {Ballero}, S. and {Calura}, F. and {Chiappini}, C. and {Fontana}, A. and {Granato}, G.~L. and {Matteucci}, F. and {Pastorini}, G. and {Pentericci}, L. and {Risaliti}, G. and {Salvati}, M. and {Silva}, L.},
        title = "{AMAZE. I. The evolution of the mass-metallicity relation at z > 3}",
      journal = {\aap},
         year = 2008,
        month = sep,
       volume = {488},
       number = {2},
        pages = {463-479},
          doi = {10.1051/0004-6361:200809678},
archivePrefix = {arXiv},
       eprint = {0806.2410},
 primaryClass = {astro-ph},
       adsurl = {https://ui.adsabs.harvard.edu/abs/2008A&A...488..463M}
}

@ARTICLE{Liu2013a,
       author = {{Liu}, Guilin and {Zakamska}, Nadia L. and {Greene}, Jenny E. and {Nesvadba}, Nicole P.~H. and {Liu}, Xin},
        title = "{Observations of feedback from radio-quiet quasars - I. Extents and morphologies of ionized gas nebulae}",
      journal = {\mnras},
         year = 2013,
        month = apr,
       volume = {430},
       number = {3},
        pages = {2327-2345},
          doi = {10.1093/mnras/stt051},
archivePrefix = {arXiv},
       eprint = {1301.1677},
 primaryClass = {astro-ph.CO},
       adsurl = {https://ui.adsabs.harvard.edu/abs/2013MNRAS.430.2327L}
}

@ARTICLE{Greene2011,
       author = {{Greene}, Jenny E. and {Zakamska}, Nadia L. and {Ho}, Luis C. and {Barth}, Aaron J.},
        title = "{Feedback in Luminous Obscured Quasars}",
      journal = {\apj},
         year = 2011,
        month = may,
       volume = {732},
       number = {1},
          eid = {9},
        pages = {9},
          doi = {10.1088/0004-637X/732/1/9},
archivePrefix = {arXiv},
       eprint = {1102.2913},
 primaryClass = {astro-ph.CO},
       adsurl = {https://ui.adsabs.harvard.edu/abs/2011ApJ...732....9G}
}

@ARTICLE{Humphrey2010,
       author = {{Humphrey}, A. and {Villar-Mart{\'\i}n}, M. and {S{\'a}nchez}, S.~F. and {Mart{\'\i}nez-Sansigre}, A. and {Delgado}, R. Gonz{\'a}lez and {P{\'e}rez}, E. and {Tadhunter}, C. and {P{\'e}rez-Torres}, M.~A.},
        title = "{Integral-field spectroscopy of type II QSOs at z = 0.3-0.4}",
      journal = {\mnras},
         year = 2010,
        month = oct,
       volume = {408},
       number = {1},
        pages = {L1-L5},
          doi = {10.1111/j.1745-3933.2010.00906.x},
archivePrefix = {arXiv},
       eprint = {1006.3751},
 primaryClass = {astro-ph.CO},
       adsurl = {https://ui.adsabs.harvard.edu/abs/2010MNRAS.408L...1H}
}

@ARTICLE{Husemann2013,
       author = {{Husemann}, B. and {Wisotzki}, L. and {S{\'a}nchez}, S.~F. and {Jahnke}, K.},
        title = "{The properties of the extended warm ionised gas around low-redshift QSOs and the lack of extended high-velocity outflows}",
      journal = {\aap},
         year = 2013,
        month = jan,
       volume = {549},
          eid = {A43},
        pages = {A43},
          doi = {10.1051/0004-6361/201220076},
archivePrefix = {arXiv},
       eprint = {1210.0566},
 primaryClass = {astro-ph.CO},
       adsurl = {https://ui.adsabs.harvard.edu/abs/2013A&A...549A..43H}
}

@ARTICLE{Bennert2006,
       author = {{Bennert}, N. and {Jungwiert}, B. and {Komossa}, S. and {Haas}, M. and {Chini}, R.},
        title = "{Size and properties of the narrow-line region in Seyfert-2 galaxies from spatially-resolved optical spectroscopy}",
      journal = {\aap},
         year = 2006,
        month = sep,
       volume = {456},
       number = {3},
        pages = {953-966},
          doi = {10.1051/0004-6361:20065319},
archivePrefix = {arXiv},
       eprint = {astro-ph/0607636},
 primaryClass = {astro-ph},
       adsurl = {https://ui.adsabs.harvard.edu/abs/2006A&A...456..953B}
}

@ARTICLE{Fraquelli2003,
       author = {{Fraquelli}, Henrique A. and {Storchi-Bergmann}, T. and {Levenson}, N.~A.},
        title = "{Extended gas in Seyfert 2 galaxies: implications for the nuclear source}",
      journal = {\mnras},
         year = 2003,
        month = may,
       volume = {341},
       number = {2},
        pages = {449-463},
          doi = {10.1046/j.1365-8711.2003.06397.x},
       adsurl = {https://ui.adsabs.harvard.edu/abs/2003MNRAS.341..449F}
}

@ARTICLE{Wylezalek2022,
       author = {{Wylezalek}, Dominika and {Vayner}, Andrey and {Rupke}, David S.~N. and {Zakamska}, Nadia L. and {Veilleux}, Sylvain and {Ishikawa}, Yuzo and {Bertemes}, Caroline and {Liu}, Weizhe and {Barrera-Ballesteros}, Jorge K. and {Chen}, Hsiao-Wen and {Goulding}, Andy D. and {Greene}, Jenny E. and {Hainline}, Kevin N. and {Hamann}, Fred and {Heckman}, Timothy and {Johnson}, Sean D. and {Lutz}, Dieter and {L{\"u}tzgendorf}, Nora and {Mainieri}, Vincenzo and {Maiolino}, Roberto and {Nesvadba}, Nicole P.~H. and {Ogle}, Patrick and {Sturm}, Eckhard},
        title = "{First Results from the JWST Early Release Science Program Q3D: Turbulent Times in the Life of a z   3 Extremely Red Quasar Revealed by NIRSpec IFU}",
      journal = {\apjl},
         year = 2022,
        month = nov,
       volume = {940},
       number = {1},
          eid = {L7},
        pages = {L7},
          doi = {10.3847/2041-8213/ac98c3},
archivePrefix = {arXiv},
       eprint = {2210.10074},
 primaryClass = {astro-ph.GA},
       adsurl = {https://ui.adsabs.harvard.edu/abs/2022ApJ...940L...7W}
}

@ARTICLE{Gardner2023,
       author = {{Gardner}, Jonathan P. and {Mather}, John C. and {Abbott}, Randy and {Abell}, James S. and {Abernathy}, Mark and {Abney}, Faith E. and {Abraham}, John G. and {Abraham}, Roberto and {Abul-Huda}, Yasin M. and {Acton}, Scott and {Adams}, Cynthia K. and {Adams}, Evan and {Adler}, David S. and {Adriaensen}, Maarten and {Aguilar}, Jonathan Albert and {Ahmed}, Mansoor and {Ahmed}, Nasif S. and {Ahmed}, Tanjira and {Albat}, R{\"u}deger and {Albert}, Lo{\"\i}c and {Alberts}, Stacey and {Aldridge}, David and {Allen}, Mary Marsha and {Allen}, Shaune S. and {Altenburg}, Martin and {Altunc}, Serhat and {Alvarez}, Jose Lorenzo and {{\'A}lvarez-M{\'a}rquez}, Javier and {Alves de Oliveira}, Catarina and {Ambrose}, Leslie L. and {Anandakrishnan}, Satya M. and {Andersen}, Gregory C. and {Anderson}, Harry James and {Anderson}, Jay and {Anderson}, Kristen and {Anderson}, Sara M. and {Aprea}, Julio and {Archer}, Benita J. and {Arenberg}, Jonathan W. and {Argyriou}, Ioannis and {Arribas}, Santiago and {Artigau}, {\'E}tienne and {Arvai}, Amanda Rose and {Atcheson}, Paul and {Atkinson}, Charles B. and {Averbukh}, Jesse and {Aymergen}, Cagatay and {Bacinski}, John J. and {Baggett}, Wayne E. and {Bagnasco}, Giorgio and {Baker}, Lynn L. and {Balzano}, Vicki Ann and {Banks}, Kimberly A. and {Baran}, David A. and {Barker}, Elizabeth A. and {Barrett}, Larry K. and {Barringer}, Bruce O. and {Barto}, Allison and {Bast}, William and {Baudoz}, Pierre and {Baum}, Stefi and {Beatty}, Thomas G. and {Beaulieu}, Mathilde and {Bechtold}, Kathryn and {Beck}, Tracy and {Beddard}, Megan M. and {Beichman}, Charles and {Bellagama}, Larry and {Bely}, Pierre and {Berger}, Timothy W. and {Bergeron}, Louis E. and {Bernier}, Antoine-Darveau and {Bertch}, Maria D. and {Beskow}, Charlotte and {Betz}, Laura E. and {Biagetti}, Carl P. and {Birkmann}, Stephan and {Bjorklund}, Kurt F. and {Blackwood}, James D. and {Blazek}, Ronald Paul and {Blossfeld}, Stephen and {Bluth}, Marcel and {Boccaletti}, Anthony and {Boegner}, Martin E., Jr. and {Bohlin}, Ralph C. and {Boia}, John Joseph and {B{\"o}ker}, Torsten and {Bonaventura}, N. and {Bond}, Nicholas A. and {Bosley}, Kari Ann and {Boucarut}, Rene A. and {Bouchet}, Patrice and {Bouwman}, Jeroen and {Bower}, Gary and {Bowers}, Ariel S. and {Bowers}, Charles W. and {Boyce}, Leslye A. and {Boyer}, Christine T. and {Boyer}, Martha L. and {Boyer}, Michael and {Boyer}, Robert and {Bradley}, Larry D. and {Brady}, Gregory R. and {Brandl}, Bernhard R. and {Brannen}, Judith L. and {Breda}, David and {Bremmer}, Harold G. and {Brennan}, David and {Bresnahan}, Pamela A. and {Bright}, Stacey N. and {Broiles}, Brian J. and {Bromenschenkel}, Asa and {Brooks}, Brian H. and {Brooks}, Keira J. and {Brown}, Bob and {Brown}, Bruce and {Brown}, Thomas M. and {Bruce}, Barry W. and {Bryson}, Jonathan G. and {Bujanda}, Edwin D. and {Bullock}, Blake M. and {Bunker}, A.~J. and {Bureo}, Rafael and {Burt}, Irving J. and {Bush}, James Aaron and {Bushouse}, Howard A. and {Bussman}, Marie C. and {Cabaud}, Olivier and {Cale}, Steven and {Calhoon}, Charles D. and {Calvani}, Humberto and {Canipe}, Alicia M. and {Caputo}, Francis M. and {Cara}, Mihai and {Carey}, Larkin and {Case}, Michael Eli and {Cesari}, Thaddeus and {Cetorelli}, Lee D. and {Chance}, Don R. and {Chandler}, Lynn and {Chaney}, Dave and {Chapman}, George N. and {Charlot}, S. and {Chayer}, Pierre and {Cheezum}, Jeffrey I. and {Chen}, Bin and {Chen}, Christine H. and {Cherinka}, Brian and {Chichester}, Sarah C. and {Chilton}, Zachary S. and {Chittiraibalan}, Dharini and {Clampin}, Mark and {Clark}, Charles R. and {Clark}, Kerry W. and {Clark}, Stephanie M. and {Claybrooks}, Edward E. and {Cleveland}, Keith A. and {Cohen}, Andrew L. and {Cohen}, Lester M. and {Col{\'o}n}, Knicole D. and {Coleman}, Benee L. and {Colina}, Luis and {Comber}, Brian J. and {Comeau}, Thomas M. and {Comer}, Thomas and {Conde Reis}, Alain and {Connolly}, Dennis C. and {Conroy}, Kyle E. and {Contos}, Adam R. and {Contreras}, James and {Cook}, Neil J. and {Cooper}, James L. and {Cooper}, Rachel Aviva and {Correia}, Michael F. and {Correnti}, Matteo and {Cossou}, Christophe and {Costanza}, Brian F. and {Coulais}, Alain and {Cox}, Colin R. and {Coyle}, Ray T. and {Cracraft}, Misty M. and {Crew}, Keith A. and {Curtis}, Gary J. and {Cusveller}, Bianca and {Da Costa Maciel}, Cleyciane and {Dailey}, Christopher T. and {Daugeron}, Fr{\'e}d{\'e}ric and {Davidson}, Greg S. and {Davies}, James E. and {Davis}, Katherine Anne and {Davis}, Michael S. and {Day}, Ratna and {de Chambure}, Daniel and {de Jong}, Pauline and {De Marchi}, Guido and {Dean}, Bruce H. and {Decker}, John E. and {Delisa}, Amy S. and {Dell}, Lawrence C. and {Dellagatta}, Gail and {Dembinska}, Franciszka and {Demosthenes}, Sandor and {Dencheva}, Nadezhda M. and {Deneu}, Philippe and {DePriest}, William W. and {Deschenes}, Jeremy and {Dethienne}, Nathalie and {Detre}, {\"O}rs Hunor and {Diaz}, Rosa Izela and {Dicken}, Daniel and {DiFelice}, Audrey S. and {Dillman}, Matthew and {Disharoon}, Maureen O. and {Dixon}, William V. and {Doggett}, Jesse B. and {Dominguez}, Keisha L. and {Donaldson}, Thomas S. and {Doria-Warner}, Cristina M. and {Santos}, Tony Dos and {Doty}, Heather and {Douglas}, Robert E., Jr. and {Doyon}, Ren{\'e} and {Dressler}, Alan and {Driggers}, Jennifer and {Driggers}, Phillip A. and {Dunn}, Jamie L. and {DuPrie}, Kimberly C. and {Dupuis}, Jean and {Durning}, John and {Dutta}, Sanghamitra B. and {Earl}, Nicholas M. and {Eccleston}, Paul and {Ecobichon}, Pascal and {Egami}, Eiichi and {Ehrenwinkler}, Ralf and {Eisenhamer}, Jonathan D. and {Eisenhower}, Michael and {Eisenstein}, Daniel J. and {El Hamel}, Zaky and {Elie}, Michelle L. and {Elliott}, James and {Elliott}, Kyle Wesley and {Engesser}, Michael and {Espinoza}, N{\'e}stor and {Etienne}, Odessa and {Etxaluze}, Mireya and {Evans}, Leah and {Fabreguettes}, Luce and {Falcolini}, Massimo and {Falini}, Patrick R. and {Fatig}, Curtis and {Feeney}, Matthew and {Feinberg}, Lee D. and {Fels}, Raymond and {Ferdous}, Nazma and {Ferguson}, Henry C. and {Ferrarese}, Laura and {Ferreira}, Marie-H{\'e}l{\'e}ne and {Ferruit}, Pierre and {Ferry}, Malcolm and {Filippazzo}, Joseph Charles and {Firre}, Daniel and {Fix}, Mees and {Flagey}, Nicolas and {Flanagan}, Kathryn A. and {Fleming}, Scott W. and {Florian}, Michael and {Flynn}, James R. and {Foiadelli}, Luca and {Fontaine}, Mark R. and {Fontanella}, Erin Marie and {Forshay}, Peter Randolph and {Fortner}, Elizabeth A. and {Fox}, Ori D. and {Framarini}, Alexandro P. and {Francisco}, John I. and {Franck}, Randy and {Franx}, Marijn and {Franz}, David E. and {Friedman}, Scott D. and {Friend}, Katheryn E. and {Frost}, James R. and {Fu}, Henry and {Fullerton}, Alexander W. and {Gaillard}, Lionel and {Galkin}, Sergey and {Gallagher}, Ben and {Galyer}, Anthony D. and {Garc{\'\i}a Mar{\'\i}n}, Macarena and {Gardner}, Lisa E. and {Garland}, Dennis and {Garrett}, Bruce Albert and {Gasman}, Danny and {G{\'a}sp{\'a}r}, Andr{\'a}s and {Gastaud}, Ren{\'e} and {Gaudreau}, Daniel and {Gauthier}, Peter Timothy and {Geers}, Vincent and {Geithner}, Paul H. and {Gennaro}, Mario and {Gerber}, John and {Gereau}, John C. and {Giampaoli}, Robert and {Giardino}, Giovanna and {Gibbons}, Paul C. and {Gilbert}, Karoline and {Gilman}, Larry and {Girard}, Julien H. and {Giuliano}, Mark E. and {Gkountis}, Konstantinos and {Glasse}, Alistair and {Glassmire}, Kirk Zachary and {Glauser}, Adrian Michael and {Glazer}, Stuart D. and {Goldberg}, Joshua and {Golimowski}, David A. and {Gonzaga}, Shireen P. and {Gordon}, Karl D. and {Gordon}, Shawn J. and {Goudfrooij}, Paul and {Gough}, Michael J. and {Graham}, Adrian J. and {Grau}, Christopher M. and {Green}, Joel David and {Greene}, Gretchen R. and {Greene}, Thomas P. and {Greenfield}, Perry E. and {Greenhouse}, Matthew A. and {Greve}, Thomas R. and {Greville}, Edgar M. and {Grimaldi}, Stefano and {Groe}, Frank E. and {Groebner}, Andrew and {Grumm}, David M. and {Grundy}, Timothy and {G{\"u}del}, Manuel and {Guillard}, Pierre and {Guldalian}, John and {Gunn}, Christopher A. and {Gurule}, Anthony and {Gutman}, Irvin Meyer and {Guy}, Paul D. and {Guyot}, Benjamin and {Hack}, Warren J. and {Haderlein}, Peter and {Hagan}, James B. and {Hagedorn}, Andria and {Hainline}, Kevin and {Haley}, Craig and {Hami}, Maryam and {Hamilton}, Forrest Clifford and {Hammann}, Jeffrey and {Hammel}, Heidi B. and {Hanley}, Christopher J. and {Hansen}, Carl August and {Hardy}, Bruce and {Harnisch}, Bernd and {Harr}, Michael Hunter and {Harris}, Pamela and {Hart}, Jessica Ann and {Hartig}, George F. and {Hasan}, Hashima and {Hashim}, Kathleen Marie and {Hashimoto}, Ryan and {Haskins}, Sujee J. and {Hawkins}, Robert Edward and {Hayden}, Brian and {Hayden}, William L. and {Healy}, Mike and {Hecht}, Karen and {Heeg}, Vince J. and {Hejal}, Reem and {Helm}, Kristopher A. and {Hengemihle}, Nicholas J. and {Henning}, Thomas and {Henry}, Alaina and {Henry}, Ronald L. and {Henshaw}, Katherine and {Hernandez}, Scarlin and {Herrington}, Donald C. and {Heske}, Astrid and {Hesman}, Brigette Emily and {Hickey}, David L. and {Hilbert}, Bryan N. and {Hines}, Dean C. and {Hinz}, Michael R. and {Hirsch}, Michael and {Hitcho}, Robert S. and {Hodapp}, Klaus and {Hodge}, Philip E. and {Hoffman}, Melissa and {Holfeltz}, Sherie T. and {Holler}, Bryan Jason and {Hoppa}, Jennifer Rose and {Horner}, Scott and {Howard}, Joseph M. and {Howard}, Richard J. and {Huber}, Jean M. and {Hunkeler}, Joseph S. and {Hunter}, Alexander and {Hunter}, David Gavin and {Hurd}, Spencer W. and {Hurst}, Brendan J. and {Hutchings}, John B. and {Hylan}, Jason E. and {Ignat}, Luminita Ilinca and {Illingworth}, Garth and {Irish}, Sandra M. and {Isaacs}, John C., III and {Jackson}, Wallace C., Jr. and {Jaffe}, Daniel T. and {Jahic}, Jasmin and {Jahromi}, Amir and {Jakobsen}, Peter and {James}, Bryan and {James}, John C. and {James}, LeAndrea Rae and {Jamieson}, William Brian and {Jandra}, Raymond D. and {Jayawardhana}, Ray and {Jedrzejewski}, Robert and {Jeffers}, Basil S. and {Jensen}, Peter and {Joanne}, Egges and {Johns}, Alan T. and {Johnson}, Carl A. and {Johnson}, Eric L. and {Johnson}, Patricia and {Johnson}, Phillip Stephen and {Johnson}, Thomas K. and {Johnson}, Timothy W. and {Johnstone}, Doug and {Jollet}, Delphine and {Jones}, Danny P. and {Jones}, Gregory S. and {Jones}, Olivia C. and {Jones}, Ronald A. and {Jones}, Vicki and {Jordan}, Ian J. and {Jordan}, Margaret E. and {Jue}, Reginald and {Jurkowski}, Mark H. and {Justis}, Grant and {Justtanont}, Kay and {Kaleida}, Catherine C. and {Kalirai}, Jason S. and {Kalmanson}, Phillip Cabrales and {Kaltenegger}, Lisa and {Kammerer}, Jens and {Kan}, Samuel K. and {Kanarek}, Graham Childs and {Kao}, Shaw-Hong and {Karakla}, Diane M. and {Karl}, Hermann and {Kassin}, Susan A. and {Kauffman}, David D. and {Kavanagh}, Patrick and {Kelley}, Leigh L. and {Kelly}, Douglas M. and {Kendrew}, Sarah and {Kennedy}, Herbert V. and {Kenny}, Deborah A. and {Keski-Kuha}, Ritva A. and {Keyes}, Charles D. and {Khan}, Ali and {Kidwell}, Richard C. and {Kimble}, Randy A. and {King}, James S. and {King}, Richard C. and {Kinzel}, Wayne M. and {Kirk}, Jeffrey R. and {Kirkpatrick}, Marc E. and {Klaassen}, Pamela and {Klingemann}, Lana and {Klintworth}, Paul U. and {Knapp}, Bryan Adam and {Knight}, Scott and {Knollenberg}, Perry J. and {Knutsen}, Daniel Mark and {Koehler}, Robert and {Koekemoer}, Anton M. and {Kofler}, Earl T. and {Kontson}, Vicki L. and {Kovacs}, Aiden Rose and {Kozhurina-Platais}, Vera and {Krause}, Oliver and {Kriss}, Gerard A. and {Krist}, John and {Kristoffersen}, Monica R. and {Krogel}, Claudia and {Krueger}, Anthony P. and {Kulp}, Bernard A. and {Kumari}, Nimisha and {Kwan}, Sandy W. and {Kyprianou}, Mark and {Labador}, Aurora Gadiano and {Labiano}, {\'A}lvaro and {Lafreni{\`e}re}, David and {Lagage}, Pierre-Olivier and {Laidler}, Victoria G. and {Laine}, Benoit and {Laird}, Simon and {Lajoie}, Charles-Philippe and {Lallo}, Matthew D. and {Lam}, May Yen and {LaMassa}, Stephanie Marie and {Lambros}, Scott D. and {Lampenfield}, Richard Joseph and {Lander}, Matthew Ed and {Langston}, James Hutton and {Larson}, Kirsten and {Larson}, Melora and {LaVerghetta}, Robert Joseph and {Law}, David R. and {Lawrence}, Jon F. and {Lee}, David W. and {Lee}, Janice and {Lee}, Yat-Ning Paul and {Leisenring}, Jarron and {Leveille}, Michael Dunlap and {Levenson}, Nancy A. and {Levi}, Joshua S. and {Levine}, Marie B. and {Lewis}, Dan and {Lewis}, Jake and {Lewis}, Nikole and {Libralato}, Mattia and {Lidon}, Norbert and {Liebrecht}, Paula Louisa and {Lightsey}, Paul and {Lilly}, Simon and {Lim}, Frederick C. and {Lim}, Pey Lian and {Ling}, Sai-Kwong and {Link}, Lisa J. and {Link}, Miranda Nicole and {Lipinski}, Jamie L. and {Liu}, XiaoLi and {Lo}, Amy S. and {Lobmeyer}, Lynette and {Logue}, Ryan M. and {Long}, Chris A. and {Long}, Douglas R. and {Long}, Ilana D. and {Long}, Knox S. and {L{\'o}pez-Caniego}, Marcos and {Lotz}, Jennifer M. and {Love-Pruitt}, Jennifer M. and {Lubskiy}, Michael and {Luers}, Edward B. and {Luetgens}, Robert A. and {Luevano}, Annetta J. and {Lui}, Sarah Marie G. Flores and {Lund}, James M., III and {Lundquist}, Ray A. and {Lunine}, Jonathan and {L{\"u}tzgendorf}, Nora and {Lynch}, Richard J. and {MacDonald}, Alex J. and {MacDonald}, Kenneth and {Macias}, Matthew J. and {Macklis}, Keith I. and {Maghami}, Peiman and {Maharaja}, Rishabh Y. and {Maiolino}, Roberto and {Makrygiannis}, Konstantinos G. and {Malla}, Sunita Giri and {Malumuth}, Eliot M. and {Manjavacas}, Elena and {Marini}, Andrea and {Marrione}, Amanda and {Marston}, Anthony and {Martel}, Andr{\'e} R. and {Martin}, Didier and {Martin}, Peter G. and {Martinez}, Kristin L. and {Maschmann}, Marc and {Masci}, Gregory L. and {Masetti}, Margaret E. and {Maszkiewicz}, Michael and {Matthews}, Gary and {Matuskey}, Jacob E. and {McBrayer}, Glen A. and {McCarthy}, Donald W. and {McCaughrean}, Mark J. and {McClare}, Leslie A. and {McClare}, Michael D. and {McCloskey}, John C. and {McClurg}, Taylore D. and {McCoy}, Martin and {McElwain}, Michael W. and {McGregor}, Roy D. and {McGuffey}, Douglas B. and {McKay}, Andrew G. and {McKenzie}, William K. and {McLean}, Brian and {McMaster}, Matthew and {McNeil}, Warren and {De Meester}, Wim and {Mehalick}, Kimberly L. and {Meixner}, Margaret and {Mel{\'e}ndez}, Marcio and {Menzel}, Michael P. and {Menzel}, Michael T. and {Merz}, Matthew and {Mesterharm}, David D. and {Meyer}, Michael R. and {Meyett}, Michele L. and {Meza}, Luis E. and {Midwinter}, Calvin and {Milam}, Stefanie N. and {Miller}, Jay Todd and {Miller}, William C. and {Miskey}, Cherie L. and {Misselt}, Karl and {Mitchell}, Eileen P. and {Mohan}, Martin and {Montoya}, Emily E. and {Moran}, Michael J. and {Morishita}, Takahiro and {Moro-Mart{\'\i}n}, Amaya and {Morrison}, Debra L. and {Morrison}, Jane and {Morse}, Ernie C. and {Moschos}, Michael and {Moseley}, S.~H. and {Mosier}, Gary E. and {Mosner}, Peter and {Mountain}, Matt and {Muckenthaler}, Jason S. and {Mueller}, Donald G. and {Mueller}, Migo and {Muhiem}, Daniella and {M{\"u}hlmann}, Prisca and {Mullally}, Susan Elizabeth and {Mullen}, Stephanie M. and {Munger}, Alan J. and {Murphy}, Jess and {Murray}, Katherine T. and {Muzerolle}, James C. and {Mycroft}, Matthew and {Myers}, Andrew and {Myers}, Carey R. and {Myers}, Fred Richard R. and {Myers}, Richard and {Myrick}, Kaila and {Nagle}, Adrian F., IV and {Nayak}, Omnarayani and {Naylor}, Bret and {Neff}, Susan G. and {Nelan}, Edmund P. and {Nella}, John and {Nguyen}, Duy Tuong and {Nguyen}, Michael N. and {Nickson}, Bryony and {Nidhiry}, John Joseph and {Niedner}, Malcolm B. and {Nieto-Santisteban}, Maria and {Nikolov}, Nikolay K. and {Nishisaka}, Mary Ann and {Noriega-Crespo}, Alberto and {Nota}, Antonella and {O'Mara}, Robyn C. and {Oboryshko}, Michael and {O'Brien}, Marcus B. and {Ochs}, William R. and {Offenberg}, Joel D. and {Ogle}, Patrick Michael and {Ohl}, Raymond G. and {Olmsted}, Joseph Hamden and {Osborne}, Shannon Barbara and {O'Shaughnessy}, Brian Patrick and {{\"O}stlin}, G{\"o}ran and {O'Sullivan}, Brian and {Otor}, O. Justin and {Ottens}, Richard and {Ouellette}, Nathalie N. -Q. and {Outlaw}, Daria J. and {Owens}, Beverly A. and {Pacifici}, Camilla and {Page}, James Christophe and {Paranilam}, James G. and {Park}, Sang and {Parrish}, Keith A. and {Paschal}, Laura and {Patapis}, Polychronis and {Patel}, Jignasha and {Patrick}, Keith and {Pattishall}, Robert A., Jr. and {Paul}, Douglas William and {Paul}, Shirley J. and {Pauly}, Tyler Andrew and {Pavlovsky}, Cheryl M. and {Pe{\~n}a-Guerrero}, Maria and {Pedder}, Andrew H. and {Peek}, Matthew Weldon and {Pelham}, Patricia A. and {Penanen}, Konstantin and {Perriello}, Beth A. and {Perrin}, Marshall D. and {Perrine}, Richard F. and {Perrygo}, Chuck and {Peslier}, Muriel and {Petach}, Michael and {Peterson}, Karla A. and {Pfarr}, Tom and {Pierson}, James M. and {Pietraszkiewicz}, Martin and {Pilchen}, Guy and {Pipher}, Judy L. and {Pirzkal}, Norbert and {Pitman}, Joseph T. and {Player}, Danielle M. and {Plesha}, Rachel and {Plitzke}, Anja and {Pohner}, John A. and {Poletis}, Karyn Konstantin and {Pollizzi}, Joseph A. and {Polster}, Ethan and {Pontius}, James T. and {Pontoppidan}, Klaus and {Porges}, Susana C. and {Potter}, Gregg D. and {Prescott}, Stephen and {Proffitt}, Charles R. and {Pueyo}, Laurent and {Quispe Neira}, Irma Aracely and {Radich}, Armando and {Rager}, Reiko T. and {Rameau}, Julien and {Ramey}, Deborah D. and {Ramos Alarcon}, Rafael and {Rampini}, Riccardo and {Rapp}, Robert and {Rashford}, Robert A. and {Rauscher}, Bernard J. and {Ravindranath}, Swara and {Rawle}, Timothy and {Rawlings}, Tynika N. and {Ray}, Tom and {Regan}, Michael W. and {Rehm}, Brian and {Rehm}, Kenneth D. and {Reid}, Neill and {Reis}, Carl A. and {Renk}, Florian and {Reoch}, Tom B. and {Ressler}, Michael and {Rest}, Armin W. and {Reynolds}, Paul J. and {Richon}, Joel G. and {Richon}, Karen V. and {Ridgaway}, Michael and {Riedel}, Adric Richard and {Rieke}, George H. and {Rieke}, Marcia J. and {Rifelli}, Richard E. and {Rigby}, Jane R. and {Riggs}, Catherine S. and {Ringel}, Nancy J. and {Ritchie}, Christine E. and {Rix}, Hans-Walter and {Robberto}, Massimo and {Robinson}, Gregory L. and {Robinson}, Michael S. and {Robinson}, Orion and {Rock}, Frank W. and {Rodriguez}, David R. and {Rodr{\'\i}guez del Pino}, Bruno and {Roellig}, Thomas and {Rohrbach}, Scott O. and {Roman}, Anthony J. and {Romelfanger}, Frederick J. and {Romo}, Felipe P., Jr. and {Rosales}, Jose J. and {Rose}, Perry and {Roteliuk}, Anthony F. and {Roth}, Marc N. and {Rothwell}, Braden Quinn and {Rouzaud}, Sylvain and {Rowe}, Jason and {Rowlands}, Neil and {Roy}, Arpita and {Royer}, Pierre and {Rui}, Chunlei and {Rumler}, Peter and {Rumpl}, William and {Russ}, Melissa L. and {Ryan}, Michael B. and {Ryan}, Richard M. and {Saad}, Karl and {Sabata}, Modhumita and {Sabatino}, Rick and {Sabbi}, Elena and {Sabelhaus}, Phillip A. and {Sabia}, Stephen and {Sahu}, Kailash C. and {Saif}, Babak N. and {Salvignol}, Jean-Christophe and {Samara-Ratna}, Piyal and {Samuelson}, Bridget S. and {Sanders}, Felicia A. and {Sappington}, Bradley and {Sargent}, B.~A. and {Sauer}, Arne and {Savadkin}, Bruce J. and {Sawicki}, Marcin and {Schappell}, Tina M. and {Scheffer}, Caroline and {Scheithauer}, Silvia and {Scherer}, Ron and {Schiff}, Conrad and {Schlawin}, Everett and {Schmeitzky}, Olivier and {Schmitz}, Tyler S. and {Schmude}, Donald J. and {Schneider}, Analyn and {Schreiber}, J{\"u}rgen and {Schroeven-Deceuninck}, Hilde and {Schultz}, John J. and {Schwab}, Ryan and {Schwartz}, Curtis H. and {Scoccimarro}, Dario and {Scott}, John F. and {Scott}, Michelle B. and {Seaton}, Bonita L. and {Seely}, Bruce S. and {Seery}, Bernard and {Seidleck}, Mark and {Sembach}, Kenneth and {Shanahan}, Clare Elizabeth and {Shaughnessy}, Bryan and {Shaw}, Richard A. and {Shay}, Christopher Michael and {Sheehan}, Even and {Sheth}, Kartik and {Shih}, Hsin-Yi and {Shivaei}, Irene and {Siegel}, Noah and {Sienkiewicz}, Matthew G. and {Simmons}, Debra D. and {Simon}, Bernard P. and {Sirianni}, Marco and {Sivaramakrishnan}, Anand and {Slade}, Jeffrey E. and {Sloan}, G.~C. and {Slocum}, Christine E. and {Slowinski}, Steven E. and {Smith}, Corbett T. and {Smith}, Eric P. and {Smith}, Erin C. and {Smith}, Koby and {Smith}, Robert and {Smith}, Stephanie J. and {Smolik}, John L. and {Soderblom}, David R. and {Sohn}, Sangmo Tony and {Sokol}, Jeff and {Sonneborn}, George and {Sontag}, Christopher D. and {Sooy}, Peter R. and {Soummer}, Remi and {Southwood}, Dana M. and {Spain}, Kay and {Sparmo}, Joseph and {Speer}, David T. and {Spencer}, Richard and {Sprofera}, Joseph D. and {Stallcup}, Scott S. and {Stanley}, Marcia K. and {Stansberry}, John A. and {Stark}, Christopher C. and {Starr}, Carl W. and {Stassi}, Diane Y. and {Steck}, Jane A. and {Steeley}, Christine D. and {Stephens}, Matthew A. and {Stephenson}, Ralph J. and {Stewart}, Alphonso C. and {Stiavelli}, Massimo and {}, Stockman, Hervey Jr. and {Strada}, Paolo and {Straughn}, Amber N. and {Streetman}, Scott and {Strickland}, David Kendal and {Strobele}, Jingping F. and {Stuhlinger}, Martin and {Stys}, Jeffrey Edward and {Such}, Miguel and {Sukhatme}, Kalyani and {Sullivan}, Joseph F. and {Sullivan}, Pamela C. and {Sumner}, Sandra M. and {Sun}, Fengwu and {Sunnquist}, Benjamin Dale and {Swade}, Daryl Allen and {Swam}, Michael S. and {Swenton}, Diane F. and {Swoish}, Robby A. and {Tam Litten}, Oi In and {Tamas}, Laszlo and {Tao}, Andrew and {Taylor}, David K. and {Taylor}, Joanna M. and {te Plate}, Maurice and {Van Tea}, Mason and {Teague}, Kelly K. and {Telfer}, Randal C. and {Temim}, Tea and {Texter}, Scott C. and {Thatte}, Deepashri G. and {Thompson}, Christopher Lee and {Thompson}, Linda M. and {Thomson}, Shaun R. and {Thronson}, Harley and {Tierney}, C.~M. and {Tikkanen}, Tuomo and {Tinnin}, Lee and {Tippet}, William Thomas and {Todd}, Connor William and {Tran}, Hien D. and {Trauger}, John and {Trejo}, Edwin Gregorio and {Vinh Truong}, Justin Hoang and {Tsukamoto}, Christine L. and {Tufail}, Yasir and {Tumlinson}, Jason and {Tustain}, Samuel and {Tyra}, Harrison and {Ubeda}, Leonardo and {Underwood}, Kelli and {Uzzo}, Michael A. and {Vaclavik}, Steven and {Valenduc}, Frida and {Valenti}, Jeff A. and {Van Campen}, Julie and {van de Wetering}, Inge and {Van Der Marel}, Roeland P. and {van Haarlem}, Remy and {Vandenbussche}, Bart and {van Dishoeck}, Ewine F. and {Vanterpool}, Dona D. and {Vernoy}, Michael R. and {Vila Costas}, Maria Bego{\~n}a and {Volk}, Kevin and {Voorzaat}, Piet and {Voyton}, Mark F. and {Vydra}, Ekaterina and {Waddy}, Darryl J. and {Waelkens}, Christoffel and {Wahlgren}, Glenn Michael and {Walker}, Frederick E., Jr. and {Wander}, Michel and {Warfield}, Christine K. and {Warner}, Gerald and {Wasiak}, Francis C. and {Wasiak}, Matthew F. and {Wehner}, James and {Weiler}, Kevin R. and {Weilert}, Mark and {Weiss}, Stanley B. and {Wells}, Martyn and {Welty}, Alan D. and {Wheate}, Lauren and {Wheeler}, Thomas P. and {White}, Christy L. and {Whitehouse}, Paul and {Whiteleather}, Jennifer Margaret and {Whitman}, William Russell and {Williams}, Christina C. and {Willmer}, Christopher N.~A. and {Willott}, Chris J. and {Willoughby}, Scott P. and {Wilson}, Andrew and {Wilson}, Debra and {Wilson}, Donna V. and {Windhorst}, Rogier and {Wislowski}, Emily Christine and {Wolfe}, David J. and {Wolfe}, Michael A. and {Wolff}, Schuyler and {Wondel}, Amancio and {Woo}, Cindy and {Woods}, Robert T. and {Worden}, Elaine and {Workman}, William and {Wright}, Gillian S. and {Wu}, Carl and {Wu}, Chi-Rai and {Wun}, Dakin D. and {Wymer}, Kristen B. and {Yadetie}, Thomas and {Yan}, Isabelle C. and {Yang}, Keith C. and {Yates}, Kayla L. and {Yeager}, Christopher R. and {Yerger}, Ethan John and {Young}, Erick T. and {Young}, Gary and {Yu}, Gene and {Yu}, Susan and {Zak}, Dean S. and {Zeidler}, Peter and {Zepp}, Robert and {Zhou}, Julia and {Zincke}, Christian A. and {Zonak}, Stephanie and {Zondag}, Elisabeth},
        title = "{The James Webb Space Telescope Mission}",
      journal = {\pasp},
         year = 2023,
        month = jun,
       volume = {135},
       number = {1048},
          eid = {068001},
        pages = {068001},
          doi = {10.1088/1538-3873/acd1b5},
archivePrefix = {arXiv},
       eprint = {2304.04869},
 primaryClass = {astro-ph.IM},
       adsurl = {https://ui.adsabs.harvard.edu/abs/2023PASP..135f8001G}
}

@INPROCEEDINGS{Perrin2014,
       author = {{Perrin}, Marshall D. and {Sivaramakrishnan}, Anand and {Lajoie}, Charles-Philippe and {Elliott}, Erin and {Pueyo}, Laurent and {Ravindranath}, Swara and {Albert}, Lo{\"\i}c.},
        title = "{Updated point spread function simulations for JWST with WebbPSF}",
    booktitle = {Space Telescopes and Instrumentation 2014: Optical, Infrared, and Millimeter Wave},
         year = 2014,
       editor = {{Oschmann}, Jacobus M., Jr. and {Clampin}, Mark and {Fazio}, Giovanni G. and {MacEwen}, Howard A.},
       series = {Society of Photo-Optical Instrumentation Engineers (SPIE) Conference Series},
       volume = {9143},
        month = aug,
          eid = {91433X},
        pages = {91433X},
          doi = {10.1117/12.2056689},
       adsurl = {https://ui.adsabs.harvard.edu/abs/2014SPIE.9143E..3XP}
}

@ARTICLE{Giavalisco2004,
       author = {{Giavalisco}, M. and {Ferguson}, H.~C. and {Koekemoer}, A.~M. and {Dickinson}, M. and {Alexander}, D.~M. and {Bauer}, F.~E. and {Bergeron}, J. and {Biagetti}, C. and {Brandt}, W.~N. and {Casertano}, S. and {Cesarsky}, C. and {Chatzichristou}, E. and {Conselice}, C. and {Cristiani}, S. and {Da Costa}, L. and {Dahlen}, T. and {de Mello}, D. and {Eisenhardt}, P. and {Erben}, T. and {Fall}, S.~M. and {Fassnacht}, C. and {Fosbury}, R. and {Fruchter}, A. and {Gardner}, J.~P. and {Grogin}, N. and {Hook}, R.~N. and {Hornschemeier}, A.~E. and {Idzi}, R. and {Jogee}, S. and {Kretchmer}, C. and {Laidler}, V. and {Lee}, K.~S. and {Livio}, M. and {Lucas}, R. and {Madau}, P. and {Mobasher}, B. and {Moustakas}, L.~A. and {Nonino}, M. and {Padovani}, P. and {Papovich}, C. and {Park}, Y. and {Ravindranath}, S. and {Renzini}, A. and {Richardson}, M. and {Riess}, A. and {Rosati}, P. and {Schirmer}, M. and {Schreier}, E. and {Somerville}, R.~S. and {Spinrad}, H. and {Stern}, D. and {Stiavelli}, M. and {Strolger}, L. and {Urry}, C.~M. and {Vandame}, B. and {Williams}, R. and {Wolf}, C.},
        title = "{The Great Observatories Origins Deep Survey: Initial Results from Optical and Near-Infrared Imaging}",
      journal = {\apjl},
         year = 2004,
        month = jan,
       volume = {600},
       number = {2},
        pages = {L93-L98},
          doi = {10.1086/379232},
archivePrefix = {arXiv},
       eprint = {astro-ph/0309105},
 primaryClass = {astro-ph},
       adsurl = {https://ui.adsabs.harvard.edu/abs/2004ApJ...600L..93G}
}

@ARTICLE{Vayner2024,
       author = {{Vayner}, Andrey and {Zakamska}, Nadia L. and {Ishikawa}, Yuzo and {Sankar}, Swetha and {Wylezalek}, Dominika and {Rupke}, David S.~N. and {Veilleux}, Sylvain and {Bertemes}, Caroline and {Barrera-Ballesteros}, Jorge K. and {Chen}, Hsiao-Wen and {Diachenko}, Nadiia and {Goulding}, Andy D. and {Greene}, Jenny E. and {Hainline}, Kevin N. and {Hamann}, Fred and {Heckman}, Timothy and {Johnson}, Sean D. and {Grace Lim}, Hui Xian and {Liu}, Weizhe and {Lutz}, Dieter and {L{\"u}tzgendorf}, Nora and {Mainieri}, Vincenzo and {McCrory}, Ryan and {Murphree}, Grey and {Nesvadba}, Nicole P.~H. and {Ogle}, Patrick and {Sturm}, Eckhard and {Whitesell}, Lillian},
        title = "{First Results from the JWST Early Release Science Program Q3D: Powerful Quasar-driven Galactic Scale Outflow at z = 3}",
      journal = {\apj},
         year = 2024,
        month = jan,
       volume = {960},
       number = {2},
          eid = {126},
        pages = {126},
          doi = {10.3847/1538-4357/ad0be9},
archivePrefix = {arXiv},
       eprint = {2307.13751},
 primaryClass = {astro-ph.GA},
       adsurl = {https://ui.adsabs.harvard.edu/abs/2024ApJ...960..126V}
}

@ARTICLE{Springel2005,
       author = {{Springel}, Volker and {Di Matteo}, Tiziana and {Hernquist}, Lars},
        title = "{Modelling feedback from stars and black holes in galaxy mergers}",
      journal = {\mnras},
         year = 2005,
        month = aug,
       volume = {361},
       number = {3},
        pages = {776-794},
          doi = {10.1111/j.1365-2966.2005.09238.x},
archivePrefix = {arXiv},
       eprint = {astro-ph/0411108},
 primaryClass = {astro-ph},
       adsurl = {https://ui.adsabs.harvard.edu/abs/2005MNRAS.361..776S}
}

@inproceedings{Fitzpatrick2014,
author = {Michael J. Fitzpatrick and Knut Olsen and Frossie Economou and Elizabeth B. Stobie and T. C. Beers and Mark Dickinson and Patrick Norris and Abi Saha and Robert Seaman and David R. Silva and Robert A. Swaters and Brian Thomas and Francisco Valdes},
title = {{The NOAO Data Laboratory: a conceptual overview}},
volume = {9149},
booktitle = {Observatory Operations: Strategies, Processes, and Systems V},
editor = {Alison B. Peck and Chris R. Benn and Robert L. Seaman},
organization = {International Society for Optics and Photonics},
publisher = {SPIE},
pages = {91491T},
year = {2014},
doi = {10.1117/12.2057445},
URL = {https://doi.org/10.1117/12.2057445}
}

@ARTICLE{harris2020,
       author = {{Harris}, Charles R. and {Millman}, K. Jarrod and {van der Walt}, St{\'e}fan J. and {Gommers}, Ralf and {Virtanen}, Pauli and {Cournapeau}, David and {Wieser}, Eric and {Taylor}, Julian and {Berg}, Sebastian and {Smith}, Nathaniel J. and {Kern}, Robert and {Picus}, Matti and {Hoyer}, Stephan and {van Kerkwijk}, Marten H. and {Brett}, Matthew and {Haldane}, Allan and {del R{\'\i}o}, Jaime Fern{\'a}ndez and {Wiebe}, Mark and {Peterson}, Pearu and {G{\'e}rard-Marchant}, Pierre and {Sheppard}, Kevin and {Reddy}, Tyler and {Weckesser}, Warren and {Abbasi}, Hameer and {Gohlke}, Christoph and {Oliphant}, Travis E.},
        title = "{Array programming with NumPy}",
      journal = {\nat},
         year = 2020,
        month = sep,
       volume = {585},
       number = {7825},
        pages = {357-362},
          doi = {10.1038/s41586-020-2649-2},
archivePrefix = {arXiv},
       eprint = {2006.10256},
 primaryClass = {cs.MS},
       adsurl = {https://ui.adsabs.harvard.edu/abs/2020Natur.585..357H}
}

@ARTICLE{astropy2013,
       author = {{Astropy Collaboration} and {Robitaille}, Thomas P. and {Tollerud}, Erik J. and {Greenfield}, Perry and {Droettboom}, Michael and {Bray}, Erik and {Aldcroft}, Tom and {Davis}, Matt and {Ginsburg}, Adam and {Price-Whelan}, Adrian M. and {Kerzendorf}, Wolfgang E. and {Conley}, Alexander and {Crighton}, Neil and {Barbary}, Kyle and {Muna}, Demitri and {Ferguson}, Henry and {Grollier}, Fr{\'e}d{\'e}ric and {Parikh}, Madhura M. and {Nair}, Prasanth H. and {Unther}, Hans M. and {Deil}, Christoph and {Woillez}, Julien and {Conseil}, Simon and {Kramer}, Roban and {Turner}, James E.~H. and {Singer}, Leo and {Fox}, Ryan and {Weaver}, Benjamin A. and {Zabalza}, Victor and {Edwards}, Zachary I. and {Azalee Bostroem}, K. and {Burke}, D.~J. and {Casey}, Andrew R. and {Crawford}, Steven M. and {Dencheva}, Nadia and {Ely}, Justin and {Jenness}, Tim and {Labrie}, Kathleen and {Lim}, Pey Lian and {Pierfederici}, Francesco and {Pontzen}, Andrew and {Ptak}, Andy and {Refsdal}, Brian and {Servillat}, Mathieu and {Streicher}, Ole},
        title = "{Astropy: A community Python package for astronomy}",
      journal = {\aap},
         year = 2013,
        month = oct,
       volume = {558},
          eid = {A33},
        pages = {A33},
          doi = {10.1051/0004-6361/201322068},
archivePrefix = {arXiv},
       eprint = {1307.6212},
 primaryClass = {astro-ph.IM},
       adsurl = {https://ui.adsabs.harvard.edu/abs/2013A&A...558A..33A}
}

@ARTICLE{astropy2018,
       author = {{Astropy Collaboration} and {Price-Whelan}, A.~M. and {Sip{\H{o}}cz}, B.~M. and {G{\"u}nther}, H.~M. and {Lim}, P.~L. and {Crawford}, S.~M. and {Conseil}, S. and {Shupe}, D.~L. and {Craig}, M.~W. and {Dencheva}, N. and {Ginsburg}, A. and {VanderPlas}, J.~T. and {Bradley}, L.~D. and {P{\'e}rez-Su{\'a}rez}, D. and {de Val-Borro}, M. and {Aldcroft}, T.~L. and {Cruz}, K.~L. and {Robitaille}, T.~P. and {Tollerud}, E.~J. and {Ardelean}, C. and {Babej}, T. and {Bach}, Y.~P. and {Bachetti}, M. and {Bakanov}, A.~V. and {Bamford}, S.~P. and {Barentsen}, G. and {Barmby}, P. and {Baumbach}, A. and {Berry}, K.~L. and {Biscani}, F. and {Boquien}, M. and {Bostroem}, K.~A. and {Bouma}, L.~G. and {Brammer}, G.~B. and {Bray}, E.~M. and {Breytenbach}, H. and {Buddelmeijer}, H. and {Burke}, D.~J. and {Calderone}, G. and {Cano Rodr{\'\i}guez}, J.~L. and {Cara}, M. and {Cardoso}, J.~V.~M. and {Cheedella}, S. and {Copin}, Y. and {Corrales}, L. and {Crichton}, D. and {D'Avella}, D. and {Deil}, C. and {Depagne}, {\'E}. and {Dietrich}, J.~P. and {Donath}, A. and {Droettboom}, M. and {Earl}, N. and {Erben}, T. and {Fabbro}, S. and {Ferreira}, L.~A. and {Finethy}, T. and {Fox}, R.~T. and {Garrison}, L.~H. and {Gibbons}, S.~L.~J. and {Goldstein}, D.~A. and {Gommers}, R. and {Greco}, J.~P. and {Greenfield}, P. and {Groener}, A.~M. and {Grollier}, F. and {Hagen}, A. and {Hirst}, P. and {Homeier}, D. and {Horton}, A.~J. and {Hosseinzadeh}, G. and {Hu}, L. and {Hunkeler}, J.~S. and {Ivezi{\'c}}, {\v{Z}}. and {Jain}, A. and {Jenness}, T. and {Kanarek}, G. and {Kendrew}, S. and {Kern}, N.~S. and {Kerzendorf}, W.~E. and {Khvalko}, A. and {King}, J. and {Kirkby}, D. and {Kulkarni}, A.~M. and {Kumar}, A. and {Lee}, A. and {Lenz}, D. and {Littlefair}, S.~P. and {Ma}, Z. and {Macleod}, D.~M. and {Mastropietro}, M. and {McCully}, C. and {Montagnac}, S. and {Morris}, B.~M. and {Mueller}, M. and {Mumford}, S.~J. and {Muna}, D. and {Murphy}, N.~A. and {Nelson}, S. and {Nguyen}, G.~H. and {Ninan}, J.~P. and {N{\"o}the}, M. and {Ogaz}, S. and {Oh}, S. and {Parejko}, J.~K. and {Parley}, N. and {Pascual}, S. and {Patil}, R. and {Patil}, A.~A. and {Plunkett}, A.~L. and {Prochaska}, J.~X. and {Rastogi}, T. and {Reddy Janga}, V. and {Sabater}, J. and {Sakurikar}, P. and {Seifert}, M. and {Sherbert}, L.~E. and {Sherwood-Taylor}, H. and {Shih}, A.~Y. and {Sick}, J. and {Silbiger}, M.~T. and {Singanamalla}, S. and {Singer}, L.~P. and {Sladen}, P.~H. and {Sooley}, K.~A. and {Sornarajah}, S. and {Streicher}, O. and {Teuben}, P. and {Thomas}, S.~W. and {Tremblay}, G.~R. and {Turner}, J.~E.~H. and {Terr{\'o}n}, V. and {van Kerkwijk}, M.~H. and {de la Vega}, A. and {Watkins}, L.~L. and {Weaver}, B.~A. and {Whitmore}, J.~B. and {Woillez}, J. and {Zabalza}, V. and {Astropy Contributors}},
        title = "{The Astropy Project: Building an Open-science Project and Status of the v2.0 Core Package}",
      journal = {\aj},
         year = 2018,
        month = sep,
       volume = {156},
       number = {3},
          eid = {123},
        pages = {123},
          doi = {10.3847/1538-3881/aabc4f},
archivePrefix = {arXiv},
       eprint = {1801.02634},
 primaryClass = {astro-ph.IM},
       adsurl = {https://ui.adsabs.harvard.edu/abs/2018AJ....156..123A}
}

@ARTICLE{astropy2022,
       author = {{Astropy Collaboration} and {Price-Whelan}, Adrian M. and {Lim}, Pey Lian and {Earl}, Nicholas and {Starkman}, Nathaniel and {Bradley}, Larry and {Shupe}, David L. and {Patil}, Aarya A. and {Corrales}, Lia and {Brasseur}, C.~E. and {N{\"o}the}, Maximilian and {Donath}, Axel and {Tollerud}, Erik and {Morris}, Brett M. and {Ginsburg}, Adam and {Vaher}, Eero and {Weaver}, Benjamin A. and {Tocknell}, James and {Jamieson}, William and {van Kerkwijk}, Marten H. and {Robitaille}, Thomas P. and {Merry}, Bruce and {Bachetti}, Matteo and {G{\"u}nther}, H. Moritz and {Aldcroft}, Thomas L. and {Alvarado-Montes}, Jaime A. and {Archibald}, Anne M. and {B{\'o}di}, Attila and {Bapat}, Shreyas and {Barentsen}, Geert and {Baz{\'a}n}, Juanjo and {Biswas}, Manish and {Boquien}, M{\'e}d{\'e}ric and {Burke}, D.~J. and {Cara}, Daria and {Cara}, Mihai and {Conroy}, Kyle E. and {Conseil}, Simon and {Craig}, Matthew W. and {Cross}, Robert M. and {Cruz}, Kelle L. and {D'Eugenio}, Francesco and {Dencheva}, Nadia and {Devillepoix}, Hadrien A.~R. and {Dietrich}, J{\"o}rg P. and {Eigenbrot}, Arthur Davis and {Erben}, Thomas and {Ferreira}, Leonardo and {Foreman-Mackey}, Daniel and {Fox}, Ryan and {Freij}, Nabil and {Garg}, Suyog and {Geda}, Robel and {Glattly}, Lauren and {Gondhalekar}, Yash and {Gordon}, Karl D. and {Grant}, David and {Greenfield}, Perry and {Groener}, Austen M. and {Guest}, Steve and {Gurovich}, Sebastian and {Handberg}, Rasmus and {Hart}, Akeem and {Hatfield-Dodds}, Zac and {Homeier}, Derek and {Hosseinzadeh}, Griffin and {Jenness}, Tim and {Jones}, Craig K. and {Joseph}, Prajwel and {Kalmbach}, J. Bryce and {Karamehmetoglu}, Emir and {Ka{\l}uszy{\'n}ski}, Miko{\l}aj and {Kelley}, Michael S.~P. and {Kern}, Nicholas and {Kerzendorf}, Wolfgang E. and {Koch}, Eric W. and {Kulumani}, Shankar and {Lee}, Antony and {Ly}, Chun and {Ma}, Zhiyuan and {MacBride}, Conor and {Maljaars}, Jakob M. and {Muna}, Demitri and {Murphy}, N.~A. and {Norman}, Henrik and {O'Steen}, Richard and {Oman}, Kyle A. and {Pacifici}, Camilla and {Pascual}, Sergio and {Pascual-Granado}, J. and {Patil}, Rohit R. and {Perren}, Gabriel I. and {Pickering}, Timothy E. and {Rastogi}, Tanuj and {Roulston}, Benjamin R. and {Ryan}, Daniel F. and {Rykoff}, Eli S. and {Sabater}, Jose and {Sakurikar}, Parikshit and {Salgado}, Jes{\'u}s and {Sanghi}, Aniket and {Saunders}, Nicholas and {Savchenko}, Volodymyr and {Schwardt}, Ludwig and {Seifert-Eckert}, Michael and {Shih}, Albert Y. and {Jain}, Anany Shrey and {Shukla}, Gyanendra and {Sick}, Jonathan and {Simpson}, Chris and {Singanamalla}, Sudheesh and {Singer}, Leo P. and {Singhal}, Jaladh and {Sinha}, Manodeep and {Sip{\H{o}}cz}, Brigitta M. and {Spitler}, Lee R. and {Stansby}, David and {Streicher}, Ole and {{\v{S}}umak}, Jani and {Swinbank}, John D. and {Taranu}, Dan S. and {Tewary}, Nikita and {Tremblay}, Grant R. and {de Val-Borro}, Miguel and {Van Kooten}, Samuel J. and {Vasovi{\'c}}, Zlatan and {Verma}, Shresth and {de Miranda Cardoso}, Jos{\'e} Vin{\'\i}cius and {Williams}, Peter K.~G. and {Wilson}, Tom J. and {Winkel}, Benjamin and {Wood-Vasey}, W.~M. and {Xue}, Rui and {Yoachim}, Peter and {Zhang}, Chen and {Zonca}, Andrea and {Astropy Project Contributors}},
        title = "{The Astropy Project: Sustaining and Growing a Community-oriented Open-source Project and the Latest Major Release (v5.0) of the Core Package}",
      journal = {\apj},
         year = 2022,
        month = aug,
       volume = {935},
       number = {2},
          eid = {167},
        pages = {167},
          doi = {10.3847/1538-4357/ac7c74},
archivePrefix = {arXiv},
       eprint = {2206.14220},
 primaryClass = {astro-ph.IM},
       adsurl = {https://ui.adsabs.harvard.edu/abs/2022ApJ...935..167A}
}

@article{Nikutta2020,
title = {Data Lab—A community science platform},
journal = {Astronomy and Computing},
volume = {33},
pages = {100411},
year = {2020},
issn = {2213-1337},
doi = {https://doi.org/10.1016/j.ascom.2020.100411},
url = {https://www.sciencedirect.com/science/article/pii/S2213133720300652},
author = {R. Nikutta and M. Fitzpatrick and A. Scott and B.A. Weaver}
}

@ARTICLE{Juneau2021_jupyter,
  author={Juneau, Stéphanie and Olsen, Knut and Nikutta, Robert and Jacques, Alice and Bailey, Stephen},
  journal={Computing in Science \& Engineering}, 
  title={Jupyter-Enabled Astrophysical Analysis Using Data-Proximate Computing Platforms}, 
  year={2021},
  volume={23},
  number={2},
  pages={15-25},
  doi={10.1109/MCSE.2021.3057097}}

@misc{JADES_doi,
  doi = {10.17909/8TDJ-8N28},
  url = {http://archive.stsci.edu/doi/resolve/resolve.html?doi=10.17909/8tdj-8n28},
  author = {{Rieke,  Marcia} and {Robertson,  Brant} and {Tacchella,  Sandro} and {Willmer,  Christopher} and {Johnson,  Ben} and {Carniani,  Stefano} and {Bunker,  Andy} and {Willott,  Chris}},
  title = {Data from the JWST Advanced Deep Extragalactic Survey (JADES)},
  publisher = {STScI/MAST},
  year = {2023}
}

@misc{JADES_DR2_doi,
  doi = {10.17909/8TDJ-8N28},
  url = {http://archive.stsci.edu/doi/resolve/resolve.html?doi=10.17909/8tdj-8n28},
  author = {{Rieke,  Marcia} and {Robertson,  Brant} and {Tacchella,  Sandro} and {Willmer,  Christopher} and {Johnson,  Ben} and {Carniani,  Stefano} and {Bunker,  Andy} and {Willott,  Chris}},
  title = {Data from the JWST Advanced Deep Extragalactic Survey (JADES)},
  publisher = {STScI/MAST},
  year = {2023}
}

@misc{JEMS_doi,
  doi = {10.17909/FSC4-DT61},
  url = {http://archive.stsci.edu/doi/resolve/resolve.html?doi=10.17909/fsc4-dt61},
  author = {{Williams,  Christina} and {Tacchella,  Sandro} and {Maseda,  Michael}},
  title = {Data from the JWST Extragalactic Medium-band Survey (JEMS)},
  publisher = {STScI/MAST},
  year = {2023}
}

@misc{FRESCO_doi,
  doi = {10.17909/GDYC-7G80},
  url = {http://archive.stsci.edu/doi/resolve/resolve.html?doi=10.17909/gdyc-7g80},
  author = {{Oesch,  Pascal} and {Magee,  Dan}},
  title = {The JWST FRESCO Survey},
  publisher = {STScI/MAST},
  year = {2023}
}

@ARTICLE{Lyu2024,
       author = {{Lyu}, Jianwei and {Alberts}, Stacey and {Rieke}, George H. and {Shivaei}, Irene and {P{\'e}rez-Gonz{\'a}lez}, Pablo G. and {Sun}, Fengwu and {Hainline}, Kevin N. and {Baum}, Stefi and {Bonaventura}, Nina and {Bunker}, Andrew J. and {Egami}, Eiichi and {Eisenstein}, Daniel J. and {Florian}, Michael and {Ji}, Zhiyuan and {Johnson}, Benjamin D. and {Morrison}, Jane and {Rieke}, Marcia and {Robertson}, Brant and {Rujopakarn}, Wiphu and {Tacchella}, Sandro and {Scholtz}, Jan and {Willmer}, Christopher N.~A.},
        title = "{Active Galactic Nuclei Selection and Demographics: A New Age with JWST/MIRI}",
      journal = {\apj},
         year = 2024,
        month = may,
       volume = {966},
       number = {2},
          eid = {229},
        pages = {229},
          doi = {10.3847/1538-4357/ad3643},
archivePrefix = {arXiv},
       eprint = {2310.12330},
 primaryClass = {astro-ph.GA},
       adsurl = {https://ui.adsabs.harvard.edu/abs/2024ApJ...966..229L}
}

@ARTICLE{Martinez-Ramirez2024,
       author = {{Mart{\'\i}nez-Ram{\'\i}rez}, L.~N. and {Calistro Rivera}, G. and {Lusso}, E. and {Bauer}, F.~E. and {Nardini}, E. and {Buchner}, J. and {Brown}, M.~J.~I. and {Pineda}, J.~C.~B. and {Temple}, M.~J. and {Banerji}, M. and {Stalevski}, M. and {Hennawi}, J.~F.},
        title = "{AGNFITTER-RX: Modeling the radio-to-X-ray spectral energy distributions of AGNs}",
      journal = {\aap},
         year = 2024,
        month = aug,
       volume = {688},
          eid = {A46},
        pages = {A46},
          doi = {10.1051/0004-6361/202449329},
archivePrefix = {arXiv},
       eprint = {2405.12111},
 primaryClass = {astro-ph.GA},
       adsurl = {https://ui.adsabs.harvard.edu/abs/2024A&A...688A..46M}
}

@ARTICLE{Rivera2016,
       author = {{Calistro Rivera}, Gabriela and {Lusso}, Elisabeta and {Hennawi}, Joseph F. and {Hogg}, David W.},
        title = "{AGNfitter: A Bayesian MCMC Approach to Fitting Spectral Energy Distributions of AGNs}",
      journal = {\apj},
         year = 2016,
        month = dec,
       volume = {833},
       number = {1},
          eid = {98},
        pages = {98},
          doi = {10.3847/1538-4357/833/1/98},
archivePrefix = {arXiv},
       eprint = {1606.05648},
 primaryClass = {astro-ph.GA},
       adsurl = {https://ui.adsabs.harvard.edu/abs/2016ApJ...833...98C}
}

@ARTICLE{Curtis-Lake2025,
       author = {{Curtis-Lake}, Emma and {Cameron}, Alex J. and {Bunker}, Andrew J. and {Scholtz}, Jan and {Carniani}, Stefano and {Parlanti}, Eleonora and {D'Eugenio}, Francesco and {Jakobsen}, Peter and {Willmer}, Christopher N.~A. and {Arribas}, Santiago and {Baker}, William M. and {Charlot}, St{\'e}phane and {Chevallard}, Jacopo and {Circosta}, Chiara and {Curti}, Mirko and {Eisenstein}, Daniel J. and {Hainline}, Kevin and {Ji}, Zhiyuan and {Johnson}, Benjamin D. and {Jones}, Gareth C. and {Maiolino}, Roberto and {Maseda}, Michael V. and {P{\'e}rez-Gonz{\'a}lez}, Pablo G. and {Rawle}, Tim and {Rieke}, Marcia and {Rinaldi}, Pierluigi and {Robertson}, Brant and {Rodr{\'\i}gez Del Pino}, Bruno and {Saxena}, Aayush and {Shivaei}, Irene and {Smit}, Renske and {Tacchella}, Sandro and {{\"U}bler}, Hannah and {Venturi}, Giacomo and {Williams}, Christina C. and {Willott}, Chris and {Duan}, Qiao},
        title = "{JADES Data Release 4 Paper I: Sample Selection, Observing Strategy and Redshifts of the complete spectroscopic sample}",
      journal = {arXiv e-prints},
         year = 2025,
        month = oct,
          eid = {arXiv:2510.01033},
        pages = {arXiv:2510.01033},
          doi = {10.48550/arXiv.2510.01033},
archivePrefix = {arXiv},
       eprint = {2510.01033},
 primaryClass = {astro-ph.GA},
       adsurl = {https://ui.adsabs.harvard.edu/abs/2025arXiv251001033C}
}

@ARTICLE{Scholtz2025,
       author = {{Scholtz}, J. and {Carniani}, S. and {Parlanti}, E. and {D'Eugenio}, F. and {Curtis-Lake}, E. and {Jakobsen}, P. and {Bunker}, A.~J. and {Cameron}, A.~J. and {Arribas}, S. and {Baker}, W.~M. and {Charlot}, S. and {Chevellard}, J. and {Circosta}, C. and {Curti}, M. and {Duan}, Q. and {Eisenstein}, D.~J. and {Hainline}, K. and {Ji}, Z. and {Johnson}, B.~D. and {Jones}, G.~C. and {Kumari}, N. and {Maiolino}, R. and {Maseda}, M.~V. and {Perna}, M. and {P{\'e}rez-Gonz{\'a}lez}, P.~G. and {Rawle}, T. and {Rieke}, M. and {Rinaldi}, P. and {Robertson}, B. and {Saxena}, A. and {Shivaei}, I. and {Silcock}, M.~S. and {Sun}, Y. and {Rodr{\'\i}guez Del Pino}, B. and {Tacchella}, S. and {{\"U}bler}, H. and {Venturi}, G. and {Williams}, C.~C. and {Willmer}, C.~N.~A. and {Willott}, C. and {Witstok}, J.},
        title = "{JADES Data Release 4 -- Paper II: Data reduction, analysis and emission-line fluxes of the complete spectroscopic sample}",
      journal = {arXiv e-prints},
         year = 2025,
        month = oct,
          eid = {arXiv:2510.01034},
        pages = {arXiv:2510.01034},
          doi = {10.48550/arXiv.2510.01034},
archivePrefix = {arXiv},
       eprint = {2510.01034},
 primaryClass = {astro-ph.GA},
       adsurl = {https://ui.adsabs.harvard.edu/abs/2025arXiv251001034S}
}

@ARTICLE{Lebowitz2025,
       author = {{Lebowitz}, S. and {Hainline}, K. and {Juneau}, S. and {Lyu}, J. and {Williams}, C.~C. and {Alberts}, S. and {Fan}, X. and {Rieke}, M.},
        title = "{JWST NIRCam Simulations and Observations of AGN Ionization Cones in Cosmic Noon Galaxies}",
      journal = {\apj},
         year = 2025,
        month = may,
       volume = {984},
       number = {1},
          eid = {13},
        pages = {13},
          doi = {10.3847/1538-4357/adc07c},
archivePrefix = {arXiv},
       eprint = {2501.05512},
 primaryClass = {astro-ph.GA},
       adsurl = {https://ui.adsabs.harvard.edu/abs/2025ApJ...984...13L}
}

@ARTICLE{Boquien,
       author = {{Boquien}, M. and {Burgarella}, D. and {Roehlly}, Y. and {Buat}, V. and {Ciesla}, L. and {Corre}, D. and {Inoue}, A.~K. and {Salas}, H.},
        title = "{CIGALE: a python Code Investigating GALaxy Emission}",
      journal = {\aap},
         year = 2019,
        month = feb,
       volume = {622},
          eid = {A103},
        pages = {A103},
          doi = {10.1051/0004-6361/201834156},
archivePrefix = {arXiv},
       eprint = {1811.03094},
 primaryClass = {astro-ph.GA},
       adsurl = {https://ui.adsabs.harvard.edu/abs/2019A&A...622A.103B}
}

@ARTICLE{Temple2021,
       author = {{Temple}, Matthew J. and {Hewett}, Paul C. and {Banerji}, Manda},
        title = "{Modelling type 1 quasar colours in the era of Rubin and Euclid}",
      journal = {\mnras},
         year = 2021,
        month = nov,
       volume = {508},
       number = {1},
        pages = {737-754},
          doi = {10.1093/mnras/stab2586},
archivePrefix = {arXiv},
       eprint = {2109.04472},
 primaryClass = {astro-ph.GA},
       adsurl = {https://ui.adsabs.harvard.edu/abs/2021MNRAS.508..737T}
}

@ARTICLE{Stalevski2016,
       author = {{Stalevski}, Marko and {Ricci}, Claudio and {Ueda}, Yoshihiro and {Lira}, Paulina and {Fritz}, Jacopo and {Baes}, Maarten},
        title = "{The dust covering factor in active galactic nuclei}",
      journal = {\mnras},
         year = 2016,
        month = may,
       volume = {458},
       number = {3},
        pages = {2288-2302},
          doi = {10.1093/mnras/stw444},
archivePrefix = {arXiv},
       eprint = {1602.06954},
 primaryClass = {astro-ph.GA},
       adsurl = {https://ui.adsabs.harvard.edu/abs/2016MNRAS.458.2288S}
}

@ARTICLE{Lusso2017,
       author = {{Lusso}, E. and {Risaliti}, G.},
        title = "{Quasars as standard candles. I. The physical relation between disc and coronal emission}",
      journal = {\aap},
         year = 2017,
        month = jun,
       volume = {602},
          eid = {A79},
        pages = {A79},
          doi = {10.1051/0004-6361/201630079},
archivePrefix = {arXiv},
       eprint = {1703.05299},
 primaryClass = {astro-ph.HE},
       adsurl = {https://ui.adsabs.harvard.edu/abs/2017A&A...602A..79L}
}

@ARTICLE{Baan2006,
       author = {{Baan}, W.~A. and {Kl{\"o}ckner}, H.-R.},
        title = "{Radio properties of FIR-megamaser nuclei}",
      journal = {\aap},
         year = 2006,
        month = apr,
       volume = {449},
       number = {2},
        pages = {559-568},
          doi = {10.1051/0004-6361:20042331},
       adsurl = {https://ui.adsabs.harvard.edu/abs/2006A&A...449..559B}
}

@ARTICLE{Schreiber2018,
       author = {{Schreiber}, C. and {Elbaz}, D. and {Pannella}, M. and {Ciesla}, L. and {Wang}, T. and {Franco}, M.},
        title = "{Dust temperature and mid-to-total infrared color distributions for star-forming galaxies at 0 < z < 4}",
      journal = {\aap},
         year = 2018,
        month = jan,
       volume = {609},
          eid = {A30},
        pages = {A30},
          doi = {10.1051/0004-6361/201731506},
archivePrefix = {arXiv},
       eprint = {1710.10276},
 primaryClass = {astro-ph.GA},
       adsurl = {https://ui.adsabs.harvard.edu/abs/2018A&A...609A..30S}
}

@ARTICLE{Bruzual2003,
       author = {{Bruzual}, G. and {Charlot}, S.},
        title = "{Stellar population synthesis at the resolution of 2003}",
      journal = {\mnras},
         year = 2003,
        month = oct,
       volume = {344},
       number = {4},
        pages = {1000-1028},
          doi = {10.1046/j.1365-8711.2003.06897.x},
archivePrefix = {arXiv},
       eprint = {astro-ph/0309134},
 primaryClass = {astro-ph},
       adsurl = {https://ui.adsabs.harvard.edu/abs/2003MNRAS.344.1000B}
}

@ARTICLE{Hainline2012,
       author = {{Hainline}, Kevin N. and {Shapley}, Alice E. and {Greene}, Jenny E. and {Steidel}, Charles C. and {Reddy}, Naveen A. and {Erb}, Dawn K.},
        title = "{Stellar Populations of Ultraviolet-selected Active Galactic Nuclei Host Galaxies at z \raisebox{-0.5ex}\textasciitilde 2-3}",
      journal = {\apj},
         year = 2012,
        month = nov,
       volume = {760},
       number = {1},
          eid = {74},
        pages = {74},
          doi = {10.1088/0004-637X/760/1/74},
archivePrefix = {arXiv},
       eprint = {1206.3308},
 primaryClass = {astro-ph.CO},
       adsurl = {https://ui.adsabs.harvard.edu/abs/2012ApJ...760...74H}
}

@ARTICLE{Kewley2013,
       author = {{Kewley}, Lisa J. and {Dopita}, Michael A. and {Leitherer}, Claus and {Dav{\'e}}, Romeel and {Yuan}, Tiantian and {Allen}, Mark and {Groves}, Brent and {Sutherland}, Ralph},
        title = "{Theoretical Evolution of Optical Strong Lines across Cosmic Time}",
      journal = {\apj},
         year = 2013,
        month = sep,
       volume = {774},
       number = {2},
          eid = {100},
        pages = {100},
          doi = {10.1088/0004-637X/774/2/100},
archivePrefix = {arXiv},
       eprint = {1307.0508},
 primaryClass = {astro-ph.CO},
       adsurl = {https://ui.adsabs.harvard.edu/abs/2013ApJ...774..100K}
}

@ARTICLE{Lamperti2017,
       author = {{Lamperti}, Isabella and {Koss}, Michael and {Trakhtenbrot}, Benny and {Schawinski}, Kevin and {Ricci}, Claudio and {Oh}, Kyuseok and {Landt}, Hermine and {Riffel}, Rog{\'e}rio and {Rodr{\'\i}guez-Ardila}, Alberto and {Gehrels}, Neil and {Harrison}, Fiona and {Masetti}, Nicola and {Mushotzky}, Richard and {Treister}, Ezequiel and {Ueda}, Yoshihiro and {Veilleux}, Sylvain},
        title = "{BAT AGN Spectroscopic Survey - IV: Near-Infrared Coronal Lines, Hidden Broad Lines, and Correlation with Hard X-ray Emission}",
      journal = {\mnras},
         year = 2017,
        month = may,
       volume = {467},
       number = {1},
        pages = {540-572},
          doi = {10.1093/mnras/stx055},
archivePrefix = {arXiv},
       eprint = {1701.02755},
 primaryClass = {astro-ph.GA},
       adsurl = {https://ui.adsabs.harvard.edu/abs/2017MNRAS.467..540L}
}

@ARTICLE{Cleri2022,
       author = {{Cleri}, Nikko J. and {Trump}, Jonathan R. and {Backhaus}, Bren E. and {Momcheva}, Ivelina and {Papovich}, Casey and {Simons}, Raymond and {Weiner}, Benjamin and {Estrada-Carpenter}, Vicente and {Finkelstein}, Steven L. and {Giavalisco}, Mauro and {Ji}, Zhiyuan and {Jung}, Intae and {Matharu}, Jasleen and {Martinez}, Felix and {Sturm}, Megan R.},
        title = "{CLEAR: Paschen-{\ensuremath{\beta}} Star Formation Rates and Dust Attenuation of Low-redshift Galaxies}",
      journal = {\apj},
         year = 2022,
        month = apr,
       volume = {929},
       number = {1},
          eid = {3},
        pages = {3},
          doi = {10.3847/1538-4357/ac5a4c},
archivePrefix = {arXiv},
       eprint = {2009.00617},
 primaryClass = {astro-ph.GA},
       adsurl = {https://ui.adsabs.harvard.edu/abs/2022ApJ...929....3C}
}

@ARTICLE{Haffner2009,
       author = {{Haffner}, L.~M. and {Dettmar}, R.-J. and {Beckman}, J.~E. and {Wood}, K. and {Slavin}, J.~D. and {Giammanco}, C. and {Madsen}, G.~J. and {Zurita}, A. and {Reynolds}, R.~J.},
        title = "{The warm ionized medium in spiral galaxies}",
      journal = {Reviews of Modern Physics},
         year = 2009,
        month = jul,
       volume = {81},
       number = {3},
        pages = {969-997},
          doi = {10.1103/RevModPhys.81.969},
archivePrefix = {arXiv},
       eprint = {0901.0941},
 primaryClass = {astro-ph.GA},
       adsurl = {https://ui.adsabs.harvard.edu/abs/2009RvMP...81..969H}
}

@ARTICLE{Liu2024,
       author = {{Liu}, Zhaoran and {Morishita}, Takahiro and {Kodama}, Tadayuki},
        title = "{Characterizing Dust Extinction and Spatially Resolved Paschen-$α$ Emission within 97 Galaxies at $1<z<1.6$ with JWST NIRCam Slitless Spectroscopy}",
      journal = {arXiv e-prints},
         year = 2024,
        month = jun,
          eid = {arXiv:2406.11188},
        pages = {arXiv:2406.11188},
          doi = {10.48550/arXiv.2406.11188},
archivePrefix = {arXiv},
       eprint = {2406.11188},
 primaryClass = {astro-ph.GA},
       adsurl = {https://ui.adsabs.harvard.edu/abs/2024arXiv240611188L}
}

@ARTICLE{Nelson2016,
       author = {{Nelson}, Erica June and {van Dokkum}, Pieter G. and {F{\"o}rster Schreiber}, Natascha M. and {Franx}, Marijn and {Brammer}, Gabriel B. and {Momcheva}, Ivelina G. and {Wuyts}, Stijn and {Whitaker}, Katherine E. and {Skelton}, Rosalind E. and {Fumagalli}, Mattia and {Hayward}, Christopher C. and {Kriek}, Mariska and {Labb{\'e}}, Ivo and {Leja}, Joel and {Rix}, Hans-Walter and {Tacconi}, Linda J. and {van der Wel}, Arjen and {van den Bosch}, Frank C. and {Oesch}, Pascal A. and {Dickey}, Claire and {Ulf Lange}, Johannes},
        title = "{Where Stars Form: Inside-out Growth and Coherent Star Formation from HST H{\ensuremath{\alpha}} Maps of 3200 Galaxies across the Main Sequence at 0.7 < z < 1.5}",
      journal = {\apj},
         year = 2016,
        month = sep,
       volume = {828},
       number = {1},
          eid = {27},
        pages = {27},
          doi = {10.3847/0004-637X/828/1/27},
archivePrefix = {arXiv},
       eprint = {1507.03999},
 primaryClass = {astro-ph.GA},
       adsurl = {https://ui.adsabs.harvard.edu/abs/2016ApJ...828...27N}
}

@ARTICLE{Reddy2023,
       author = {{Reddy}, Naveen A. and {Topping}, Michael W. and {Sanders}, Ryan L. and {Shapley}, Alice E. and {Brammer}, Gabriel},
        title = "{Paschen-line Constraints on Dust Attenuation and Star Formation at z   1-3 with JWST/NIRSpec}",
      journal = {\apj},
         year = 2023,
        month = may,
       volume = {948},
       number = {2},
          eid = {83},
        pages = {83},
          doi = {10.3847/1538-4357/acc869},
archivePrefix = {arXiv},
       eprint = {2301.07249},
 primaryClass = {astro-ph.GA},
       adsurl = {https://ui.adsabs.harvard.edu/abs/2023ApJ...948...83R}
}

@BOOK{Osterbrock2006,
       author = {{Osterbrock}, Donald E. and {Ferland}, Gary J.},
        title = "{Astrophysics of gaseous nebulae and active galactic nuclei}",
         year = 2006,
       adsurl = {https://ui.adsabs.harvard.edu/abs/2006agna.book.....O}
}

@ARTICLE{Larkin1998,
       author = {{Larkin}, J.~E. and {Armus}, L. and {Knop}, R.~A. and {Soifer}, B.~T. and {Matthews}, K.},
        title = "{A Near-Infrared Spectroscopic Survey of LINER Galaxies}",
      journal = {\apjs},
         year = 1998,
        month = jan,
       volume = {114},
       number = {1},
        pages = {59-72},
          doi = {10.1086/313063},
archivePrefix = {arXiv},
       eprint = {astro-ph/9708097},
 primaryClass = {astro-ph},
       adsurl = {https://ui.adsabs.harvard.edu/abs/1998ApJS..114...59L}
}

@ARTICLE{Neufeld2024,
       author = {{Neufeld}, Chloe and {van Dokkum}, Pieter and {Asali}, Yasmeen and {Covelo-Paz}, Alba and {Leja}, Joel and {Lin}, Jamie and {Matthee}, Jorryt and {Oesch}, Pascal A. and {Reddy}, Naveen A. and {Shivaei}, Irene and {Whitaker}, Katherine E. and {Wuyts}, Stijn and {Brammer}, Gabriel and {Marchesini}, Danilo and {Maseda}, Michael V. and {Naidu}, Rohan P. and {Nelson}, Erica J. and {Velichko}, Anna and {Weibel}, Andrea and {Xiao}, Mengyuan},
        title = "{FRESCO: The Paschen-{\ensuremath{\alpha}} Star-forming Sequence at Cosmic Noon}",
      journal = {\apj},
         year = 2024,
        month = sep,
       volume = {972},
       number = {2},
          eid = {156},
        pages = {156},
          doi = {10.3847/1538-4357/ad6158},
archivePrefix = {arXiv},
       eprint = {2404.10816},
 primaryClass = {astro-ph.GA},
       adsurl = {https://ui.adsabs.harvard.edu/abs/2024ApJ...972..156N}
}

@ARTICLE{Kauffmann2003,
       author = {{Kauffmann}, Guinevere and {Heckman}, Timothy M. and {Tremonti}, Christy and {Brinchmann}, Jarle and {Charlot}, St{\'e}phane and {White}, Simon D.~M. and {Ridgway}, Susan E. and {Brinkmann}, Jon and {Fukugita}, Masataka and {Hall}, Patrick B. and {Ivezi{\'c}}, {\v{Z}}eljko and {Richards}, Gordon T. and {Schneider}, Donald P.},
        title = "{The host galaxies of active galactic nuclei}",
      journal = {\mnras},
         year = 2003,
        month = dec,
       volume = {346},
       number = {4},
        pages = {1055-1077},
          doi = {10.1111/j.1365-2966.2003.07154.x},
archivePrefix = {arXiv},
       eprint = {astro-ph/0304239},
 primaryClass = {astro-ph},
       adsurl = {https://ui.adsabs.harvard.edu/abs/2003MNRAS.346.1055K}
}

@ARTICLE{Bongiorno2016,
       author = {{Bongiorno}, A. and {Schulze}, A. and {Merloni}, A. and {Zamorani}, G. and {Ilbert}, O. and {La Franca}, F. and {Peng}, Y. and {Piconcelli}, E. and {Mainieri}, V. and {Silverman}, J.~D. and {Brusa}, M. and {Fiore}, F. and {Salvato}, M. and {Scoville}, N.},
        title = "{AGN host galaxy mass function in COSMOS. Is AGN feedback responsible for the mass-quenching of galaxies?}",
      journal = {\aap},
         year = 2016,
        month = apr,
       volume = {588},
          eid = {A78},
        pages = {A78},
          doi = {10.1051/0004-6361/201527436},
archivePrefix = {arXiv},
       eprint = {1601.02091},
 primaryClass = {astro-ph.GA},
       adsurl = {https://ui.adsabs.harvard.edu/abs/2016A&A...588A..78B}
}

@ARTICLE{Donley2012,
       author = {{Donley}, J.~L. and {Koekemoer}, A.~M. and {Brusa}, M. and {Capak}, P. and {Cardamone}, C.~N. and {Civano}, F. and {Ilbert}, O. and {Impey}, C.~D. and {Kartaltepe}, J.~S. and {Miyaji}, T. and {Salvato}, M. and {Sanders}, D.~B. and {Trump}, J.~R. and {Zamorani}, G.},
        title = "{Identifying Luminous Active Galactic Nuclei in Deep Surveys: Revised IRAC Selection Criteria}",
      journal = {\apj},
         year = 2012,
        month = apr,
       volume = {748},
       number = {2},
          eid = {142},
        pages = {142},
          doi = {10.1088/0004-637X/748/2/142},
archivePrefix = {arXiv},
       eprint = {1201.3899},
 primaryClass = {astro-ph.CO},
       adsurl = {https://ui.adsabs.harvard.edu/abs/2012ApJ...748..142D}
}

@ARTICLE{Harrison2017,
       author = {{Harrison}, C.~M.},
        title = "{Impact of supermassive black hole growth on star formation}",
      journal = {Nature Astronomy},
         year = 2017,
        month = jul,
       volume = {1},
          eid = {0165},
        pages = {0165},
          doi = {10.1038/s41550-017-0165},
archivePrefix = {arXiv},
       eprint = {1703.06889},
 primaryClass = {astro-ph.GA},
       adsurl = {https://ui.adsabs.harvard.edu/abs/2017NatAs...1E.165H}
}

@ARTICLE{Mainieri2011,
       author = {{Mainieri}, V. and {Bongiorno}, A. and {Merloni}, A. and {Aller}, M. and {Carollo}, M. and {Iwasawa}, K. and {Koekemoer}, A.~M. and {Mignoli}, M. and {Silverman}, J.~D. and {Bolzonella}, M. and {Brusa}, M. and {Comastri}, A. and {Gilli}, R. and {Halliday}, C. and {Ilbert}, O. and {Lusso}, E. and {Salvato}, M. and {Vignali}, C. and {Zamorani}, G. and {Contini}, T. and {Kneib}, J.-P. and {Le F{\`e}vre}, O. and {Lilly}, S. and {Renzini}, A. and {Scodeggio}, M. and {Balestra}, I. and {Bardelli}, S. and {Caputi}, K. and {Coppa}, G. and {Cucciati}, O. and {de la Torre}, S. and {de Ravel}, L. and {Franzetti}, P. and {Garilli}, B. and {Iovino}, A. and {Kampczyk}, P. and {Knobel}, C. and {Kova{\v{c}}}, K. and {Lamareille}, F. and {Le Borgne}, J.-F. and {Le Brun}, V. and {Maier}, C. and {Nair}, P. and {Pello}, R. and {Peng}, Y. and {Perez Montero}, E. and {Pozzetti}, L. and {Ricciardelli}, E. and {Tanaka}, M. and {Tasca}, L. and {Tresse}, L. and {Vergani}, D. and {Zucca}, E. and {Aussel}, H. and {Capak}, P. and {Cappelluti}, N. and {Elvis}, M. and {Fiore}, F. and {Hasinger}, G. and {Impey}, C. and {Le Floc'h}, E. and {Scoville}, N. and {Taniguchi}, Y. and {Trump}, J.},
        title = "{Black hole accretion and host galaxies of obscured quasars in XMM-COSMOS}",
      journal = {\aap},
         year = 2011,
        month = nov,
       volume = {535},
          eid = {A80},
        pages = {A80},
          doi = {10.1051/0004-6361/201117259},
archivePrefix = {arXiv},
       eprint = {1105.5395},
 primaryClass = {astro-ph.CO},
       adsurl = {https://ui.adsabs.harvard.edu/abs/2011A&A...535A..80M}
}

@ARTICLE{Reddy2018,
       author = {{Reddy}, Naveen A. and {Shapley}, Alice E. and {Sanders}, Ryan L. and {Kriek}, Mariska and {Coil}, Alison L. and {Shivaei}, Irene and {Freeman}, William R. and {Mobasher}, Bahram and {Siana}, Brian and {Azadi}, Mojegan and {Fetherolf}, Tara and {Fornasini}, Francesca M. and {Leung}, Gene and {Price}, Sedona H. and {Zick}, Tom and {Barro}, Guillermo},
        title = "{The MOSDEF Survey: Significant Evolution in the Rest-frame Optical Emission Line Equivalent Widths of Star-forming Galaxies at z = 1.4-3.8}",
      journal = {\apj},
         year = 2018,
        month = dec,
       volume = {869},
       number = {2},
          eid = {92},
        pages = {92},
          doi = {10.3847/1538-4357/aaed1e},
archivePrefix = {arXiv},
       eprint = {1811.11767},
 primaryClass = {astro-ph.GA},
       adsurl = {https://ui.adsabs.harvard.edu/abs/2018ApJ...869...92R}
}

@ARTICLE{Somerville2015,
       author = {{Somerville}, Rachel S. and {Dav{\'e}}, Romeel},
        title = "{Physical Models of Galaxy Formation in a Cosmological Framework}",
      journal = {\araa},
         year = 2015,
        month = aug,
       volume = {53},
        pages = {51-113},
          doi = {10.1146/annurev-astro-082812-140951},
archivePrefix = {arXiv},
       eprint = {1412.2712},
 primaryClass = {astro-ph.GA},
       adsurl = {https://ui.adsabs.harvard.edu/abs/2015ARA&A..53...51S}
}

@ARTICLE{Feltre2016,
       author = {{Feltre}, A. and {Charlot}, S. and {Gutkin}, J.},
        title = "{Nuclear activity versus star formation: emission-line diagnostics at ultraviolet and optical wavelengths}",
      journal = {\mnras},
         year = 2016,
        month = mar,
       volume = {456},
       number = {3},
        pages = {3354-3374},
          doi = {10.1093/mnras/stv2794},
archivePrefix = {arXiv},
       eprint = {1511.08217},
 primaryClass = {astro-ph.GA},
       adsurl = {https://ui.adsabs.harvard.edu/abs/2016MNRAS.456.3354F}
}

@ARTICLE{Harrison2012,
       author = {{Harrison}, C.~M. and {Alexander}, D.~M. and {Swinbank}, A.~M. and {Smail}, Ian and {Alaghband-Zadeh}, S. and {Bauer}, F.~E. and {Chapman}, S.~C. and {Del Moro}, A. and {Hickox}, R.~C. and {Ivison}, R.~J. and {Men{\'e}ndez-Delmestre}, Kar{\'\i}n. and {Mullaney}, J.~R. and {Nesvadba}, N.~P.~H.},
        title = "{Energetic galaxy-wide outflows in high-redshift ultraluminous infrared galaxies hosting AGN activity}",
      journal = {\mnras},
         year = 2012,
        month = oct,
       volume = {426},
       number = {2},
        pages = {1073-1096},
          doi = {10.1111/j.1365-2966.2012.21723.x},
archivePrefix = {arXiv},
       eprint = {1205.1801},
 primaryClass = {astro-ph.CO},
       adsurl = {https://ui.adsabs.harvard.edu/abs/2012MNRAS.426.1073H}
}

@ARTICLE{Carniani2015,
       author = {{Carniani}, S. and {Marconi}, A. and {Maiolino}, R. and {Balmaverde}, B. and {Brusa}, M. and {Cano-D{\'\i}az}, M. and {Cicone}, C. and {Comastri}, A. and {Cresci}, G. and {Fiore}, F. and {Feruglio}, C. and {La Franca}, F. and {Mainieri}, V. and {Mannucci}, F. and {Nagao}, T. and {Netzer}, H. and {Piconcelli}, E. and {Risaliti}, G. and {Schneider}, R. and {Shemmer}, O.},
        title = "{Ionised outflows in z \raisebox{-0.5ex}\textasciitilde 2.4 quasar host galaxies}",
      journal = {\aap},
         year = 2015,
        month = aug,
       volume = {580},
          eid = {A102},
        pages = {A102},
          doi = {10.1051/0004-6361/201526557},
archivePrefix = {arXiv},
       eprint = {1506.03096},
 primaryClass = {astro-ph.GA},
       adsurl = {https://ui.adsabs.harvard.edu/abs/2015A&A...580A.102C}
}

@ARTICLE{Vayner2021,
       author = {{Vayner}, Andrey and {Zakamska}, Nadia L. and {Riffel}, Rogemar A. and {Alexandroff}, Rachael and {Cosens}, Maren and {Hamann}, Fred and {Perrotta}, Serena and {Rupke}, David S.~N. and {Bergmann}, Thaisa Storchi and {Veilleux}, Sylvain and {Walth}, Greg and {Wright}, Shelley and {Wylezalek}, Dominika},
        title = "{Powerful winds in high-redshift obscured and red quasars}",
      journal = {\mnras},
         year = 2021,
        month = jul,
       volume = {504},
       number = {3},
        pages = {4445-4459},
          doi = {10.1093/mnras/stab1176},
archivePrefix = {arXiv},
       eprint = {2101.04688},
 primaryClass = {astro-ph.GA},
       adsurl = {https://ui.adsabs.harvard.edu/abs/2021MNRAS.504.4445V}
}

@ARTICLE{Lau2024,
       author = {{Lau}, Marie Wingyee and {Perrotta}, Serena and {Hamann}, Fred and {Gillette}, Jarred and {Rupke}, David S.~N. and {Vayner}, Andrey and {Zakamska}, Nadia L. and {Wylezalek}, Dominika},
        title = "{[O III] {\ensuremath{\lambda}}5007 emissions in extremely red quasars (ERQs) are compact}",
      journal = {\mnras},
         year = 2024,
        month = aug,
       volume = {532},
       number = {2},
        pages = {2044-2064},
          doi = {10.1093/mnras/stae1621},
archivePrefix = {arXiv},
       eprint = {2312.03917},
 primaryClass = {astro-ph.GA},
       adsurl = {https://ui.adsabs.harvard.edu/abs/2024MNRAS.532.2044L}
}

@ARTICLE{Husemann2014,
       author = {{Husemann}, B. and {Jahnke}, K. and {S{\'a}nchez}, S.~F. and {Wisotzki}, L. and {Nugroho}, D. and {Kupko}, D. and {Schramm}, M.},
        title = "{Integral field spectroscopy of nearby QSOs - I. ENLR size-luminosity relation, ongoing star formation and resolved gas-phase metallicities}",
      journal = {\mnras},
         year = 2014,
        month = sep,
       volume = {443},
       number = {1},
        pages = {755-783},
          doi = {10.1093/mnras/stu1167},
archivePrefix = {arXiv},
       eprint = {1406.4131},
 primaryClass = {astro-ph.GA},
       adsurl = {https://ui.adsabs.harvard.edu/abs/2014MNRAS.443..755H}
}

@ARTICLE{Bergmann2018,
       author = {{Storchi-Bergmann}, T. and {Dall'Agnol de Oliveira}, B. and {Longo Micchi}, L.~F. and {Schmitt}, H.~R. and {Fischer}, T.~C. and {Kraemer}, S. and {Crenshaw}, M. and {Maksym}, P. and {Elvis}, M. and {Fabbiano}, G. and {Colina}, L.},
        title = "{Bipolar Ionization Cones in the Extended Narrow-line Region of Nearby QSO2s}",
      journal = {\apj},
         year = 2018,
        month = nov,
       volume = {868},
       number = {1},
          eid = {14},
        pages = {14},
          doi = {10.3847/1538-4357/aae7cd},
archivePrefix = {arXiv},
       eprint = {1810.06246},
 primaryClass = {astro-ph.GA},
       adsurl = {https://ui.adsabs.harvard.edu/abs/2018ApJ...868...14S}
}

@ARTICLE{Fischer2018,
       author = {{Fischer}, Travis C. and {Kraemer}, S.~B. and {Schmitt}, H.~R. and {Longo Micchi}, L.~F. and {Crenshaw}, D.~M. and {Revalski}, M. and {Vestergaard}, M. and {Elvis}, M. and {Gaskell}, C.~M. and {Hamann}, F. and {Ho}, L.~C. and {Hutchings}, J. and {Mushotzky}, R. and {Netzer}, H. and {Storchi-Bergmann}, T. and {Straughn}, A. and {Turner}, T.~J. and {Ward}, M.~J.},
        title = "{Hubble Space Telescope Observations of Extended [O III]{\ensuremath{\lambda}} 5007 Emission in Nearby QSO2s: New Constraints on AGN Host Galaxy Interaction}",
      journal = {\apj},
         year = 2018,
        month = apr,
       volume = {856},
       number = {2},
          eid = {102},
        pages = {102},
          doi = {10.3847/1538-4357/aab03e},
archivePrefix = {arXiv},
       eprint = {1802.06184},
 primaryClass = {astro-ph.GA},
       adsurl = {https://ui.adsabs.harvard.edu/abs/2018ApJ...856..102F}
}

@ARTICLE{Alberts2026,
       author = {{Alberts}, Stacey and {Eisenstein}, Daniel J. and {Bunker}, Andrew J. and {Curtis-Lake}, Emma and {Duan}, Qiao and {Hainline}, Kevin and {Hausen}, Ryan and {Helton}, Jakob M. and {Ji}, Zhiyuan and {Johnson}, Benjamin D. and {Lyu}, Jianwei and {Morrison}, Jane and {Perez-Gonzalez}, Pablo G. and {Rieke}, George H. and {Rieke}, Marcia and {Rinaldi}, Pierluigi and {Robertson}, Brant and {Sun}, Yang and {Tacchella}, Sandro and {Williams}, Christina C. and {Willmer}, Christopher N.~A. and {Wu}, Zihao},
        title = "{JWST Advanced Deep Extragalactic Survey (JADES) Data Release 5: MIRI Coordinated Parallels in GOODS-S and GOODS-N}",
      journal = {arXiv e-prints},
         year = 2026,
        month = jan,
          eid = {arXiv:2601.15955},
        pages = {arXiv:2601.15955},
          doi = {10.48550/arXiv.2601.15955},
archivePrefix = {arXiv},
       eprint = {2601.15955},
 primaryClass = {astro-ph.GA},
       adsurl = {https://ui.adsabs.harvard.edu/abs/2026arXiv260115955A}
}

@ARTICLE{Whitaker2019,
       author = {{Whitaker}, Katherine E. and {Ashas}, Mohammad and {Illingworth}, Garth and {Magee}, Daniel and {Leja}, Joel and {Oesch}, Pascal and {van Dokkum}, Pieter and {Mowla}, Lamiya and {Bouwens}, Rychard and {Franx}, Marijn and {Holden}, Bradford and {Labb{\'e}}, Ivo and {Rafelski}, Marc and {Teplitz}, Harry and {Gonzalez}, Valentino},
        title = "{The Hubble Legacy Field GOODS-S Photometric Catalog}",
      journal = {\apjs},
         year = 2019,
        month = sep,
       volume = {244},
       number = {1},
          eid = {16},
        pages = {16},
          doi = {10.3847/1538-4365/ab3853},
archivePrefix = {arXiv},
       eprint = {1908.05682},
 primaryClass = {astro-ph.GA},
       adsurl = {https://ui.adsabs.harvard.edu/abs/2019ApJS..244...16W}
}

@ARTICLE{Robertson2026,
       author = {{Robertson}, Brant E. and {Johnson}, Benjamin D. and {Tacchella}, Sandro and {Eisenstein}, Daniel J. and {Hainline}, Kevin and {Alberts}, Stacey and {Arribas}, Santiago and {Baker}, William M. and {Bunker}, Andrew J. and {Cameron}, Alex J. and {Carniani}, Stefano and {Carreira}, Courtney and {Chevallard}, Jacopo and {Circosta}, Chiara and {Curtis-Lake}, Emma and {Danhaive}, A. Lola and {Duan}, Qiao and {Egami}, Eiichi and {Hausen}, Ryan and {Helton}, Jakob M. and {Ji}, Zhiyuan and {Maiolino}, Roberto and {P{\'e}rez-Gonz{\'a}lez}, Pablo G. and {Pusk{\'a}s}, D{\'a}vid and {Rieke}, Marcia and {Rinaldi}, Pierluigi and {Sun}, Fengwu and {Sun}, Yang and {{\"U}bler}, Hannah and {Trussler}, James A.~A. and {Villanueva}, Natalia C. and {Whitler}, Lily and {Williams}, Christina C. and {Willmer}, Christopher N.~A. and {Willott}, Chris and {Wu}, Zihao and {Zhu}, Yongda},
        title = "{JWST Advanced Deep Extragalactic Survey (JADES) Data Release 5: Photometric Catalog}",
      journal = {arXiv e-prints},
         year = 2026,
        month = jan,
          eid = {arXiv:2601.15956},
        pages = {arXiv:2601.15956},
          doi = {10.48550/arXiv.2601.15956},
archivePrefix = {arXiv},
       eprint = {2601.15956},
 primaryClass = {astro-ph.GA},
       adsurl = {https://ui.adsabs.harvard.edu/abs/2026arXiv260115956R}
}

@ARTICLE{Hainline2026,
       author = {{Hainline}, Kevin N. and {Eisenstein}, Daniel J. and {Whitler}, Lily and {Robertson}, Brant and {Johnson}, Benjamin D. and {Jakobsen}, Peter and {Puskas}, David and {Tacchella}, Sandro and {Helton}, Jakob M. and {Wu}, Zihao and {Arribas}, Santiago and {Baker}, William M. and {Bunker}, Andrew J. and {Cameron}, Alex J. and {Carniani}, Stefano and {Carreira}, Courtney and {Charlot}, Stephane and {Chevallard}, Jacopo and {Curtis-Lake}, Emma and {D'Eugenio}, Francesco and {Duan}, Qiao and {Egami}, Eiichi and {Hausen}, Ryan and {Ji}, Zhiyuan and {Looser}, Tobias J. and {Maiolino}, Roberto and {Mengistu}, Petra and {Perez-Gonzalez}, Pablo G. and {Rieke}, Marcia and {Rinaldi}, Pierluigi and {Sun}, Fengwu and {Trussler}, James A.~A. and {Ubler}, Hannah and {Williams}, Christina C. and {Willmer}, Christopher N.~A. and {Willott}, Chris and {Witstok}, Joris},
        title = "{JWST Advanced Deep Extragalactic Survey (JADES) Data Release 5: Photometrically Selected Galaxy Candidates at z > 8}",
      journal = {arXiv e-prints},
         year = 2026,
        month = jan,
          eid = {arXiv:2601.15959},
        pages = {arXiv:2601.15959},
          doi = {10.48550/arXiv.2601.15959},
archivePrefix = {arXiv},
       eprint = {2601.15959},
 primaryClass = {astro-ph.GA},
       adsurl = {https://ui.adsabs.harvard.edu/abs/2026arXiv260115959H}
}

@ARTICLE{DEugenio2025,
       author = {{D'Eugenio}, Francesco and {Cameron}, Alex J. and {Scholtz}, Jan and {Carniani}, Stefano and {Willott}, Chris J. and {Curtis-Lake}, Emma and {Bunker}, Andrew J. and {Parlanti}, Eleonora and {Maiolino}, Roberto and {Willmer}, Christopher N.~A. and {Jakobsen}, Peter and {Robertson}, Brant E. and {Johnson}, Benjamin D. and {Tacchella}, Sandro and {Cargile}, Phillip A. and {Rawle}, Tim and {Arribas}, Santiago and {Chevallard}, Jacopo and {Curti}, Mirko and {Egami}, Eiichi and {Eisenstein}, Daniel J. and {Kumari}, Nimisha and {Looser}, Tobias J. and {Rieke}, Marcia J. and {Rodr{\'\i}guez Del Pino}, Bruno and {Saxena}, Aayush and {{\"U}bler}, Hannah and {Venturi}, Giacomo and {Witstok}, Joris and {Baker}, William M. and {Bhatawdekar}, Rachana and {Bonaventura}, Nina and {Boyett}, Kristan and {Charlot}, Stephane and {Danhaive}, A. Lola and {Hainline}, Kevin N. and {Hausen}, Ryan and {Helton}, Jakob M. and {Ji}, Xihan and {Ji}, Zhiyuan and {Jones}, Gareth C. and {Juod{\v{z}}balis}, Ignas and {Maseda}, Michael V. and {P{\'e}rez-Gonz{\'a}lez}, Pablo G. and {Perna}, Michele and {Pusk{\'a}s}, D{\'a}vid and {Shivaei}, Irene and {Silcock}, Maddie S. and {Simmonds}, Charlotte and {Smit}, Renske and {Sun}, Fengwu and {Villanueva}, Natalia C. and {Williams}, Christina C. and {Zhu}, Yongda},
        title = "{JADES Data Release 3: NIRSpec/Microshutter Assembly Spectroscopy for 4000 Galaxies in the GOODS Fields}",
      journal = {\apjs},
         year = 2025,
        month = mar,
       volume = {277},
       number = {1},
          eid = {4},
        pages = {4},
          doi = {10.3847/1538-4365/ada148},
archivePrefix = {arXiv},
       eprint = {2404.06531},
 primaryClass = {astro-ph.GA},
       adsurl = {https://ui.adsabs.harvard.edu/abs/2025ApJS..277....4D}
}

@ARTICLE{Malkan1982,
       author = {{Malkan}, M.~A. and {Sargent}, W.~L.~W.},
        title = "{The ultraviolet excess of Seyfert 1 galaxies and quasars.}",
      journal = {\apj},
         year = 1982,
        month = mar,
       volume = {254},
        pages = {22-37},
          doi = {10.1086/159701},
       adsurl = {https://ui.adsabs.harvard.edu/abs/1982ApJ...254...22M}
}

@ARTICLE{zhu2025,
       author = {{Zhu}, Yongda and {Rieke}, Marcia J. and {Ji}, Zhiyuan and {Simmonds}, Charlotte and {Sun}, Fengwu and {Sun}, Yang and {Alberts}, Stacey and {Bhatawdekar}, Rachana and {Bunker}, Andrew J. and {Cargile}, Phillip A. and {Carniani}, Stefano and {de Graaff}, Anna and {Hainline}, Kevin and {Helton}, Jakob M. and {Jones}, Gareth C. and {Lyu}, Jianwei and {Rieke}, George H. and {Rinaldi}, Pierluigi and {Robertson}, Brant and {Scholtz}, Jan and {{\"U}bler}, Hannah and {Williams}, Christina C. and {Willmer}, Christopher N.~A.},
        title = "{A Systematic Search for Galaxies with Extended Emission Lines and Potential Outflows in JADES Medium-band Images}",
      journal = {\apj},
         year = 2025,
        month = jun,
       volume = {986},
       number = {2},
          eid = {162},
        pages = {162},
          doi = {10.3847/1538-4357/add2f4},
archivePrefix = {arXiv},
       eprint = {2409.11464},
 primaryClass = {astro-ph.GA},
       adsurl = {https://ui.adsabs.harvard.edu/abs/2025ApJ...986..162Z}
}

@ARTICLE{Brown2019,
       author = {{Brown}, Arianna and {Nayyeri}, Hooshang and {Cooray}, Asantha and {Ma}, Jingzhe and {Hickox}, Ryan C. and {Azadi}, Mojegan},
        title = "{Infrared Contributions of X-Ray Selected Active Galactic Nuclei in Dusty Star-forming Galaxies}",
      journal = {\apj},
         year = 2019,
        month = jan,
       volume = {871},
       number = {1},
          eid = {87},
        pages = {87},
          doi = {10.3847/1538-4357/aaf73b},
archivePrefix = {arXiv},
       eprint = {1801.02233},
 primaryClass = {astro-ph.GA},
       adsurl = {https://ui.adsabs.harvard.edu/abs/2019ApJ...871...87B}
}

@ARTICLE{Alberts2024,
       author = {{Alberts}, Stacey and {Lyu}, Jianwei and {Shivaei}, Irene and {Rieke}, George H. and {P{\'e}rez-Gonz{\'a}lez}, Pablo G. and {Bonaventura}, Nina and {Zhu}, Yongda and {Helton}, Jakob M. and {Ji}, Zhiyuan and {Morrison}, Jane and {Robertson}, Brant E. and {Stone}, Meredith A. and {Sun}, Yang and {Williams}, Christina C. and {Willmer}, Christopher N.~A.},
        title = "{SMILES Initial Data Release: Unveiling the Obscured Universe with MIRI Multiband Imaging}",
      journal = {\apj},
         year = 2024,
        month = dec,
       volume = {976},
       number = {2},
          eid = {224},
        pages = {224},
          doi = {10.3847/1538-4357/ad7396},
archivePrefix = {arXiv},
       eprint = {2405.15972},
 primaryClass = {astro-ph.GA},
       adsurl = {https://ui.adsabs.harvard.edu/abs/2024ApJ...976..224A}
}

@ARTICLE{Calzetti2000,
       author = {{Calzetti}, Daniela and {Armus}, Lee and {Bohlin}, Ralph C. and {Kinney}, Anne L. and {Koornneef}, Jan and {Storchi-Bergmann}, Thaisa},
        title = "{The Dust Content and Opacity of Actively Star-forming Galaxies}",
      journal = {\apj},
         year = 2000,
        month = apr,
       volume = {533},
       number = {2},
        pages = {682-695},
          doi = {10.1086/308692},
archivePrefix = {arXiv},
       eprint = {astro-ph/9911459},
 primaryClass = {astro-ph},
       adsurl = {https://ui.adsabs.harvard.edu/abs/2000ApJ...533..682C}
}

@ARTICLE{Danhaive2026,
       author = {{Danhaive}, A. Lola and {Tacchella}, Sandro and {McClymont}, William and {Robertson}, Brant and {Carniani}, Stefano and {Carreira}, Courtney and {Egami}, Eiichi and {Bunker}, Andrew J. and {Curtis-Lake}, Emma and {Eisenstein}, Daniel J. and {Ji}, Zhiyuan and {Johnson}, Benjamin D. and {Rieke}, Marcia and {Villanueva}, Natalia C. and {Willmer}, Christopher N.~A. and {Willot}, Chris and {Wu}, Zihao and {Zhu}, Yongda},
        title = "{Beyond the stars: Linking H{\ensuremath{\alpha}} sizes, kinematics, and star formation in galaxies at z ≍ 4 - 6 with JWST grism surveys and GEKO}",
      journal = {\mnras},
         year = 2026,
        month = mar,
          doi = {10.1093/mnras/stag437},
archivePrefix = {arXiv},
       eprint = {2510.06315},
 primaryClass = {astro-ph.GA},
       adsurl = {https://ui.adsabs.harvard.edu/abs/2026MNRAS.tmp..413D}
}

@ARTICLE{Lyu2025,
       author = {{Lyu}, Yipeng and {Magnelli}, Benjamin and {Elbaz}, David and {P{\'e}rez-Gonz{\'a}lez}, Pablo G. and {Correa}, Camila and {Daddi}, Emanuele and {G{\'o}mez-Guijarro}, Carlos and {Dunlop}, James S. and {Grogin}, Norman A. and {Koekemoer}, Anton M. and {McLeod}, Derek J. and {Lu}, Shiying},
        title = "{PRIMER: JWST/MIRI reveals the evolution of star-forming structures in galaxies at z {\ensuremath{\leq}} 2.5}",
      journal = {\aap},
         year = 2025,
        month = jan,
       volume = {693},
          eid = {A313},
        pages = {A313},
          doi = {10.1051/0004-6361/202451067},
archivePrefix = {arXiv},
       eprint = {2406.11571},
 primaryClass = {astro-ph.GA},
       adsurl = {https://ui.adsabs.harvard.edu/abs/2025A&A...693A.313L}
}

@ARTICLE{Martorano2024,
       author = {{Martorano}, Marco and {van der Wel}, Arjen and {Baes}, Maarten and {Bell}, Eric F. and {Brammer}, Gabriel and {Franx}, Marijn and {Nersesian}, Angelos},
        title = "{The Size─Mass Relation at Rest-frame 1.5 {\ensuremath{\mu}}m from JWST/NIRCam in the COSMOS-WEB and PRIMER-COSMOS Fields}",
      journal = {\apj},
         year = 2024,
        month = sep,
       volume = {972},
       number = {2},
          eid = {134},
        pages = {134},
          doi = {10.3847/1538-4357/ad5c6a},
archivePrefix = {arXiv},
       eprint = {2406.17756},
 primaryClass = {astro-ph.GA},
       adsurl = {https://ui.adsabs.harvard.edu/abs/2024ApJ...972..134M}
}
\bibliographystyle{aasjournal}

\end{document}